\documentclass{article}

\usepackage{PRIMEarxiv}

\usepackage[utf8]{inputenc} 
\usepackage[T1]{fontenc}    
\usepackage{hyperref}       
\usepackage{url}            
\usepackage{booktabs}       
\usepackage{amsfonts}       
\usepackage{nicefrac}       
\usepackage{microtype}      
\usepackage{lipsum}
\usepackage{xcolor}
\usepackage{fancyhdr}       
\usepackage{graphicx}       
\graphicspath{{media/}}     
\usepackage{authblk}
\newcommand{\rev}[1]{\textcolor{black}{#1}}
\title{Linguistics and Human Brain: A Perspective of Computational Neuroscience 
}

\author[1,2]{Fudong Zhang}
\author[2]{Bo Chai}
\author[3]{Yujie Wu}
\author[2,*]{Wai Ting Siok}
\author[2,*]{Nizhuan Wang}

\affil[1]{Institute of AI and Robotics, College of Intelligent Robotics and Advanced Manufacturing, Fudan University, Shanghai 200433, China}
\affil[2]{Department of Language Science and Technology, The Hong Kong Polytechnic University, Hung Hom, Kowloon,
	Hong Kong SAR, China}
\affil[3]{Department of Computing,  The Hong Kong Polytechnic University, Hung Hom, Kowloon, Hong Kong SAR, China}
\affil[*]{Correspondence: wai-ting.siok@polyu.edu.hk, wangnizhuan1120@gmail.com}

\begin{document}
\maketitle

\begin{abstract}
Elucidating the language–brain relationship requires bridging the methodological gap between linguistics’ abstract theoretical frameworks and neuroscience’s empirical neural data. As an interdisciplinary cornerstone, computational neuroscience formalizes language’s hierarchical and dynamic structures into testable neural representation models through modeling, simulation, and data analysis, enabling computational dialogue between linguistic hypotheses and neural mechanisms. Recent advances in deep learning, particularly large language models (LLMs), have further advanced this inquiry: their high-dimensional representational spaces provide a new scale for probing the neural basis of linguistic processing, the model–brain alignment framework offers a principled approach to evaluating the biological plausibility of language-related theories, provided that representational correspondence is interpreted together with behavioral, temporal, causal, and biological constraints.

This review synthesizes interdisciplinary progress from a computational neuroscience perspective. First, it outlines the core connotations of major linguistic frameworks (generative grammar, functional linguistics, and cognitive linguistics), their cross-cultural and evolutionary characteristics, and key challenges for neural alignment, including limited quantitative mechanisms, poor accessibility of abstract constructs to neural measures, and insufficient treatment of dynamics and plasticity. Second, it introduces the methodological foundations of linguistics–neuroscience dialogue, focusing on four technical pillars: neural activity measurement (e.g., fMRI, EEG, MEG, fNIRS, ECoG, SEEG), linguistic numerical representation, the evolution of language models from statistical approaches to LLMs, and neural coding frameworks that link model representations to brain signals, illustrated with a model–brain alignment case study. Third, it summarizes major findings, ranging from early computational insights into predictability and structural processing to recent LLM-driven progress in cross-modal interaction, inter-brain coupling, hierarchical computation, learning strategy sensitivity, and language plasticity. Finally, the review discusses current limitations—including functional alignment without structural homology, constraints on real-time validation, biased research coverage, and narrow evaluation metrics—and proposes future directions, such as exploring whether spiking neural network--based language models can improve biological plausibility in settings requiring temporally precise and event-driven neural modeling, developing cognitive-level alignment frameworks integrating memory, causality, and metacognition, and extending clinical applications. 

In summary, this work aims to advance a comprehensive, mechanistic understanding of the language-brain relationship and promote computational neuroscience as a generative theoretical framework for testable neuro-computational accounts of language.
\end{abstract}

\begin{center}
\footnotesize This paper has been accepted for publication in \textit{Cognitive Neurodynamics}, 2026.
\end{center}

\section{Introduction}
Language is conceptualized as a multi-layered abstract symbolic system encompassing distinct yet interconnected structures ranging from sounds (phonetics and phonology) to word formation (morphology), sentence structure (syntax), and meaning (semantics) \cite{chomsky2002syntactic,chomsky1993lectures}. Using a finite set of discrete elements and combinatorial rules, it can generate an infinite array of expressions, enabling the flexible and precise transmission of meaning -- a feature known as recursion \cite{boboqulovna2025linguistic,fitch2010evolution,hauser2002faculty,chomsky2014minimal}. Linguistics seeks to formalize these implicit rules and structures in order to uncover the cognitive architecture underlying human linguistic competence. In parallel, neuroscience investigates how coordinated neural activity across brain circuits supports language production and comprehension \cite{indefrey2004spatial,silbert2014coupled}. A persistent interdisciplinary challenge, however, arises from the methodological and explanatory divide between these fields. Abstract linguistic theories are often formulated as symbolic, hierarchical systems that are difficult to map directly onto the dynamic, distributed patterns of neural activity observed through neuroimaging or electrophysiology, partly due to the inherent limitations of these techniques. Conversely, neural data alone often lack the computational interpretability needed to account for the structured, rule-governed nature of language. This gap between theoretical description and empirical evidence limits a comprehensive understanding of language and its neural basis.

Beyond methodological mismatches, a more fundamental difficulty lies in the nature of the neural system itself. The neural system for language is a complex, adaptive, and dynamic network, composed of billions of neurons that exhibit intricate connectivity and continuously evolving plasticity \cite{friederici2011brain,christiansen2008language}. Within this network, linguistic information is processed through hierarchical, parallel, and recurrent interactions, supported by bidirectional inter-regional connections that are dynamically regulated by contexts and task demands \cite{demonet2005renewal,nasios2019broca}. In formal terms from mathematics and systems science, such a system is described as an adaptive complex dynamic system \cite{lansing2003complex} where global behavior cannot be reduced to the sum of its local parts, and it often exhibits nonlinear, self-organizing, and emergent properties. Consequently, neither purely linguistic models nor isolated neural observations can fully explain the integrated mechanisms of human language processing.

As a well-founded discipline dedicated to decoding neural mechanisms, computational neuroscience serves as a crucial bridge between linguistics and neuroscience. It integrates linguistics, neuroscience, computer science, and systems theory to convert formal linguistic hypotheses into testable computational models, which are then tested against neural data \cite{hopfield1982neural,hale2022neurocomputational}. The core methodology involves constructing models that represent linguistic structure while simulating neural dynamics, which are then validated or revised using brain imaging and electrophysiological evidence \cite{lopopolo2024cognitive,kriegeskorte2018cognitive}. Ever since generative linguistics posited a neural basis for linguistic competence \cite{lenneberg1967biological}, computational modeling has become a central tool for evaluating the neural plausibility of linguistic theories.

Recent advances in artificial intelligence, particularly in deep learning and Large Language Models (LLMs), have accelerated this integration further. Researchers can now explore linguistic structures and their correspondence with brain networks within higher-dimensional representational spaces \cite{sun2020neural,shain2020fmri,wang2020fine}. LLMs provide computational platforms for examining modern linguistic phenomena such as semantic integration, long-distance dependencies, and predictive processing. They also support the emerging framework of model–brain alignment, wherein internal model representations are used to predict and explain aspects of neural responses during language processing \cite{deniz2019representation,anderson2017predicting,koehler2025compositional}. Importantly, such predictive alignment is best interpreted as evidence that LLMs capture representational dimensions relevant to human language processing, while further behavioral, temporal, causal, and biological evidence is needed to support stronger mechanistic claims. This development marks a shift from early conceptual modeling toward quantitative mappings between large-scale linguistic data and high-resolution neural signals.

This paper provides a systematic review of how computational neuroscience serves as a methodological bridge between linguistics and neuroscience. Section 2 surveys modern linguistic frameworks and their development across languages, identifying the core challenges that have historically hindered their alignment with neural data. Section 3 introduces key methodological foundations, including neural data acquisition, word embeddings, modern language models, and neural coding frameworks. Section 4 reviews applications of classical computational neuroscience models to language comprehension, followed by a synthesis of recent progress in LLM-driven neural alignment, covering domains including cross-modal representation, cross-brain coupling, hierarchical computation, and learning mechanisms. Section 5 analyzes current methodological limitations and theoretical challenges, and outlines promising future research directions, such as enhancing biological plausibility, developing cognitive-level alignment frameworks, and expanding clinical and brain–computer interface applications. Finally, Section 6 summarizes the main conclusions and discusses the long-term significance of computational neuroscience for fostering a deeper integration of linguistics and neuroscience.

\section{Theoretical Foundations of Linguistics}

\begin{figure}[!tbhp]
	\centering
	\includegraphics[width=1\textwidth]{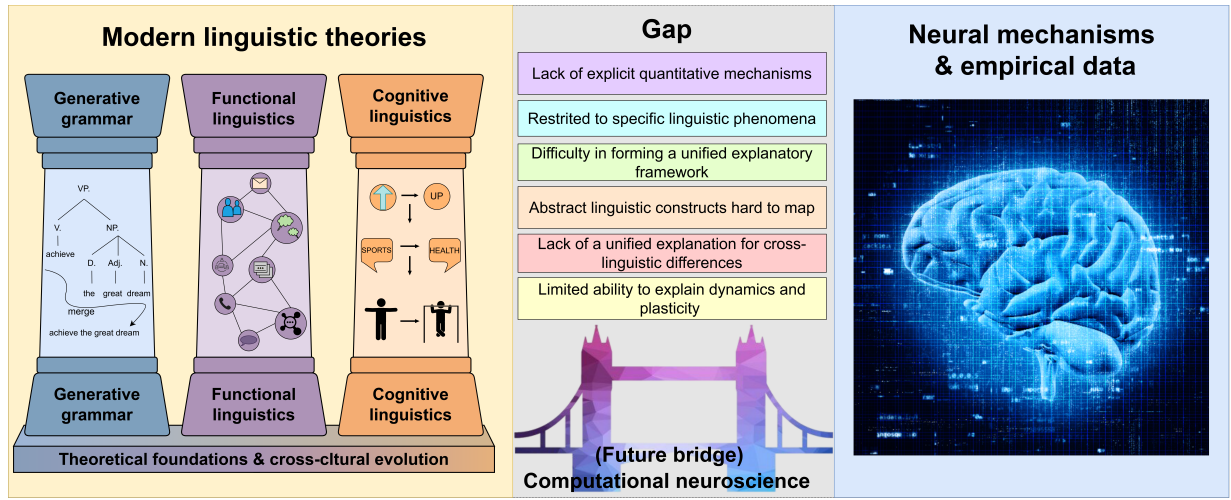}
	\caption{Modern Linguistic Theories and Challenges for Bridging Linguistics and Neuroscience.}
	\label{fig1_1}
\end{figure}

Language has long been the core research object defining the intersection of linguistics and neuroscience. Over time, linguistic theories have continued to develop through modeling, cross-cultural research, and empirical investigation. Yet they have also gradually revealed structural limitations that hinder their direct alignment with the neural mechanisms of the brain. Modern linguistic theories offer clear, organized ways to describe the structure and function of language. But when faced with the diversity of languages across cultures, the dynamic nature of everyday language use, and growing evidence from brain science, their ability to fully explain language remains limited \cite{evans2009myth,ding2016cortical,mitchell2008predicting,grossberg2000complementary}. In this section, we review major linguistic perspectives and their evolution across cultural contexts, identifying the key challenges that motivate interdisciplinary integration. This review provides the theoretical background for later discussions on computational neuroscience as a bridging framework, as outlined in Fig. \ref{fig1_1}.
\subsection{Frameworks of Modern Linguistic Theories}

Modern linguistics comprises several influential theoretical frameworks. Among the most influential are generative grammar, functional linguistics, and cognitive linguistics, each offering a distinct perspective on language structure and use.

Generative grammar, most prominently associated with Noam Chomsky, posits linguistic competence as an innate, biologically grounded component of the human cognitive system \cite{chomsky2002syntactic}. This framework is founded on the hypothesis of a Universal Grammar, arguing that children's acquisition of complex grammar must be guided by innate principles, given the impoverished and limited nature of their linguistic input. This "poverty of the stimulus" argument is taken as primary evidence for internally constrained, domain-specific grammatical principles. The distinction between deep structure (abstract syntactic relations) and surface structure (observable linguistic forms) illustrates how underlying representations are transformed into observable linguistic expressions \cite{chomsky2014aspects,jackendoff1972semantic}. Later developments, most notably the Minimalist Program, sought to reduce linguistic theory to a minimal set of core computational operations, with Merge as the fundamental process, to enhance its cognitive and biological plausibility \cite{lasnik2002minimalist}. Nevertheless, these highly abstract, formal constructs have proven difficult to map onto specific, observable patterns of neural activity, which remains a source of sustained debate in the neurobiology of language. 

In contrast, functional linguistics treats language primarily as a system for social communication. Its central tenet is that linguistic structure is shaped and constrained by communicative needs and contextual factors \cite{urban1981language}. A major framework within this tradition, Systemic Functional Linguistics, posits that language simultaneously fulfills three core metafunctions: the ideational (for construing experience), the interpersonal (for enacting social relations), and the textual (for organizing discourse). Within this framework, linguistic choices are fundamentally motivated by a speaker's communicative goals and the dynamics of social interaction \cite{halliday2013halliday}. Structural differences between language modalities, such as the prevalence of ellipsis in spoken conversation versus the structural completeness expected in formal writing, are analyzed as adaptive outcomes of the distinct functional demands of each modality \cite{eggins2004introduction}. While functional linguistics has been highly influential in discourse analysis, sociolinguistics, and cross-cultural communication studies, its strong emphasis on usage, context, and meaning poses significant challenges for strict formalization. This inherent difficulty with formal modeling, in turn, complicates the establishment of direct, mechanistic links between functional explanations and underlying neural processes. 

Cognitive linguistics explicitly rejects the generative conception of language as an autonomous, innate cognitive module. Instead, it argues that language is grounded in, and emerges from, general cognitive processes such as categorization, metaphorical mapping, and mental simulation \cite{lakoff2008metaphors}. Within this framework, Conceptual Metaphor Theory proposes that abstract domains (e.g., time, mind) are understood through systematic mappings from more concrete, embodied experiences, and that grammatical structures often reflect these underlying conceptual representations \cite{kovecses2010metaphor}. Embodied cognition accounts extend this view, proposing that language comprehension is inherently tied to the reactivation of sensorimotor experiences and dynamic mental simulation \cite{barsalou2010grounded}. While this perspective offers a powerfully unified account of language and cognition, many of its core constructs, such as conceptual metaphors or mental simulations, are difficult to operationalize and quantify experimentally. This lack of form specification consequently limits the framework's ability to generate precise, testable predictions that can be directly compared with neural data. 

\rev{Overall, generative grammar, functional linguistics, and cognitive linguistics should not be viewed merely as three descriptive traditions, but as distinct hypotheses about the computational organization of language and its possible neural implementation. Generative grammar emphasizes abstract, structure-dependent operations and therefore predicts that neural language systems should be sensitive to hierarchical syntactic relations, recursion, and long-distance dependencies beyond surface-level word sequence statistics \cite{brennan2012syntactic,ding2016cortical}. Functional linguistics, by contrast, predicts that language-related neural activity should be strongly modulated by communicative goals, discourse context, pragmatic demands, and speaker–listener interaction, suggesting that language processing depends on broader networks for contextual integration, attention, prediction, and social cognition \cite{stephens2010speaker,hasson2012brain}. Cognitive linguistics further differs from both approaches by treating linguistic meaning as grounded in domain-general conceptual and sensorimotor systems; it therefore predicts distributed and multimodal neural representations involving perception, action, affect, memory, and conceptual simulation.}

\rev{These differences have direct implications for computational neuroscience. A generative account would be supported if formal syntactic hierarchy or structural complexity explains neural variance beyond lexical predictability and semantic similarity. A functional account would gain support if discourse-level, pragmatic, and interactional variables improve prediction of neural responses during naturalistic communication. A cognitive-linguistic account would be strengthened if embodied, sensorimotor, or multimodal features improve alignment with neural activity during semantic comprehension. Thus, the central question is not simply whether linguistic theories can be mapped onto the brain, but which theoretical constructs can be operationalized as computational variables and tested against neural data \cite{caucheteux2022brains}. This perspective motivates the subsequent discussion of neural measurement, language representation, and model--brain alignment as tools for moving from descriptive linguistic theory toward comparative, testable neurocomputational explanation.}

\subsection{Cross-Cultural Evolution of Linguistic Theories}

The existence of thousands of languages worldwide exhibits remarkable diversity in linguistic structure and expression. This diversity has continuously driven the revision and expansion of linguistic theory, while also underscoring the challenge of constructing a unified explanatory framework, raising questions about whether genuinely universal linguistic features can be abstracted at all \cite{newmeyer2005possible,croft2002typology}.

Early generative grammar was primarily developed on the basis of a small number of languages. To account for typological variation, it later introduced parameter-setting mechanisms to explain differences in word order, morphological structure, and subject realization \cite{chomsky1993lectures}. However, for languages with highly complex structures or those that differ substantially from Indo-European languages, parameter-based explanations remain controversial. This suggests that a limited set of parameters may be insufficient to capture the full richness of linguistic diversity \cite{newmeyer2005possible}.

In functional linguistics, cross-linguistic variation is examined in relation to the communicative and cultural contexts in which language is used. Honorific systems, context-dependent expressions, and discourse-level differences are treated not as peripheral but as core objects of grammatical analysis \cite{halliday2013halliday}. For example, cultures differ in their emphasis on politeness, indirectness, and social hierarchy, which in turn shape linguistic forms \cite{bartlett2017routledge}. Although this perspective helps explain cultural variation, its strong reliance on specific cultural contexts makes it difficult to extract universally applicable principles.

Cognitive linguistics examines the interaction between universal embodied experience and culturally specific conceptualizations \cite{lakoff2008metaphors}. Many spatial and emotional metaphors are shared across languages (e.g., time is conceived along a front-back axis, and emotion is metaphorically mapped onto a vertical axis, with happiness represented as "up" and sadness as "down"), suggesting shared cognitive constraints. At the same time, domains such as spatial reference (with systems based on body coordinates like left/right/front/back or on cardinal directions like north/south/east/west), color categorization (with the number and boundaries of basic colour terms varying across languages), and kinship terminology (with varying distinctions between nuclear and extended family) exhibit clear cultural specificity \cite{kovecses2005metaphor}. The coexistence of universality and diversity indicates that language is shaped by both biological constraints and cultural experience.

Cross-linguistic research thus demonstrates that language structure reflects shared cognitive foundations as well as cultural history \cite{christiansen2018creating}. However, this twofold origin poses a challenge for determining which linguistic properties stem from universal neural mechanisms and which are products of cultural transmission. Evidence from cross-script studies of developmental dyslexia further illustrates this complexity: reading impairments in alphabetic versus logographic systems engage partially distinct neural circuits, suggesting that the neural correlates of dyslexia reflect both universal constraints and writing-system-specific adaptations \cite{siok2004biological}. Such nuanced findings underscore the need for closer integration between theoretical linguistics and neurobiologically grounded models of language. These considerations also have important implications for model--brain alignment research. \rev{If alignment studies are based predominantly on a small number of high-resource languages, it becomes difficult to determine whether observed correspondences reflect general principles of human language processing or properties tied to particular linguistic types, writing systems, or sociocultural contexts \cite{joshi2020state}. From this perspective, typological diversity is not only a theoretical issue for linguistics, but also a necessary condition for evaluating the scope and generalizability of computational accounts of the language--brain relationship. Recent multilingual neuroimaging resources and studies make this issue increasingly testable. For example, multilingual naturalistic fMRI datasets based on comparable narrative materials allow researchers to examine whether neural language representations are shared across languages or shaped by language-specific morphosyntactic and writing-system properties \cite{li2022petit,chen2026bilingual}. Such datasets are especially valuable because they move cross-linguistic neuroscience beyond isolated word or sentence contrasts toward more ecologically valid language comprehension. Cross-language brain decoding further provides a computational framework for testing whether neural models trained in one language can generalize to another, thereby evaluating whether semantic and syntactic representations are language-general or language-specific. These approaches are directly relevant to debates about universal grammar, typological diversity, and the neural generalizability of model--brain alignment.
}

\subsection{Challenges in Modern Linguistic Theories}

The theoretical distinctions outlined above also clarify why the neural alignment of linguistic theories remains difficult: each framework highlights different explanatory variables, but these variables differ substantially in how easily they can be formalized, quantified, and linked to measurable neural signals. With advances in neuroimaging techniques and computational methods, language research has increasingly moved beyond purely theoretical analysis toward empirical integration with neural data \cite{hale2022neurocomputational}. In this interdisciplinary context, major linguistic theories have revealed several structural challenges. These challenges limit direct links between linguistic constructs and neural mechanisms while motivating the development of new computational and neural modeling approaches \cite{van2023cleaning}. Broadly speaking, these difficulties include a lack of quantitative mechanisms, limited research scope, and poor integrability of findings.

\begin{enumerate}
\item[i.]\textbf{Lack of explicit quantitative mechanisms}

Most linguistic and cognitive language studies remain at the level of phenomenological description and qualitative explanation. The computational mechanisms underlying language processing are often underspecified. For example, during lexical processing, the N400 event-related potential (ERP) is frequently observed, and numerous studies have shown that its amplitude is modulated by contextual information \cite{dambacher2006frequency,brothers2015effects}. However, such findings typically demonstrate the existence of contextual effects without clarifying the underlying computations.

Several interpretations have been proposed in the cognitive science literature. Some link the N400 to lexical predictability, suggesting that more predictable words elicit smaller responses. Others interpret it as reflecting semantic integration difficulty \cite{michaelov2024strong}. Until recently, these accounts remained largely conceptual, lacking specification of how prediction and integration are implemented computationally within neural circuits. While recent computational models have begun to address this gap, no unifying theory has yet achieved consensus, and the challenge of precise theory-data correspondence remains central to the field.

\item[ii.]\textbf{Research restricted to specific linguistic phenomena}

Language research often relies on highly controlled experimental designs to ensure reliability and interpretability \cite{verbeke2024role}. Such designs typically focus on decontextualized or localized linguistic phenomena using tightly constrained, uniform stimuli. As a result, findings may not generalize to the richness and variability of natural language use.

A further challenge lies in the long-standing misalignment of research scales between linguistics and neuroscience \cite{cutler2005twenty}. Linguistic studies often examine fine-grained structural distinctions, whereas neuroscience research typically focuses on broader functional organization across brain regions. Even in interdisciplinary work, experiments predominantly rely on simplified or prototypical materials, limiting ecological validity and generalizability.

\item[iii.]\textbf{Difficulty in forming a unified explanatory framework}

While controlled experiments are effective at identifying local effects, they also contribute to fragmented findings. Individual studies typically target specific phenomena, yet language itself resists decomposition into a small set of independent units. Differences in task design, stimulus materials, and measurement techniques further hinder direct integration across studies. For instance, distinct types of linguistic stimuli may elicit divergent activation patterns or electrophysiological responses. However, aggregating such results rarely yields a coherent account of how the brain incrementally constructs the meaning of a complete sentence \cite{hickok2007cortical}.

\item[iv.]\textbf{Abstract linguistic constructs are difficult to map onto neural measures}

Many central linguistic concepts, such as recursion, thematic roles, and conceptual metaphor, are highly abstract and lack clear operational definitions, making them difficult to translate into measurable neural variables \cite{poeppel2016maps}. Moreover, neural activity is distributed across large-scale networks rather than confined to single regions, further complicating attempts to establish simple, one-to-one mappings between linguistic functions and specific neural indicators \cite{drijvers2025language}.

\item[v.]\textbf{Lack of a unified explanation for cross-linguistic differences}

There is no consensus on whether cross-linguistic structural differences arise primarily from innate mechanisms or from cultural learning and experience. These divergent theoretical positions make it difficult for neuroimaging studies to determine whether observed processing differences reflect biological predispositions or experiential adaptation. This uncertainty hinders the development of a unified neurocognitive model of language.

\item[vi.]\textbf{Limited ability to explain dynamics and plasticity in language processing}

Major linguistic theories are largely based on static sentence analysis and struggle to account for the dynamic aspects of real-world communication, including real-time prediction, contextual updating, and social interaction processes under naturalistic conditions \cite{pickering2004toward,pickering2018predicting}. In addition, human language abilities show substantial plasticity during child acquisition, second language learning, and recovery after brain injury. These theoretical frameworks lack a unified account of these developmental and adaptive changes, as well as of individual differences in language processing.
\end{enumerate}

Taken together, abstract linguistic theories alone are insufficient for establishing direct links between language and neural mechanisms. Bridging this gap requires translating linguistic constructs into computable models and testing them quantitatively against neural data. Computational neuroscience has emerged in this context as a key platform for connecting linguistic theory with brain mechanisms. By integrating neural data, computational modeling, and large-scale models of language processing, this field is promoting a shift in language research from descriptive accounts toward mechanistic explanations and laying a critical foundation for subsequent work on brain-inspired models and advances in language processing.

\begin{figure}[!htbp]
	\centering
	\includegraphics[width=1\textwidth]{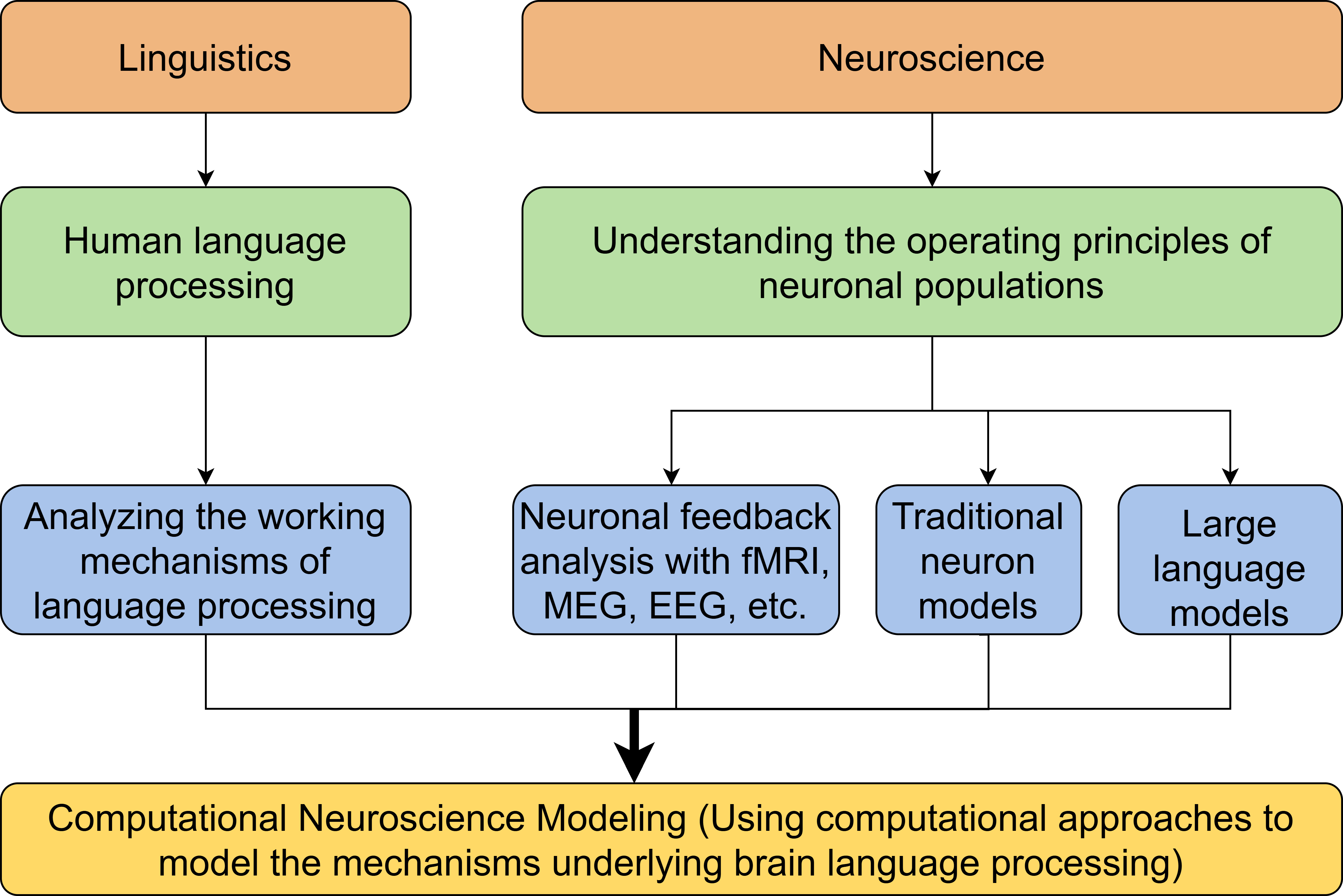}
	\caption{The Association between Linguistics and Neuroscience from the Perspective of Computational Neuroscience.}
	\label{fig2_1}
\end{figure}
\section{Computational Neuroscience Methods for Linguistics}
Computational neuroscience provides a methodological bridge between formal linguistic theory and empirical neural data. As illustrated in Fig.~\ref{fig2_1}, it transforms abstract linguistic structures into computational representations that can be quantitatively compared with neural activity through a set of interrelated tools and analytical frameworks. This section introduces the core components of this bridge. We first outline the methodology of computational neuroscience and then describe four technical pillars: \textbf{Measurement Methods} (neural activity recording), \textbf{Word Embeddings} (numerical representations of language), \textbf{Evolving Language Models}, and \textbf{Neural Coding} (analytical frameworks that link model representations to brain signals). Finally, we present an integrative case study to illustrate how these components operate jointly and support model--brain alignment research in linguistics and neuroscience.

\subsection{Methodology in Computational Neuroscience}
As an interdisciplinary field, computational neuroscience adopts diverse and integrative methodological strategies. Rather than pursuing a single unifying theory, it typically develops modular and composable models to study \rev{complex systems such as language and the brain} \cite{healey2015toward}. Model evaluation emphasizes explanatory and predictive power with respect to neural data, formal parsimony, and the ability to generate testable hypotheses \cite{levenstein2023role}. From this perspective, modeling approaches to language research have traditionally been grouped into three categories:
\begin{itemize}
	\item \textbf{Descriptive models}, which quantify observed neural activity, including spike train statistics and population decoding methods \cite{dayan2005theoretical,rieke1999spikes,olshausen2004sparse,pouget2013probabilistic,frank2015erp}.
	\item \textbf{Normative theories}, which explain neural computation through functional optimization principles, such as efficient coding and Bayesian inference \cite{barlow1961possible,simoncelli2001natural,knill2004bayesian,kording2004bayesian,heilbron2022hierarchy}.
	\item \textbf{Biological simulation models}, which aim to reproduce neural structure and dynamics at varying levels of biological detail \cite{markram2006blue,eliasmith2012large,brennan2016abstract}.
\end{itemize}

The emergence of deep learning models, particularly LLMs, challenges this traditional classification. LLMs are not designed to fit neural data or obey biological constraints. Nevertheless, their strong correspondence with human language behavior, representational structure, and neural responses positions them as a distinctive mesoscale computational reference frame \cite{kriegeskorte2018cognitive,schrimpf2021neural,an2025hierarchical,goldstein2022shared,caucheteux2022brains,wang2024computational,cai2024large,caucheteux2023evidence,radford2018improving,oniani2024emerging}. Their central contribution lies in offering an explicit, hierarchical representational space that enables systematic investigation of how linguistic information may be organized in the brain \cite{doerig2025high}.

These developments motivate a new framework of multi-scale integration. Within this framework, LLMs characterize linguistic structure at the computational level, cognitive interpretability is introduced at the algorithmic level, and biological constraints are progressively incorporated at the implementation level. Accordingly, the primary research goal shifts from fitting neural data toward establishing principled connections between expressive computational models and neurobiological mechanisms. This shift supports progress from black-box modeling toward mechanistic explanation.
\begin{table}[htbp]
	\centering
	\caption{Core components and their roles in model-brain alignment research.}
	\label{tableIV}

	\begin{tabular}{p{3.0cm}p{3.0cm}p{2.5cm}p{3.0cm}}
		\toprule 
		\textbf{Core Components} & \textbf{Core Forms/Technologies} & \textbf{Core Functions} & \textbf{Roles in Alignment} \\
		\midrule 
		Measurement Methods & fMRI, EEG, MEG & Recording neural activity signals & Providing neural reference data \\
		\midrule 
		Word Embedding & High-dimensional semantic vector mapping & Numerical representation of linguistic units & Laying the foundation for linguistic representation \\
		\midrule 
		Evolving Language Models & n-gram, Transformer architecture, etc. & Learning intrinsic linguistic rules & Providing computational reference frame \\
		\midrule 
		Neural Coding & Regression/Neural network models & Establishing representation-neural mapping & Quantifying model-brain matching degree \\
		\bottomrule 
	\end{tabular}
\end{table}

Guided by this methodological perspective, the following sections introduce the four technical pillars of this framework in detail and clarify their respective roles in model–brain alignment. Their respective roles and interactions in model--brain alignment research are summarized in Table~\ref{tableIV}.

\subsection{Brain Activity Measurement for Linguistic Tasks}
Testing linguistic theories within a computational framework requires reliable measurements of neural activity during language processing. Research on the neural basis of language therefore employs multiple techniques that capture neural responses across distinct spatial, temporal, and representational scales \cite{hickok2007cortical,friederici2013language,turker2023cortical,karunathilake2025neural,gross2019magnetoencephalography}. Because language processing emerges from dynamic interactions across distributed brain regions and multiple scales, no single method is sufficient. Instead, measurement techniques are selected and combined according to the research focus, such as spatial localization, temporal resolution, or representational specificity.
\begin{table}[htbp]
	\centering
	\caption{Comparison of characteristics and applicable scenarios of mainstream techniques for detecting neural mechanisms of language.}
	\label{tableIII}
	
	\begin{tabular}{p{2.0cm}p{2.0cm}p{2.5cm}p{2.5cm}p{1.5cm}}
		\toprule
		\textbf{Measurement Techniques} & \textbf{Spatial Resolution} & \textbf{Temporal Resolution} & \textbf{Core Advantages} &  \textbf{Limitations} \\
		\midrule
		fMRI     & Millimeter-level    & Second-level          & Precise brain region localization                     & Unable to capture millisecond-scale dynamics \\
		\midrule
		EEG      & Centimeter-level    & Millisecond-level     & Capturing temporal dynamics                      & Ambiguous spatial localization \\
		\midrule
		MEG      & Sub-centimeter-level& Millisecond-level     & Balanced spatiotemporal resolution                        & High equipment cost \\
		\midrule
		fNIRS    & Centimeter-level    & Hundred-millisecond-level & \rev{Robustness to motion artifacts }                  & Limited penetration depth \\
		\midrule
		ECoG     & Millimeter-level    & Sub-millisecond-level & High SNR, fine cortical spatial precision       & Invasive, only for clinical patients \\
		\midrule
		SEEG     & Millimeter-level    & Sub-millisecond-level & Deep structure 3D localization, high precision   & Invasive, surgical risk, limited coverage \\
		\bottomrule
	\end{tabular}
\end{table}

In this section, we introduce six widely used neural recording modalities in language research. Their core characteristics and methodological trade-offs are summarized in Table~\ref{tableIII}.

\begin{enumerate}
	\item[i.] \textbf{fMRI (functional Magnetic Resonance Imaging)}: fMRI is a core non-invasive technique for measuring brain function based on the Blood Oxygen Level–Dependent (BOLD) effect \cite{worsley1996unified,ogawa1990brain}. Neural activation increases metabolic oxygen consumption, which triggers compensatory increases in local cerebral blood flow. The resulting influx of oxygenated blood exceeds immediate neuronal demand, reducing local deoxyhemoglobin concentration. Because deoxyhemoglobin is paramagnetic whereas oxyhemoglobin is diamagnetic, these changes modify local magnetic resonance signals, allowing indirect imaging of regional neural activation.  

	The main advantage of fMRI is its high spatial resolution, which enables accurate localization of cortical and subcortical regions involved in language processing, including left perisylvian areas (Broca’s area, Wernicke’s area, supramarginal gyrus, angular gyrus) and subcortical structures such as the caudate nucleus and thalamus. It can also distinguish spatial activation patterns associated with different linguistic levels such as phonological, lexical, semantic, and syntactic processing \cite{binder1997human,price2012review,blank2014functional,vigneau2006meta,wu2023occupation}. Its primary limitation is low temporal resolution: the BOLD response is delayed by approximately 4–8 seconds, with signal peaks typically occurring 6–8 seconds after neuronal activation. Consequently, millisecond-scale processes such as rapid semantic integration cannot be directly captured \cite{glover1999deconvolution,matthews2004functional}. In addition, fMRI measurements are sensitive to physiological noise from cardiac and respiratory activity, which complicates experiments involving infants or patients with language disorders. A further limitation is that the BOLD signal is an indirect correlate of neural activity, reflecting hemodynamic changes rather than neuronal firing itself. Overall, fMRI offers high spatial resolution but limited temporal resolution \cite{bhattasali2020alice}.
	
	\item[ii.] \textbf{EEG (Electroencephalography)}: EEG records weak electric field fluctuations generated by the synchronous firing of large neuronal populations using scalp electrode arrays \cite{li2025tale}. Its key advantage is millisecond-level temporal resolution, which enables precise tracking of rapid neural dynamics during language processing, including responses related to semantic and syntactic analysis. However, electrical signals are attenuated and spatially blurred when they pass through brain tissue, skull, and scalp, resulting in poor spatial localization and significant difficulty in identifying precise neural sources \cite{nunez2006electric,makeig1995independent}. EEG is therefore highly sensitive to synchronized cortical activity but provides limited spatial precision \cite{luck2014introduction}. This positions EEG as a complement to fMRI, which offers high spatial resolution but limited temporal resolution.

A commonly used EEG-based analysis approach in language research is the Event-Related Potential (ERP). ERP is obtained by time-locking EEG signals to stimulus onset and averaging across repeated presentations, thereby extracting stimulus-related neural responses while reducing ongoing background activity and noise \cite{kutas1980reading}. Because ERP is derived from EEG rather than recorded independently, it inherits EEG's millisecond temporal resolution while sharing its spatial limitations. ERP has played a central role in language research by characterizing the temporal course of processing from early sensory analysis to lexical, semantic, and syntactic integration. Among ERP components, the N400 and P600 are the most widely studied. The N400 is strongly associated with semantic processing and typically increases when words are semantically incongruent or unpredictable in context, whereas the P600 is primarily linked to syntactic processing and often increases in response to syntactic violations or elevated structural complexity \cite{swaab2012language,osterhout1992event,kutas2011thirty}.
	
	\item[iii.] \textbf{MEG (Magnetoencephalography)}: MEG measures weak magnetic fields generated by neuronal electrical activity using highly sensitive sensors -- traditionally Superconducting Quantum Interference Devices (SQUIDs) \cite{cohen1968magnetoencephalography,hamalainen1993magnetoencephalography} and, more recently, Optically Pumped Magnetometers (OPMs). Synchronous neuronal currents produce magnetic fields that can be detected outside the head, providing an indirect measure of neural activity. Like EEG, MEG offers millisecond temporal resolution and can track rapid processing stages from phonetic perception to syntactic and semantic integration.  

	Compared with EEG, MEG generally provides better spatial localization because magnetic fields are minimally distorted by skull and scalp tissues, reducing uncertainties associated with EEG inverse problem \cite{hansen2010meg}. This combination of high temporal resolution and improved (though still limited) spatial accuracy makes MEG particularly suitable for studies requiring precise spatiotemporal characterization of language processes, such as phonetic-to-lexical conversion and syntactic processing \cite{salmelin2009electromagnetic,schmidt2009second}. However, MEG is primarily sensitive to tangential cortical sources (e.g., in sulci such as the superior temporal and intraparietal sulci) but is largely insensitive to radial sources (e.g., gyral crowns of the inferior frontal gyrus, angular gyrus, and motor cortex) and to subcortical structures. Source localization also remains an ill-posed inverse problem, and spatial resolution, though better than EEG, remains limited to the order of millimeters to centimeters.

	\item[iv.] \textbf{fNIRS (functional Near-Infrared Spectroscopy)}: fNIRS is a non-invasive optical imaging technique that measures brain activity through changes in the absorption of near-infrared light by oxygenated and deoxygenated hemoglobin \cite{quaresima2012brief,tsunashima2012measurement}. Near-infrared light emitted from scalp-mounted optodes penetrates superficial cortical tissue and is differentially absorbed by oxyhemoglobin and deoxyhemoglobin. Detectors capture diffusely reflected light, and concentration changes are estimated using the modified Beer--Lambert law, allowing indirect inference of neural activity. Like fMRI, fNIRS relies on hemodynamic responses associated with underlying neural activation.  

	fNIRS offers several practical advantages. The equipment is portable and does not require shielded environments, enabling experiments in more natural communication settings. It is also generally less sensitive to motion artifacts than fMRI and, in some contexts, more robust than EEG, making it particularly suitable for studies involving infants, children, or patients with language disorders, as well as for tasks involving overt speech or facial movements \cite{villringer1997non}. Its temporal resolution lies between fMRI and EEG/MEG, allowing tracking of sub-second hemodynamic changes during tasks such as lexical access and sentence comprehension. Spatial resolution is typically on the order of centimeters, sufficient for localizing broad cortical language regions, though not for resolving fine-grained functional organization \cite{ferrari2012brief,scholkmann2014review}.  

	Limitations include lower spatial resolution than fMRI and limited penetration depth, which precludes reliable measurement of subcortical structures. fNIRS signals are also influenced by individual physiological factors such as scalp thickness, hair properties, and superficial (extracerebral) blood flow, requiring careful experimental design to ensure comparability across participants \cite{leon2012functional,pinti2020present}. Although whole-head coverage is now technically feasible with modular high-density arrays, it remains less common than localized cortical applications due to increased demands on hardware, setup time, and data processing, fNIRS is therefore more suitable for targeted investigations of cortical regions rather than comprehensive brain mapping \cite{scholkmann2014review}.
    \item[v.] \textbf{ECoG (Electrocorticography)}: ECoG is an invasive neurophysiological technique that records electrical potentials directly from the exposed cortical surface using subdural electrode grids or strips placed during surgery \cite{hill2012recording}. Unlike scalp EEG, ECoG avoids signal attenuation and distortion by the skull and scalp, yielding high signal amplitude, excellent spatial resolution, and sub-millisecond temporal precision \cite{kanth2020electrocorticogram}. These characteristics make it uniquely capable of resolving fine-grained, region-specific cortical dynamics underlying language processing, including articulatory planning, phonemic encoding, lexical access, and syntactic binding.
ECoG is widely used in presurgical mapping for patients with epilepsy or brain tumors, allowing direct localization of essential language areas such as Broca’s region, superior temporal gyrus, and inferior parietal cortex with high clinical and scientific reliability \cite{de2025intraoperative}. High-frequency broadband  power modulations in ECoG are particularly sensitive to local neuronal population activity and correlate strongly with task-specific language computations. Key limitations include its invasive nature, restricted clinical eligibility, limited spatial coverage determined by electrode placement, and inherent ethical constraints that preclude use in healthy participants.
\item[vi.] \textbf{SEEG (Stereo‐Electroencephalography)}: SEEG is an invasive, stereotactic technique that uses multiple thin depth electrodes implanted through small burr holes to record electrical activity from deep brain structures and cortical regions, including those buried in sulci, while traversing white matter tracts. This enables precise three-dimensional mapping of the epileptogenic zone \cite{youngerman2019stereoelectroencephalography}. Unlike ECoG, which samples from the cortical surface, SEEG provides three-dimensional access to both cortical and subcortical targets critical for language control, semantic memory, and pragmatic integration \cite{aron2021language}. It combines sub-millisecond temporal resolution with precise three‐dimensional localization, enabling characterization of spatiotemporal dynamics with language networks during both overt and covert processing \cite{zhao2024spatiotemporal}. SEEG is extensively used in epilepsy presurgical evaluation to map language-related networks and identify epileptogenic zones while preserving essential functions. Similar to ECoG, SEEG is restricted to clinical populations, involves surgical risk, and offers limited whole-brain sampling density \cite{parvizi2018promises}. Nevertheless, its unique ability to record from deep and hidden structures makes SEEG irreplaceable for investigating subcortical‐cortical circuits supporting language comprehension, production, and monitoring.
\end{enumerate}
Although less commonly available in language research, single-unit and multi-unit recordings obtained through microelectrode arrays provide the most direct access to spike-resolved neural activity. These methods offer exceptionally fine temporal precision and can reveal how individual neurons or small neuronal populations encode stimulus features, sequence information, and task-related dynamics. In the context of language research, such recordings remain rare because they are typically restricted to highly specialized clinical or experimental settings and often provide limited spatial coverage. Nevertheless, they are methodologically important because they offer the empirical scale most directly relevant for evaluating spike-based computational models, including spiking neural networks (SNNs) \cite{bellec2021fitting}, and can therefore complement field-level and population-level measures such as EEG, MEG, ECoG, and fMRI.

\subsection{Word Embeddings}
Numerical representations of language provide a critical bridge between linguistics, computer science, and neuroscience. Word embeddings, also referred to as distributed word representations, have become a foundational tool in Natural Language Processing (NLP) and in computational studies of neural language mechanisms \cite{mikolov2013efficient}. The central idea is to map discrete words onto a continuous, high-dimensional vector space, where semantic similarity is reflected in geometric relations among vectors. This mapping enables symbolic linguistic units to be processed quantitatively by computational models. Unlike traditional one-hot encoding, which merely indicates word identity and captures no semantic relationships, word embeddings represent words through distributed, low-dimensional numerical features learned from distributional patterns in text. These representations capture rich semantic and syntactic associations and have significantly improved performance across a wide range of language tasks, from lexical similarity judgments to machine translation and brain encoding models \cite{harris1954distributional}.

Word embeddings are typically learned using machine learning or deep learning models and are grounded in the distributional hypothesis, which holds that word meaning is derived from patterns of contextual usage \cite{sahlgren2008distributional}. Based on training objectives and model architectures, mainstream approaches fall into two broad categories. The first comprises static embedding models such as Word2Vec and GloVe. Word2Vec learns word vectors by predicting local context using either the Skip-gram or CBOW architecture, while GloVe integrates global word-co-occurrence statistics with local context information. Both methods generate fixed vector representations that capture semantic similarity and analogical relations among words \cite{pennington2014glove,church2017word2vec}. The second category consists of context-dependent embeddings produced by modern Transformer-based pre-trained language models. These models dynamically produce context-sensitive representations, addressing polysemy (a limitation static embeddings cannot resolve) and substantially enhancing semantic representation quality \cite{koroteev2021bert,achiam2023gpt}. All such approaches rely on large-scale unlabeled corpora and learn patterns of word co-occurrence that place semantically related words in close proximity within the vector space, providing effective foundations for a wide range of downstream language tasks.

Researchers have extended lexical embeddings to larger linguistic units such as phrases, sentences, and discourses to represent complex structures numerically.  \rev{This extension is directly related to the problem of semantic composition: how discrete lexical meanings are combined into structured representations of phrases and sentences. In model–brain alignment research, semantic composition is important because it provides a computationally tractable bridge between linguistic theories of combinatorial meaning and neural measurements of sentence-level comprehension.} Existing approaches can be grouped into three types. The first uses simple arithmetic composition, typically averaging or summing constituent word vectors \cite{mitchell2010composition}. These methods are computationally efficient and were widely used in early sentence-level tasks. However, by treating words as independent units, they ignore word order, semantic roles, and syntactic structure. As a result, expressions with opposite meanings but identical vocabularies may receive similar representations, limiting semantic precision \cite{pagliardini2018unsupervised,cichy2019deep}. The second category employs structured composition methods, such as tensor product representations \cite{smolensky1990tensor}. These approaches combine word embeddings through tensor operations to encode syntactic and combinatorial relations explicitly, preserving hierarchical and relational information within linguistic structures. Although this framework offers principled structured representations, high-dimensional tensor operations significantly increase computational cost and limit large-scale applications \cite{mitchell2008vector,milajevs2014evaluating,clark2008compositional}.  The third category includes task-driven dynamic representation methods that learn sentence encoders through supervised downstream tasks \cite{conneau2017supervised}. The assumption is that optimizing task performance encourages models to encode semantic and structural information automatically. However, empirical studies show that models often exploit superficial statistical patterns rather than genuine semantic structure, leading to reduced robustness and poor generalization under syntactic variation or domain shifts \cite{naik2018stress,belinkov2019analysis}.

Word embeddings play a central role for three main reasons. First, they connect computational linguistics with neurolinguistics \cite{landauer1997solution}. The distributional principle underlying embeddings is broadly consistent with the brain’s distributed representation of lexical semantics, and similarities in embedding space often correspond to similarities in neural activity patterns measured with techniques such as fMRI and MEG \cite{mitchell2008predicting}. Understanding these representations therefore facilitates direct comparison between computational and neural semantic structures. Second, embeddings provide the numerical input foundation for modern language models. Natural language is symbolic and discrete, and must be transformed into continuous numerical representations before it can be processed by neural networks. From early static embeddings to the context-dependent representations of LLMs, embedding quality strongly influences model performance \cite{devlin2019bert}. Since later sections examine deep learning models for neural simulation, reviewing embedding principles clarifies how linguistic information enters computational models. Third, research on extending embeddings to complex expressions informs studies of neural representations of higher-level linguistic structure \cite{baroni2010nouns}. This work addresses not only lexical processing but also sentence- and discourse-level representations. The strengths and weaknesses of existing extension methods, particularly regarding syntactic preservation and semantic robustness, mirror challenges in modeling hierarchical neural language processing. Reviewing these methods therefore provides conceptual guidance for investigating how the brain encodes complex linguistic structures.

In summary, word embeddings and their extensions provide the core framework for numerical language representation. They have driven rapid progress in NLP while also offering essential tools and theoretical insights for interdisciplinary research on the neural mechanisms of language.

\subsection{The Evolving Language Models}
Language models are central tools at the intersection of computational neuroscience and linguistics. By learning statistical regularities from large text corpora, they extract latent semantic and syntactic information and generate outputs that approximate human language conventions. Their core computational objective is to predict lexical sequences based on preceding or surrounding context.  

Research on language models is important on both theoretical and practical grounds. Theoretically, their development parallels efforts to understand the neurocognitive mechanisms underlying human language processing. Evaluating how well these models capture contextual dependencies and hierarchical structure provides quantitative benchmarks for studying the computational operations of the brain’s language network. Practically, language models form the backbone of many NLP applications, including machine translation, text generation, semantic analysis. Improving their alignment with human cognition is therefore widely regarded as an important step toward more general and robust artificial intelligence systems.

\begin{table}[htbp]
	\centering
	\caption{Characteristic comparison of the evolving language models.}
	\label{tableI}
	
	\begin{tabular}{p{2.5cm}p{2.0cm}p{2.0cm}p{2.0cm}p{2.5cm}}
		\toprule
		\textbf{Model Type} & \textbf{Core Architecture} & \textbf{Linguistic Hierarchy Modeling Capability} & \textbf{Representative Models} & \textbf{Limitations} \\
		\midrule
		n-gram & Statistical co-occurrence probability & No hierarchical structure & Trigram model & Poor generalization ability \\
		\midrule
		PCFG & Explicit modeling of syntactic rules & Supports long-distance dependency modeling &  Collins parser, Berkeley parser & Weak rule generalization \\
		\midrule
		RNNs & Sequential recurrence, gating mechanisms & Implicitly captures long-distance dependencies & LSTM, GRU & Low serial computing efficiency \\
		\midrule
		Transformer/LLMs & Attention mechanism, parallel computation & Supports ultra-long context processing & GPT, BERT & Lack of biological structural constraints \\

		\bottomrule
	\end{tabular}
\end{table}

The development of language models has progressed from statistical approaches to neural architectures, with many design principles influenced by insights from neuroscience. Major model families include n-gram models, Probabilistic Context-Free Grammars (PCFG), Recurrent Neural Networks (RNNs), Transformer architectures, pre-trained  LLMs and LLM-based \rev{agents}. Successive generations have improved contextual modeling and representation capacity, ultimately enabling the emergence of LLMs. Their main characteristics are summarized in Table~\ref{tableI}, and briefly introduced below.

\begin{enumerate}
	\item[i.] \textbf{n-gram}: n-gram models represent the earliest form of language modeling. They predict a word based on the co-occurrence statistics of a fixed number of preceding words. For example, a trigram model conditions predictions on the previous two tokens \cite{martin2009speech}. Although computationally simple and efficient, these models cannot represent hierarchical structure or capture long-distance dependencies, which limits their ability to model complex linguistic phenomena.
	
	\item[ii.] \textbf{PCFG}: Probabilistic Context-Free Grammars explicitly model hierarchical syntactic structure and can, in principle, capture certain long-distance dependencies. However, both PCFG and n-gram approaches suffer from limited generalization: they cannot exploit semantic similarity across words or structural similarity across contexts. As a result, they have been largely replaced by neural network models, which learn distributed representations and achieve substantially better generalization performance \cite{jelinek1991computation}.
	
	\item[iii.] \textbf{RNNs}: RNNs represent the first generation of neural language models. They encode contextual information by mapping words into embedding spaces and updating hidden states sequentially as each new input is processed \cite{elman1990finding}. Although RNNs do not explicitly encode hierarchical structure, their recurrent connections allow them, in principle, to capture contextual dependencies, including long-distance relations. In practice, however, standard RNNs suffer from vanishing gradients, which limits their ability to learn very long-range dependencies. Long Short-Term Memory (LSTM) networks are a widely used variant of RNNs designed to mitigate the vanishing gradient problem through gating mechanisms. This architecture substantially improves long-range dependency modeling and became the dominant approach in early neural language modeling systems \cite{hochreiter1997long}. Beyond their historical role in the evolution of language models, RNNs are especially relevant to research on temporally unfolding language processing. Because they update internal states incrementally as linguistic input arrives, they provide a natural computational framework for modeling online comprehension, prediction, and memory updating under sequential constraints. In this respect, recurrent models occupy an important intermediate position between earlier statistical approaches, which captured local sequential regularities, and later Transformer-based systems, which achieved broader contextual integration through attention. RNNs therefore helped establish the idea that language representations could be modeled as continuously updated internal states rather than only as fixed symbolic structures or short-range conditional probabilities. This sequential and state-based character also made RNNs particularly relevant to earlier brain-alignment research. Before the rise of contemporary Transformer-based LLMs, recurrent architectures provided one of the first neural modeling frameworks that could be compared more directly with temporally resolved neural data during naturalistic language processing. Their emphasis on incremental context accumulation and dynamic hidden-state updating made them conceptually well suited to studies of word-by-word comprehension, temporally unfolding prediction, and neural encoding during speech and sentence processing \cite{toneva2019interpreting,brennan2016naturalistic,moses2016neural}. Although later Transformer-based approaches surpassed RNNs in representational scale and performance, recurrent models remain an important conceptual bridge linking classical sequential models with contemporary large-scale alignment frameworks.
	
	\item[iv.] \textbf{Transformers}: Transformer architectures have become the dominant framework for language modeling. Unlike RNNs, which process inputs sequentially, Transformers employ attention mechanisms that allow each token to interact with all others simultaneously \cite{vaswani2017attention}. This enables parallel computation and efficient modeling of global context. While Transformers do not explicitly encode linguistic hierarchy, dynamic attention weighting and scalable parallel computation allow them to model long-range dependencies effectively in practice. These properties have also enabled large-scale training on massive corpora, forming the technical foundation for modern LLMs.
	
	\item[v.] \textbf{Pre-trained LLMs}: Pre-trained LLMs represent the current framework in NLP \cite{brown2020language}. Their core training strategy follows a two-stage pipeline: large-scale self-supervised pre-training on massive unlabeled corpora, followed by fine-tuning or instruction adaptation for downstream tasks. Pre-training allows models to acquire syntactic patterns, semantic associations, and long-range dependencies from large textual datasets, while fine-tuning adapts these capabilities to specific applications. Owing to large parameter scales and extensive training data, these models significantly surpass earlier approaches in language understanding and generation. A key advantage is the separation between general language learning and task adaptation, enabling transfer across tasks without retraining from scratch.

	Current LLM systems can be broadly categorized based on architecture and training objectives:

	\begin{itemize}
		\item \textit{Autoregressive LLMs}: These models generate text sequentially from left to right by predicting each token based on preceding context. The GPT series is the most prominent example, with parameter scales growing from millions to trillions and training corpora spanning diverse web and literary sources \cite{radford2018improving,achiam2023gpt,brown2020language,radford2019language,ouyang2022training}. Such models excel at text generation, dialogue, and long-form continuation tasks. Related systems include ERNIE Bot and the GLM family. Their primary strength lies in fluent generation, although strictly unidirectional modeling limits access to full bidirectional context \cite{sun2019ernie,du2022glm,zeng2025glm,hong2025glm}.
		
		\item \textit{Autoencoding Pre-trained Models}: These models use masked token prediction and thus implement bidirectional language modeling. BERT is the most influential example, achieving major improvements in tasks requiring semantic understanding, including classification and reading comprehension \cite{devlin2019bert}. Variants such as RoBERTa, ALBERT, and SpanBERT further refine training strategies and architectures. However, such models are less suitable for long-form text generation \cite{liu2019roberta,lan2019albert,joshi2020spanbert}.
		
		\item \textit{Encoder–Decoder Hybrid LLMs}: These models combine bidirectional encoding with autoregressive decoding, balancing semantic understanding and generation. Representative systems include T5 and multilingual variants such as mT5. They perform well on tasks requiring both comprehension and generation, such as translation and summarization \cite{radford2019language,raffel2020exploring,xue2021mt5}.
		
		\item \textit{Multimodal Ultra-large LLMs}: Recent systems extend text-only models to incorporate visual, audio, and video inputs, enabling multimodal reasoning and generation. Models such as Gemini integrate cross-modal information and expand capabilities from text processing to multimodal cognition and reasoning, marking an important direction toward more general AI systems \cite{team2023gemini,team2024gemini,comanici2025gemini}.
        
	\end{itemize}

\end{enumerate}

\subsection{Neural Coding}
Neural encoding, neural decoding, and Representational Similarity Analysis (RSA) are central analytical frameworks for linking computational models with neural mechanisms of language processing. Together, they provide standardized tools for quantifying correspondence between computational representations and neural activity patterns \cite{kriegeskorte2008representational,naselaris2011encoding}. Broadly defined, neural coding methods construct mappings between stimulus representations and neural responses using statistical or machine learning models, allowing neural activity elicited by stimuli to be predicted computationally \cite{mitchell2008predicting}. In language research, stimuli include not only spoken or written inputs but also internally generated semantic and syntactic representations. These stimuli are typically represented numerically using language models trained on large corpora, enabling quantitative comparison with neural recordings \cite{kietzmann2017deep}.

RSA provides a complementary framework for model--brain alignment by comparing the internal similarity structure of representations rather than directly predicting neural responses. In a typical RSA analysis, researchers first compute Representational Dissimilarity Matrices (RDMs) for model-derived features and for neural activity patterns across a shared set of stimuli, and then evaluate the correspondence between these matrices using correlation-based measures \cite{kriegeskorte2008representational,nili2014toolbox}. This approach is particularly useful when the goal is to compare representational geometry across different models, brain regions, or temporal stages of processing without assuming a direct one-to-one mapping between specific model units and neural signals. In contrast to neural encoding, which emphasizes predictive mappings from model representations to brain activity, RSA emphasizes the relational structure of representations and is therefore widely used to assess whether computational models and neural systems organize linguistic information in similar ways. \rev{At the same time, RSA remains a measure of functional correspondence rather than mechanistic identity, and thus should be interpreted as evidence of representational alignment rather than proof of shared computational implementation.}

Neural coding analyses generally follow three steps. First, stimulus representations are paired with recorded neural signals to train mapping models, often using regression or shallow neural networks. Second, trained models predict neural responses to previously unseen stimuli. Third, predicted and measured responses are compared using metrics such as correlation or prediction error to evaluate model performance. Higher predictive accuracy indicates stronger alignment between model representations and neural mechanisms of language processing \cite{huth2016natural,naselaris2009bayesian}.

Neural coding plays a critical role in evaluating computational language models. Standard NLP metrics such as perplexity or task accuracy do not directly reflect biological plausibility. Neural coding instead measures whether model representations correspond to neural activity patterns, providing evidence about whether models capture cognitively relevant features rather than merely exploiting surface statistics. It therefore offers a quantitative bridge between computational modeling and neural mechanisms of language processing \cite{mitchell2008predicting,ettinger2020bert}.

Neural coding is also essential for assessing alignment between LLMs and human neural language systems \cite{goldstein2025temporal}. Although LLMs achieve impressive performance, it remains unclear whether their internal representations match brain mechanisms. Neural coding provides a quantitative test: if model-derived representations accurately predict neural responses, representational alignment is supported. Conversely, weak prediction suggests that models rely on processing strategies distinct from human cognition. Neural coding thus functions both as an evaluation tool for cognitive plausibility and as a guide for developing more brain-aligned language models \cite{xiao2025exploring}. \rev{It is also important to distinguish representational alignment from full mechanistic equivalence. High neural predictivity or representational similarity suggests that a model captures information dimensions relevant to brain language processing, such as lexical predictability, semantic structure, or contextual integration. However, such findings do not necessarily imply that the model reproduces the same algorithms, learning history, or biological implementation as the human brain. For this reason, neural coding results are most informative when combined with behavioral evidence, temporal dynamics, causal perturbation, and comparisons against simpler linguistic or statistical baselines.}

\begin{figure}[!htbp]
	\centering
	\includegraphics[width=1\textwidth]{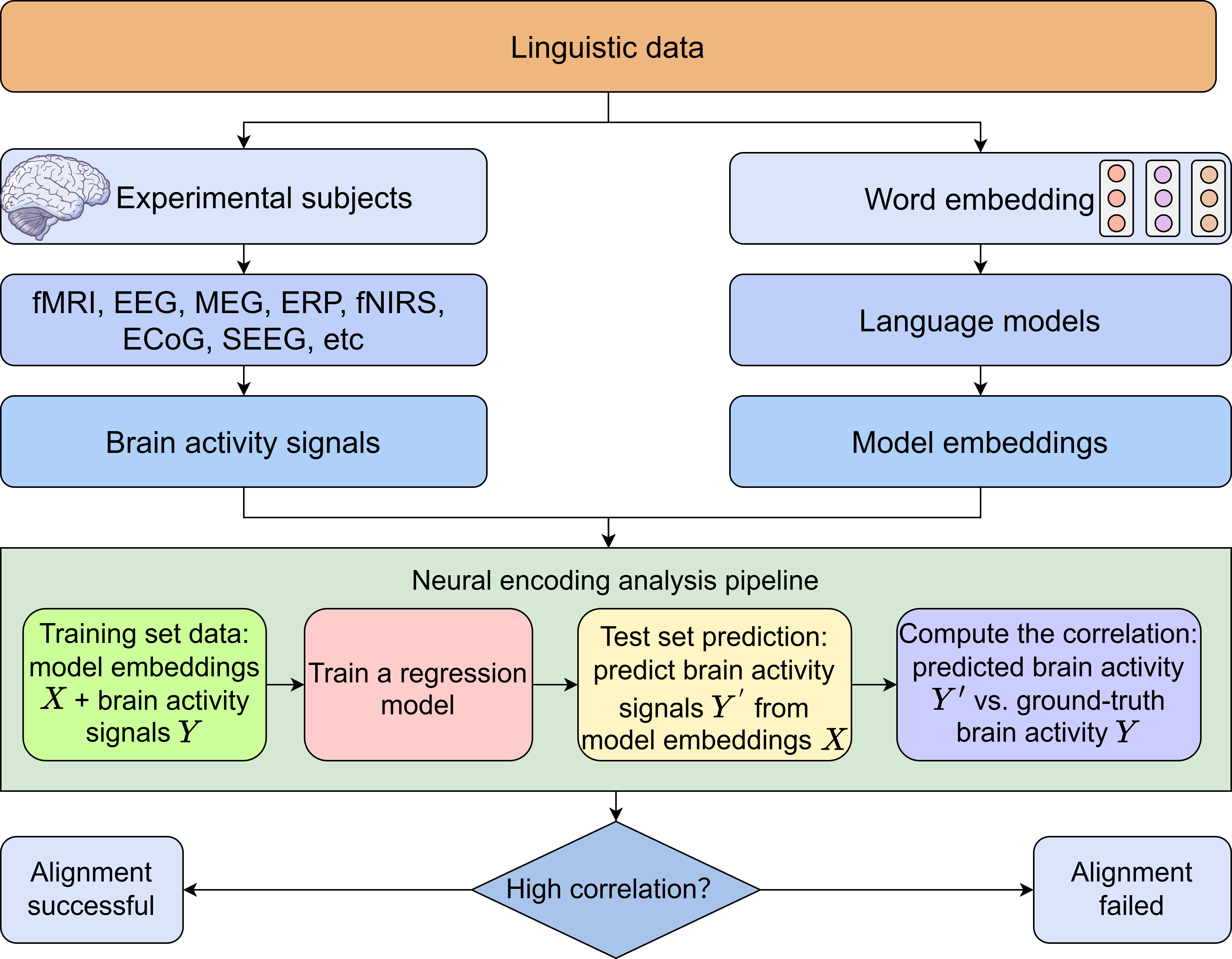}
	\caption{Computational Neuroscience Workflow of Each Component Collaboration in the Linguistic Case Study.}
	\label{fig2_2}
\end{figure}

\rev{\subsection{Model–Brain Alignment Workflow in Computational Neuroscience of Language}}
The methods introduced above function as an integrated framework rather than as isolated components (Fig.~\ref{fig2_2}). To illustrate their interaction, consider a representative alignment study testing whether LLM representations predict neural activity during sentence reading.

\begin{enumerate}
	\item[i.] \textbf{Neural data acquisition:} A participant reads sentences sequentially during fMRI scanning. Whole-brain BOLD responses are recorded and used as the empirical neural data for alignment analysis.

	\item[ii.] \textbf{Computational representation:} The same sentences are processed by a pre-trained LLM. Words are converted into embeddings and passed through multiple Transformer layers, yielding representations at different levels of abstraction. Lower layers typically emphasize lexical and syntactic information, whereas deeper layers tend to capture broader semantic structure. Activation vectors extracted from different layers serve as candidate computational representations.
	
	\item[iii.] \textbf{Model--brain mapping:} Sentence data are divided into training and test sets. For a given model layer, activation vectors from the training set are used to predict neural responses in a selected region of interest (ROI), typically with a regression-based encoding model. The trained model is then applied to the test set, and prediction accuracy is evaluated by comparing predicted and observed neural responses, for example using Pearson's correlation $r$.
	
	\item[iv.] \textbf{Alignment interpretation:} Higher prediction accuracy indicates stronger correspondence between model representations and neural activity under the tested stimulus and task conditions. Such correspondence provides useful evidence that the model captures representational features relevant to language processing, but its mechanistic interpretation depends on whether the effect generalizes across tasks, survives appropriate control models, and converges with behavioral and temporal evidence. Comparing results across layers helps identify which representational level best aligns with a given brain region. For example, stronger alignment between intermediate model layers and semantic regions would suggest that these layers capture semantic information relevant to neural processing, while stronger claims about shared mechanisms require additional evidence beyond prediction accuracy alone.
\end{enumerate}

This workflow highlights the complementary roles of the main components: neural measurements provide empirical benchmarks, language models and embeddings supply candidate representations, and neural coding quantitatively evaluates their correspondence. In this way, model--brain alignment moves beyond qualitative analogy and provides a testable framework for linking computational representations with neural mechanisms of language processing.
\section{Foundational Insights and Framework Shifts in Language-Brain Computational Research}

Existing research commonly introduces or adopts language processing models to examine the relationship between language and the brain from multiple perspectives. One major point of differentiation concerns model design motivation. Some studies employ architectures inspired by cognitive or linguistic theories, explicitly aiming to simulate particular aspects of human language processing. Other studies rely on engineering-oriented LLMs originally developed for NLP applications, which nonetheless provide high-dimensional representations and performance metrics useful for analyzing linguistic processes. Although LLMs largely originate from engineering practice, their conceptual foundations relate to parallel distributed processing (PDP), a framework initially proposed to explain neural information processing in the brain \cite{rumelhart1986parallel}. With rapid recent development, LLM-based analyses have increasingly contributed to understanding how language processing in artificial systems relates to neural mechanisms in humans.

\subsection{Early Model Exploration of Predictability and Structural Features}
Early investigations into the relationship between language processing and brain activity focused primarily on predictability, sequential and structural information, and neural encoding and decoding frameworks. These studies largely conceptualized language comprehension as an online, temporally unfolding process, in which moment-by-moment expectations and updates could be quantitatively linked to neural signals recorded with noninvasive techniques such as EEG, MEG, or fMRI \cite{brennan2016naturalistic}. At the same time, invasive intracranial electrophysiological recordings, particularly ECoG, complemented these approaches by providing high temporal and spatial resolution access to cortical dynamics during naturalistic speech perception and production \cite{oganian2019speech}. 

A representative line of work employs $n$-gram language models and related variants. These models estimate word surprisal from conditional probabilities in context, which are then used to predict neural responses \cite{hale2016information,willems2016prediction}. Owing to their simplicity and broad applicability, such methods are well suited for naturalistic narrative studies. Empirical results show that surprisal reliably predicts neural responses associated with language processing, including modulation of the N400 component, supporting the view that predictive mechanisms operate continuously during comprehension \cite{frank2015erp}. However, these models primarily capture surface-level sequential regularities and make limited commitments to the hierarchical structure of language. Consistent effects of surprisal and predictability have also been observed during continuous speech processing, including in ECoG recordings where high-gamma activity in superior temporal and frontal regions tracks word-level expectations, reinforcing the role of predictive coding at fine temporal scales \cite{moses2016neural}.

This limitation prompted subsequent work to move beyond purely sequential models by incorporating hierarchical syntactic structure. Researchers introduced phrase structure grammars, recursive syntactic models, or explicit syntactic representations to examine whether hierarchical dependencies improve explanation of neural data. Results indicate that hierarchical information contributes uniquely to explaining EEG and fMRI responses, suggesting that human language processing integrates both local lexical statistics and abstract structural dependencies \cite{shain2020fmri,brennan2019hierarchical}. Related work further suggests that such hierarchical and combinatorial structure is reflected in distributed cortical responses during both speech perception and production, a pattern that is also evident in intracranial recordings with distinct temporal dynamics across temporal and frontal language areas \cite{oganian2019speech}.

Methodologically, the growing diversity of linguistic models—ranging from sequential surprisal-based approaches to hierarchical syntactic representations—necessitated a common analytical framework for systematic comparison. In this context, the neural encoding framework emerged as a central analytical tool. This approach uses internal representations from computational models to predict neural responses, thereby enabling quantitative comparison between model-derived linguistic features and brain activity. It enables comparison across models and examination of how model architecture, training scale, or input representations affect prediction performance \cite{wehbe2014aligning,toneva2019interpreting}. Findings from encoding studies suggest that lexical semantics, syntactic structure, and semantic updating processes all contribute to model–brain correspondence. Such encoding approaches have proven particularly effective for speech data, as model-derived features can predict high-frequency cortical activity during naturalistic listening and speaking \cite{moses2016neural}.

Complementary to encoding approaches, decoding frameworks reverse the mapping by treating neural activity patterns themselves as representations of linguistic content and training classifiers to distinguish stimuli or semantic categories directly from brain data \cite{smith2005construct}. This perspective highlights the discriminability and generalization properties of neural representations, though challenges remain regarding interpretability and construct validity. Decoding studies further demonstrate that phonetic, lexical, and semantic information can be reliably recovered from cortical activity during speech, with particularly clear evidence from ECoG recordings \cite{pasley2012reconstructing}.

At the neural component level, questions about prediction, structure, and semantic updating converge most prominently on the N400, which has become a major focus connecting language models and neural signals. Studies modeling N400 responses using surprisal, semantic similarity, and semantic updating metrics demonstrate that multiple predictive factors can exert independent or complementary effects on amplitude \cite{rabovsky2018modelling,lopopolo2024tracking}. These findings suggest that the N400 reflects combined predictive and integrative processing across linguistic levels. Although the N400 is classically characterized in scalp EEG, converging intracranial evidence, including ECoG, links analogous semantic and predictive effects to localized cortical dynamics during speech comprehension \cite{broderick2018electrophysiological}.

In summary, early computational studies of language--brain alignment indicate that neural language processing is simultaneously sensitive to lexical predictability, hierarchical syntactic structure, and semantic updating. Sequential models, structural representations, and encoding and decoding frameworks capture complementary aspects of model--brain correspondence and together establish the theoretical questions and methodological foundation for later work incorporating deep learning--based models into neural alignment studies. Across recording modalities, these findings collectively underscore the continuity between computational models of language and the neural mechanisms supporting both written and spoken language processing. At the same time, these early findings also help clarify why recurrent neural models became an important intermediate stage in the subsequent development of language--brain computational research. Because many of these studies treated comprehension as an online, word-by-word, temporally unfolding process, they naturally motivated modeling approaches that could represent continuously updated internal states rather than only local transition probabilities or static structural descriptions. In this context, RNNs and later LSTMs provided a computational framework for linking sequential linguistic input with incrementally evolving representations, making them particularly relevant to temporally resolved neural data and to early efforts at neural prediction and alignment \cite{elman1990finding,hochreiter1997long,toneva2019interpreting}. From this perspective, recurrent architectures should not be viewed merely as a historical waypoint between n-gram models and Transformers. Rather, they marked an important conceptual transition: from classical sequential and surprisal-based modeling toward neural representations that unfold dynamically over time, while still remaining closely tied to incremental language processing. Contemporary Transformer-based and LLM-driven alignment studies have expanded this tradition by enabling broader contextual integration and richer representational comparisons, but they build in part on questions first sharpened by recurrent approaches. Recognizing this continuity helps make the transition from early computational models to modern model--brain alignment frameworks more conceptually explicit.

\subsection{Phonological Processing and Speech Dynamics}
Phonological and speech-related processing constitutes a foundational yet comparatively less integrated dimension of language--brain computational research. Relative to syntax, semantics, and higher-level predictive processing, phonological computation is more tightly coupled to the temporally unfolding nature of speech, linking continuous acoustic input to phonetic categorization, phonological representation, lexical access, and articulatory planning \cite{hickok2007cortical,friederici2011brain}. For this reason, phonological processing provides a particularly important testing ground for computational neuroscience: it requires models to account not only for abstract linguistic structure, but also for fast neural dynamics during online language comprehension and production.

A major body of evidence has come from methods with high temporal resolution. EEG and MEG are well suited for tracking the rapid transformation from speech input to phonological and lexical representations, while intracranial recordings such as ECoG provide finer-grained access to phonemic encoding, articulatory planning, and speech-related cortical dynamics \cite{moses2016neural}. In parallel, fMRI has helped localize broader cortical systems involved in speech and phonological processing, particularly along superior temporal and inferior frontal pathways, thereby complementing temporally precise but spatially less specific methods \cite{hickok2007cortical,price2012review}. Together, these approaches suggest that phonological processing is neither a peripheral preprocessing stage nor reducible to low-level acoustics; rather, it forms an intermediate computational bridge between sensory input and higher-order linguistic interpretation.

From a computational perspective, phonological and speech-based research has also played an important role in the development of model--brain alignment. Early work on naturalistic speech perception showed that neural responses can be predicted from temporally unfolding linguistic features, including word-level expectation, semantic dissimilarity, and speech-linked representations derived from computational models \cite{frank2015erp,broderick2018electrophysiological,brennan2016naturalistic}. Decoding studies further demonstrated that phonetic, lexical, and semantic information can be recovered from cortical activity during speech processing, especially in intracranial recordings \cite{pasley2012reconstructing,moses2016neural,khanna2024single}. At the same time, more recent alignment research has been dominated by text-based large language models and has focused more systematically on syntax, semantics, and hierarchical prediction than on phonological representations grounded in continuous speech input \cite{mischler2024contextual}. As a result, compared with higher-level language representations, speech and phonological alignment remains less coherently synthesized and methodologically less mature, despite recent efforts to integrate acoustic, speech, and language representations in naturalistic settings \cite{goldstein2025unified,gnanateja2025cortical}.

This imbalance points to an important direction for future work. A more complete account of the language--brain relationship requires integrating speech-based neural data, temporally sensitive computational models, and representational frameworks that can bridge acoustic, phonological, lexical, and higher-level linguistic structure. Strengthening this line of research would help situate phonological processing more clearly within the broader computational neuroscience agenda and provide a necessary complement to current model--brain alignment studies centered primarily on text and high-level semantics.

\subsection{LLM-Driven Framework for Neural Alignment}

Compared with the gradual advances achieved by statistical language models and early neural architectures, the emergence of LLMs has substantially accelerated interdisciplinary research at the intersection of linguistics, neuroscience, and artificial intelligence, particularly in studies examining alignment between computational models and neural mechanisms of language processing \cite{xiao2025exploring,alkhamissi2025language,proietti2025fine}. Two developments largely account for this shift. First, large-scale self-supervised pre-training on massive corpora enables models to capture broad patterns of syntax, semantics, and language use. Second, deep neural architectures provide distributed representations capable of integrating long-range contextual information \cite{maini2025beyondweb,xing2025pretrainzero,wang2025multilingual,shahriar2025inceptive}. Consequently, LLMs display behavioral performance resembling human processing across tasks involving contextual integration, syntactic analysis, semantic reasoning, and text generation. In several domains, including ambiguity resolution and anaphora interpretation, performance approaches or surpasses human benchmarks, making these models valuable tools for investigating computational aspects of language cognition \cite{beinborn2023cognitive,wang2023aligning,hu2023fine}.

More importantly, LLMs have increasingly become analytical instruments for studying neural language mechanisms \cite{singh2025evaluating,ye2025generative,lei2025large}. Earlier work frequently relied on correlations between behavioral results and brain imaging data without an intermediate computational framework. LLMs now provide explicit representational systems linking linguistic input, behavioral outcomes, and neural responses. Recent studies systematically compare hierarchical representations of LLMs with human brain imaging data to infer organizational principles of language networks, temporal dynamics of processing, and neural coding strategies underlying semantic representation \cite{lopez2025brain}. This trend has strengthened integration between linguistics and neuroscience, enabling analyses that jointly consider computational representations, behavioral responses, and neural activity. Accordingly, the remainder of this section surveys recent progress in LLM-driven alignment research across five dimensions: cross-modal representation, inter-brain neural coupling, hierarchical language computation, learning strategies and data distribution sensitivity, and language plasticity with shared representational spaces. Because the maturity of the evidence varies substantially across these domains, the following discussion distinguishes, where possible, between conclusions supported by multiple peer-reviewed studies and findings that remain recent, exploratory, or comparatively under-validated.

\subsubsection{Language Modulation of Perception: Cross-Modal Representations (Vision \& Music)}

Whether language actively shapes human perception of the external world remains a long-standing and contested issue in cognitive science and linguistics, often discussed under the framework of linguistic relativity. Early cross-linguistic studies, including work on color categorization and laboratory label-learning experiments, suggested that linguistic labels may influence categorical judgments and perceptual sensitivity to attributes such as color, shape, and texture. However, consensus remains lacking regarding the strength and interpretation of these effects \cite{whorf2012language,berlin1991basic,webster2012color,kay2006language}. Debates primarily concern three issues: the robustness and replicability of behavioral effects across populations and paradigms; the typically small effect sizes observed in many experiments; and the processing stage at which language exerts influence, whether during early perceptual encoding or only in later decision or memory stages. These unresolved issues have long complicated interpretation of language–perception interactions \cite{gilbert2006whorf,button2013power,thierry2009unconscious}.

Recent advances in vision–language models, particularly Contrastive Language–Image Pre-training (CLIP), have introduced new computational tools for addressing these debates and pushed research into a new interdisciplinary phase \cite{pmlr-v139-radford21a}. Through joint training on large-scale image–text pairs, such models learn unified representational spaces spanning visual and linguistic modalities. Their controllable training conditions allow partial disentanglement of linguistic and visual influences. Neuroimaging studies have increasingly reported that representations derived from vision--language models show stronger correspondence with neural representations in high-level visual regions such as the Ventral OccipitoTemporal Cortex (VOTC) than purely visual models, although the evidence base remains relatively recent and still limited in scope \cite{yang2025clip,liu2025data}.

Two principal interpretations of this similarity remain under discussion. One proposes that language exerts a direct modulatory effect on visual representations, reshaping VOTC representational geometry through neural pathways. The alternative common-cause hypothesis argues that language and perception share underlying relational structures, and model similarities simply reflect learning of general multimodal associations rather than language-specific effects. The central disagreement concerns whether language plays a causal role in shaping visual representations \cite{yang2025clip,chen2025language}.
To address this issue, several studies have used patients with disrupted language–vision connectivity as natural experiments. By examining how VOTC representations change following disconnection of language pathways, these studies provide evidence relevant to causal interpretations. Using diffusion tensor imaging to identify damage in white matter pathways linking language networks and VOTC, researchers compare neural responses during visual classification tasks between patients and healthy controls. Results indicate that pathway integrity significantly influences VOTC representational structure: greater disconnection is associated with representations more closely resembling purely visual models, while healthy individuals show representations incorporating linguistic relational structure \cite{chen2025combined}. Studies of sign language users further indicate that both spoken and signed languages can influence visual processing, suggesting modality-independent language effects. Although such AI–brain–lesion comparisons provide strong converging evidence, larger multi-center studies remain necessary for confirmation \cite{emmorey2014language}.

Cross-modal representational interactions also arise between language and other symbolic systems such as music \cite{patel2010music}. While neural substrates underlying low-level musical processing are relatively well characterized, mechanisms supporting representation of high-level musical semantics, including genre and emotion, remain less understood \cite{koelsch2014brain}. Meta-analytic work suggests partial neural overlap between language and music processing, particularly in auditory–motor circuits involved in phonological and interval processing, while higher-level structural processing shows greater domain specificity \cite{heard2020shared}.
The development of MusicLM and related text–music generation models offers new computational tools for studying musical semantic representation. These models learn mappings between textual descriptions and musical features, enabling quantitative comparison with neural responses. Recent fMRI studies of music listening further suggest that intermediate and higher-layer MusicLM representations may enable reconstruction of musical segments from neural signals and may predict auditory cortex responses better than generic auditory models \cite{denk2025text}.

Taken together, the current literature permits two preliminary conclusions. First, cross-modal interactions between language and perceptual systems such as vision and music exist and depend on specific neural pathways. Second, multimodal large models partially capture these interaction regularities and serve as useful computational tools for studying semantic coding in perceptual systems. Nonetheless, several open questions remain. Individual differences in linguistic modulation of perception remain poorly understood; model–brain discrepancies persist, particularly in fine perceptual detail processing; and developmental mechanisms underlying shared representations between language and music are unclear. Future research combining larger and more diverse samples with refined model comparison frameworks will be necessary to clarify the mechanisms and boundary conditions governing language modulation of perceptual systems.

\subsubsection{Inter-Brain Neural Coupling in Language Communication: Speaker-Listener Circuitry}

The central function of language is to enable transmission of information and coordination of cognition across individuals. Understanding how speech production and comprehension dynamically interact during real communication has therefore become a central topic in language neuroscience \cite{pickering2013integrated}. Early research typically employed isolated experimental frameworks, such as monologue production or passive listening, to localize neural substrates of production and comprehension separately. These studies established that speech production primarily involves the sensorimotor cortex, supplementary motor area (SMA), and inferior parietal regions responsible for phonological planning and articulation, whereas speech comprehension is centered on the superior temporal sulcus (STS) and auditory cortex, which support speech perception and early parsing \cite{hickok2004dorsal,giglio2022commonalities}. However, single-participant and non-interactive designs cannot capture real-time interactions between speakers and listeners, leaving unresolved how distinct production and comprehension systems coordinate across individuals during conversation \cite{stephens2010speaker,zada2025linguistic,redcay2019using}.

Recent advances in hyperscanning techniques and LLMs have provided complementary tools for addressing this problem. Hyperscanning allows simultaneous recording of neural activity from interacting individuals, enabling direct measurement of inter-brain correlations during communication. At the same time, LLMs enable quantitative modeling of conversational semantic content through contextual embeddings that capture both lexical meaning and long-range dependencies. This combination helps separate semantic contributions from low-level acoustic influences and supports identification of factors driving inter-brain neural coupling. As a result, joint use of hyperscanning and LLM-based representations has become an increasingly common framework for studying neural mechanisms underlying natural language interaction \cite{goldstein2025temporal,zada2025linguistic,aili2025cortical}.

Recent studies using this framework have begun to reveal several recurring patterns, although the evidence base remains comparatively limited and uneven across tasks, languages, and interaction settings. Anatomically, production and comprehension systems remain largely distinct: production-related activity concentrates in sensorimotor and parietal regions, whereas comprehension-related activity is centered in temporal auditory regions, with limited spatial overlap. Functionally, inter-brain coupling is often observed during successful communication. Neural activity in a speaker’s production network predicts activity in a listener’s comprehension network, and this coupling is not explained by acoustic similarity alone. Instead, coupling strength increases when interlocutors share aligned semantic representations of conversational content, a pattern reported across several languages and communication settings, though broader replication is still needed \cite{zada2025linguistic}.

Further work shows that coupling patterns adapt to communicative demands in both spatial and temporal dimensions. Spatially, coupling extends beyond classical language areas to include regions associated with social cognition and mentalizing, such as the temporoparietal junction (TPJ), posterior cingulate cortex (PCC), and medial prefrontal cortex (mPFC). This indicates that communication involves not only linguistic decoding but also alignment of intentions and shared knowledge between participants. Temporally, coupling often exhibits predictive characteristics: although comprehension signals typically lag production signals overall, higher-level cortical regions sometimes show anticipatory alignment with the speaker’s intended meaning before key information is fully expressed. These findings provide empirical support for predictive processing accounts of language comprehension and parallel computational principles employed in modern LLMs \cite{pickering2013integrated,okada2018neural,fujii2023neural}.

Inter-brain coupling research carries both theoretical and methodological implications. Theoretically, it challenges views that treat production and comprehension as isolated processes, instead proposing communication as functional coupling between specialized systems across individuals \cite{levelt1999theory}. Such mechanisms help explain communication success under noisy conditions and may illuminate communication impairments observed in disorders such as autism, aphasia, and schizophrenia, where altered coupling strength or instability has been reported \cite{hirsch2018cross,dumas2010inter}. Methodologically, combining LLM-based representations with hyperscanning extends alignment research beyond single-brain analyses toward dynamic interaction, reinforcing the utility of computational language models in neuroscience.

Nevertheless, important questions remain. Individual differences in coupling dynamics are poorly characterized; cross-modal communication mechanisms remain underexplored; and the extent to which LLM representations capture pragmatic and interactional aspects of communication is still unclear. Future work combining multimodal hyperscanning, conversational LLMs, and more diverse participant populations will help refine spatiotemporal models of inter-brain communication and clarify boundaries of model–brain alignment.

\subsubsection{Hierarchical Language Computation: Human Brain vs. LLMs}

Language exhibits hierarchical structure, progressing from phonological and lexical processing to syntactic and semantic integration, thereby enabling efficient information transmission \cite{chomsky2014aspects}. Neuroscientific research has consistently implicated core language regions such as the left inferior frontal gyrus (Broca’s area), superior temporal gyrus, and middle temporal gyrus in hierarchical processing. Broadly, Broca’s area contributes to syntactic construction and repair, whereas temporal regions support lexical-semantic activation and integration \cite{friederici2011brain,golestani2014brain}. Transformer-based LLMs similarly demonstrate strong performance on tasks involving hierarchical structures such as nested dependencies and long-distance agreement \cite{linzen2016assessing}. However, a central debate persists: whether LLMs genuinely implement brain-inspired hierarchical computation or merely reproduce similar behavioral outcomes through large-scale statistical learning.

To examine this issue, recent work has developed methods linking syntactic representations in models and neural activity patterns. The Hierarchical Frequency Probe approach, for example, uses frequency-domain analyses to identify populations encoding syntactic structure in both neural data and model representations \cite{an2025hierarchical,ding2016cortical}. Comparative studies using multiple LLM families alongside fMRI recordings have reported broadly convergent patterns: in models, syntactic information tends to be encoded more strongly in intermediate and higher layers, whereas lower layers primarily capture lexical features. In human cortex, lower-level syntactic information is associated with anterior temporal regions, while higher-level syntactic processing engages Broca’s area and adjacent regions \cite{friederici2011brain,pallier2011cortical}. Representational similarity analyses further indicate stronger correspondence between model representations and left-hemisphere language regions than with right-hemisphere homologues, supporting partial functional correspondence.

Temporal correspondence has also been investigated to address differences between biological sequential processing and model parallel computation. Using ECoG, researchers have recorded neural responses during natural story listening and compared time-resolved neural activity with representations from different model layers \cite{goldstein2025temporal}. Results show systematic correspondence between processing stages: shallow model layers align with early neural responses, whereas deeper layers correspond to later neural activity, particularly within core language regions. Despite hardware-level differences between brains and models, hierarchical organization thus appears functionally mappable onto neural processing time courses \cite{dehaene2011experimental}.

Research has further extended to predictive processing over longer time scales. Human language comprehension relies heavily on prediction, and recent fMRI studies have compared neural responses during narrative listening with predictions generated by models of varying complexity \cite{yu2024predicting,wang2023predictive,zhou2025hierarchical}. Findings indicate hierarchical prediction mechanisms: superior temporal regions support local lexical predictions resembling shallow model layers or n-gram models, whereas default mode network regions contribute to global predictions over sentence and discourse scales, aligning more closely with higher-level LLM representations. Sparse update prediction models appear to better capture neural dynamics at discourse boundaries, suggesting multi-time-scale predictive organization. LLMs exhibit comparable hierarchical prediction patterns when processing long texts, further reinforcing parallels between model and brain computations.

Taken together, current evidence tentatively suggests a multi-dimensional correspondence between LLMs and neural mechanisms underlying hierarchical language processing, although the robustness and mechanistic interpretation of this correspondence remain open questions. However, several open questions remain. Correspondence at pragmatic and discourse reasoning levels remains insufficiently explored; individual differences in hierarchical processing are poorly understood; and it remains unclear whether model hierarchies reflect genuine structural computation or statistical overfitting to text data. Future research combining multimodal imaging, diverse language tasks, and causal intervention methods will be necessary to clarify limits of model–brain correspondence and support both neuroscientific theory development and brain-inspired improvements in language models.

\rev{\subsubsection{Semantic Composition: Linking Linguistic Theory, Brain Activity, and LLM Representations}}

\rev{Semantic composition is a central problem for connecting linguistic theory, cognitive neuroscience, and computational language models. Human language comprehension requires more than recognizing individual words; it depends on combining lexical meanings into structured phrase-, sentence-, and discourse-level representations. This process directly links syntactic structure with semantic interpretation, making it a key test case for theories of language--brain alignment. From a linguistic perspective, semantic composition concerns how smaller units such as morphemes and words combine into larger meanings. From a neuroscience perspective, it raises the question of how the brain constructs complex meanings over time and how syntactic and semantic operations interact in regions such as the left inferior frontal gyrus, anterior temporal cortex, and posterior temporal cortex \cite{brennan2012syntactic,pallier2011cortical}. \cite{pylkkanen2019neural} emphasized that combinatory syntax and semantics should be studied as an integrated neurobiological problem rather than as separate syntactic or lexical-semantic processes.}

\rev{Recent LLM-based work provides new tools for quantifying semantic composition. \cite{gao2024measuring} proposed a Composition Score derived from internal LLM representations to estimate the degree of meaning composition during sentence comprehension. Their results showed that this model-based metric correlates with fMRI responses in brain regions associated with word frequency, structural processing, and general sensitivity to words, suggesting that LLM-derived features can capture aspects of how the human brain builds complex meaning from sequential linguistic input. This approach is important because it moves beyond comparing static word embeddings with neural activity and instead evaluates how model representations change as meanings are compositionally constructed across a sentence \cite{caucheteux2022brains}.}

\rev{At the same time, semantic composition also illustrates the interpretive limits of current model--brain alignment. A high correspondence between LLM-derived composition metrics and neural activity does not necessarily show that LLMs and the brain implement the same compositional algorithm \cite{cichy2019deep,caucheteux2022brains}. Rather, such findings indicate that LLM representations encode variables relevant to human meaning construction \cite{gao2024measuring}. Future studies should therefore compare LLM-based composition metrics with more theory-driven linguistic predictors, such as syntactic node counts, dependency relations, thematic role assignment, and semantic plausibility, while also testing whether these predictors explain neural responses beyond lexical frequency, surprisal, and contextual similarity \cite{brennan2012syntactic,shain2020fmri,kauf2024lexical}. In this sense, semantic composition provides a particularly useful bridge between formal linguistic theory, neural measurement, and computational model analysis.}

\subsubsection{Learning Strategies \& Data Distribution Sensitivity: Human Brain vs. LLMs}

In neuroscience and artificial intelligence, a central question concerns how learning systems acquire rules and adapt to new tasks: do they rely primarily on weight-based learning (memorization) or contextual learning (inductive reasoning from limited examples)? The answer bears directly on our understanding of human learning and the foundations of intelligence in artificial systems \cite{lake2017building,frankish2010dual}. The emergence of strong in-context learning capabilities in LLMs has renewed interest in this debate. Without updating parameters, LLMs can rapidly adapt to new tasks by incorporating a small number of examples into prompts, behaviorally resembling rapid human inductive reasoning. However, direct empirical evidence remains limited regarding whether the underlying mechanisms are truly homologous, restricting deeper analysis of human-like intelligence and model optimization \cite{kumar2021semantic}.

To address this issue, recent work introduced a standardized image–label associative learning framework to systematically compare humans and Transformer-based LLMs under controlled data distribution conditions \cite{pesnot2025shared}. Three distribution scenarios were designed: (1) highly diverse data, where image–label pairs rarely repeat and systems must infer general rules; (2) highly repetitive data, enabling performance gains through memory consolidation; and (3) mixed distributions that test adaptive strategy switching. Results showed strong convergence between humans and LLMs. Under diverse data, both relied primarily on contextual learning. Under repetitive data, both shifted toward memory-based strategies: humans consolidated memory traces, whereas LLMs encoded recurring patterns in internal representations, reducing computational cost. Under mixed conditions, both displayed flexible strategy use, dynamically selecting processing modes according to input structure. These findings provide behavioral and computational evidence that humans and LLMs share key statistical learning tendencies and support the human-like nature of in-context learning.

Important differences nevertheless emerged. Humans showed stronger resilience to distribution shifts, maintaining prior strategies while adapting to new ones and rapidly reinstating earlier strategies when distributions reverted. In contrast, LLMs exhibited strategy forgetting, requiring many new examples to recover previous processing modes. This divergence highlights differences in learning flexibility and in the interaction between memory stability and adaptive mechanisms.

Such comparative studies carry both theoretical and practical implications. Theoretically, they clarify that human learning involves coordinated interaction between long-term consolidation and contextual reasoning, enabling flexible adaptation while preserving stable knowledge structures \cite{russin2025parallel,dherin2025learning}. Practically, these findings suggest directions for improving in-context learning in LLMs, for example through strategy memory mechanisms or adaptive responses to distribution shifts, thereby bringing model behavior closer to human learning patterns \cite{rafiuddin2025learning,muttakhiroh2025tackling}.

Several open questions remain. Existing work has largely focused on simple associative tasks, leaving distribution sensitivity in complex and naturalistic language learning—particularly for long-range dependencies, compositional structure, and pragmatic reasoning—poorly understood. Individual differences in human learners and performance variability across models and training regimes are also underexplored, limiting insight into the generalizability of learning strategies. Moreover, the neural mechanisms underlying these differences remain unclear due to the limited integration of brain imaging, cognitive modeling, and model interpretability approaches. Future research should therefore adopt more complex, naturalistic tasks; combine behavioral experiments with multimodal neural measurements; and recruit larger, more diverse participant samples. In parallel, systematic model analyses—including ablations, architectural comparisons, and training data manipulations—are needed to identify the computational sources of learning differences. Integrating these directions will help clarify divergences between human and LLM learning strategies and advance both language neuroscience and brain-inspired artificial intelligence.

\subsubsection{Language Plasticity, Individual Differences \& Shared Representations}

In natural communication, the human language system must balance plasticity and stability, enabling adaptation across speakers and contexts while preserving consistent linguistic representations \cite{kuhl2004early,grossberg2000complementary}. On the one hand, listeners must rapidly adapt to accents, dialects, and contextual variations; on the other, phonological and semantic categories must remain stable to ensure reliable communication \cite{kraljic2006generalization}. For example, listeners can quickly adapt to unfamiliar accents without permanently altering established phonological categories \cite{clarke2004rapid}. Although this adaptive stability is well documented behaviorally, its neural implementation remains incompletely understood, particularly regarding how plasticity and stability are balanced across processing stages \cite{mcclelland1986trace,bhaya2026shared}.

Recent EEG studies have addressed this issue using controlled manipulations of acoustic cues, such as fundamental frequency (F0) and voice onset time (VOT), to simulate unfamiliar accent conditions \cite{llanos2025accented}. By degrading cue reliability, researchers examined how listeners adapt during word recognition while recording neural responses. Analyses of early perceptual components (N1, P2) and later semantic processing (N400) revealed a mechanism of selective cue downweighting. Rather than restructuring phonological categories, listeners reduced reliance on unreliable cues while maintaining processing of stable cues. Importantly, cue reweighting occurred at early perceptual stages without altering later semantic processing, thereby preserving representational stability. These findings clarify how the brain achieves adaptive flexibility without compromising core representations and offer potential guidance for improving accent adaptation mechanisms in LLM-based systems.

Parallel challenges arise from individual variability and limited sample sizes in language brain mapping, particularly in high-resolution techniques such as ECoG \cite{button2013power,lipkin2022probabilistic}. Anatomical and connectivity differences across individuals complicate extraction of common neural patterns, while clinical constraints limit participant numbers. To mitigate these issues, recent work introduced the Shared Response Model (SRM) to align neural data across individuals within a shared low-dimensional representational space \cite{bhattacharjee2026aligning}. Using ECoG recordings during story listening and reading tasks, researchers applied SRM to standardize neural responses before comparing them with hierarchical representations in LLMs. Results showed that despite anatomical variability, aligned neural representations exhibited strong cross-individual consistency during language processing. These shared representations extended beyond classical left-hemisphere language regions and were also observed in right-hemisphere homologous areas, indicating bilateral contributions to language processing.

Further validation demonstrated practical benefits: neural encoding models trained in the shared space achieved improved prediction of unseen individuals' brain activity, and semantic decoding performance substantially exceeded that of unaligned approaches. This work provides both a technical solution to data scarcity and individual variability and empirical evidence for shared neural coding principles underlying language processing. The resulting shared representational space also supports clinical translation, particularly in Brain–Computer Interface (BCI) development for language-impaired patients \cite{spalding2025shared,singh2025transfer,willett2023high,metzger2023high}.

Nevertheless, open questions remain. Mechanisms underlying individual differences in adaptive language plasticity remain poorly characterized, and the generalization limits of SRM in complex scenarios such as bilingual or dialectal processing require further testing. Moreover, correspondence between shared neural representations and hierarchical representations in LLMs remains insufficiently explored beyond semantic levels. Future studies should integrate multimodal imaging, larger and more diverse participant cohorts, and complex language tasks to refine models of language plasticity, improve cross-individual alignment techniques, and deepen brain–model correspondence analyses, thereby advancing both theoretical and clinical applications.

\subsubsection{Development, Bilingual Processing, and Language/Speech-related Disorders}

From a computational neuroscience perspective, language development provides a critical test case for understanding how linguistic representations emerge, stabilize, and reorganize across learning. Unlike most current alignment studies, which are based on adult participants and relatively mature language systems, developmental research foregrounds the gradual formation of phonological, lexical, and structural representations under strong constraints from neural plasticity and environmental input. In this sense, developmental populations are especially informative for linking learning trajectories with neural coding, because they allow researchers to examine not only where linguistic information is represented, but also how such representations are acquired and reshaped over time \cite{kuhl2004early}. Related evidence from developmental dyslexia further illustrates that language-related neural organization reflects both shared constraints and script-specific adaptations, indicating that developmental data can provide important tests of the generalizability of computational accounts across writing systems and learning histories \cite{siok2004biological}.

Language/speech disorders and related clinical populations are equally important for computational neuroscience because they reveal what aspects of language processing break down when typical neural organization is disrupted. Conditions such as aphasia, stuttering, dyslexia, autism, and post-injury language impairment should not be treated merely as downstream applications of otherwise complete theories; rather, they provide strong constraints on mechanistic explanation by highlighting which components of linguistic representation, neural coding, and cross-regional coordination are vulnerable under atypical development or neural damage. Existing work on communication impairments and clinical translation already suggests that altered coupling, reduced stability of shared representations, and degraded decoding capacity may all be relevant to understanding pathological language processing \cite{hirsch2018cross,dumas2010inter,willett2023high,metzger2023high,spalding2025shared,singh2025transfer}. Nevertheless, current evidence remains limited, and developmental, bilingual, and clinical populations are still underrepresented relative to the central role they should play in testing the explanatory scope of brain-aligned language models. Future progress will therefore depend on integrating these populations more systematically into alignment research, not only to improve translational relevance, but also to sharpen the computational theories used to explain human language more broadly.

\section{Discussion}
\subsection{Current Limitations}

Despite substantial progress at the intersection of linguistics and neuroscience, recent advances reveal several persistent limitations that constrain current understanding and call for targeted future developments:

\begin{enumerate}
	\item[i.] \textbf{Mechanistic interpretation remains limited by black-box mapping, and functional alignment does not imply structural homology}. Most current studies evaluate correspondence between LLM representations and brain activity using methods such as Representational Similarity Analysis (RSA) and neural encoding models. However, correlation-based alignment demonstrates only functional similarity and does not establish shared computational mechanisms. Existing evidence suggests that similarities between LLMs and human language processing largely reflect functional fitting rather than mechanistic equivalence. Human language acquisition is grounded in multimodal interactive experience, whereas LLMs learn from unimodal textual data, limiting the explanatory power of alignment results regarding why model representations predict neural responses \cite{cuskley2024limitations}. For example, although model attention mechanisms resemble human selective attention at a functional level, the former relies on parallel weight computation, whereas the latter depends on dynamically regulated neural circuits and neuromodulatory processes. No structural correspondence has yet been established, leaving open the possibility that observed alignment reflects statistical fitting rather than mechanistic simulation. This limitation does not diminish the value of LLM–brain alignment, but it requires a clearer distinction between what current LLMs capture well and what they leave unresolved. Existing alignment findings suggest that LLMs are particularly informative for modeling distributional semantics, lexical predictability, contextual integration, and certain forms of long-range dependency processing. These are important components of human language comprehension and help explain why LLM representations often show strong correspondence with neural activity. However, other aspects of human language remain less well captured, including embodied sensorimotor grounding, interactive pragmatics, developmental learning, bilingual adaptation, social communicative intention, neural plasticity, and biologically precise temporal dynamics. Thus, strong LLM–brain alignment should be understood as evidence that LLMs capture important representational dimensions of language processing, rather than as sufficient evidence that they reproduce the full neural mechanisms of human language. Future work should therefore distinguish predictive alignment from mechanistic explanation. Mechanistic claims become stronger when model representations correspond to interpretable cognitive operations, explain neural responses beyond simpler lexical or statistical baselines, predict behavioral patterns, and remain robust across tasks, languages, individuals, and measurement modalities. Under this view, LLM–brain alignment is not merely correlational, but neither is it automatically mechanistic; it is a valuable intermediate level of evidence for constraining theories of language processing.
	
	\item[ii.] \textbf{Technical constraints limit research depth and hinder real-time interactive verification}. In studies combining LLMs with speech/language BCIs, current information transmission rates remain insufficient for efficient interaction between neural signals and model parameters, making direct model–brain mapping difficult. Neural recordings also suffer from low signal-to-noise ratios and strong individual variability, challenging accurate decoding of dynamic correspondences between neural activity and model representations. Cross-individual alignment methods such as SRM extract group-level regularities but do not preserve individual neural coding patterns \cite{da2020channel}. As a result, most research remains limited to offline correlation analyses, preventing real-time closed-loop experiments and restricting investigation of dynamic language processing mechanisms.
	
	\item[iii.] \textbf{Research coverage remains limited in data diversity, typological breadth, and task design.}
A related limitation concerns typological diversity. Although this review has emphasized the cross-cultural evolution of linguistic theories, current model--brain alignment research still provides only limited evidence from typologically diverse languages. Most existing studies are concentrated on a small number of high-resource languages and on relatively standardized written materials. As a result, it remains difficult to determine whether reported correspondences between computational models and neural activity reflect language-general mechanisms or patterns that are partly shaped by specific morphosyntactic systems, writing systems, or culturally situated usage conventions \cite{joshi2020state}. \rev{This limitation is particularly important in light of recent multilingual and cross-language resources. Multilingual naturalistic fMRI datasets, such as the Le Petit Prince corpus, provide a basis for comparing neural responses to comparable narrative content across languages, while cross-language brain decoding studies test whether neural representations learned from one language can generalize to another. Recent computational alignment work further suggests that LLM–brain correspondence should be evaluated not only within a single high-resource language but also across languages, model scales, and training regimes. For example, \cite{gao2025increasing} compared base and instruction-tuned LLMs of different sizes against eye-tracking and fMRI responses during naturalistic reading, showing that larger model scale improved brain alignment more than instruction tuning. Such findings highlight the need to ask whether improvements in model–brain alignment reflect language-general processing principles or properties of specific languages, datasets, and model training objectives.} LLM training data are dominated by standardized written text, with limited representation of spoken language, dialectal variation, and multilingual mixing. Consequently, alignment studies may fail to capture neural mechanisms underlying natural communication. Experimental frameworks are also largely restricted to passive tasks such as reading or story listening, with comparatively little work examining active conversational interaction, pragmatic reasoning, or communicative intent transmission, despite these processes constituting the core function of language. In addition, neuroimaging studies predominantly involve healthy young adults, while data from older populations and individuals with language-related disorders such as aphasia or autism remain scarce. These biases limit generalizability and reduce potential clinical impact.
	
	\item[iv.] \textbf{Evaluation frameworks remain narrow, and measures of brain alignment are incomplete}. Current alignment studies typically rely on neural prediction accuracy as the primary metric, yet this single measure cannot fully characterize brain-inspired processing. Models may achieve high prediction accuracy through statistical shortcuts rather than genuine replication of hierarchical linguistic computations \cite{cuskley2024limitations}. For instance, brain activity during syntactic processing may be predicted using surface lexical statistics rather than structural parsing, resulting in strong predictive performance but weak mechanistic correspondence. Furthermore, evaluation frameworks rarely incorporate human-specific cognitive properties such as language plasticity, long-term memory consolidation, or adaptive strategy switching, making it difficult to assess fundamental differences between LLMs and human language systems.
\end{enumerate}

In summary, although interdisciplinary research has established productive connections between LLM development and language neural studies, substantial limitations remain in mechanistic interpretation, technical feasibility, ecological validity, and comprehensive evaluation. Future progress requires tighter integration across neuroscience, linguistics, and artificial intelligence, combined with technical innovation, richer experimental scenarios, and more comprehensive evaluation frameworks. Such advances are necessary both for clarifying model–brain correspondence and for promoting the development of brain-inspired language models alongside a deeper understanding of neural language mechanisms.

\begin{figure}[!htbp]
	\centering
	\includegraphics[width=0.88\textwidth]{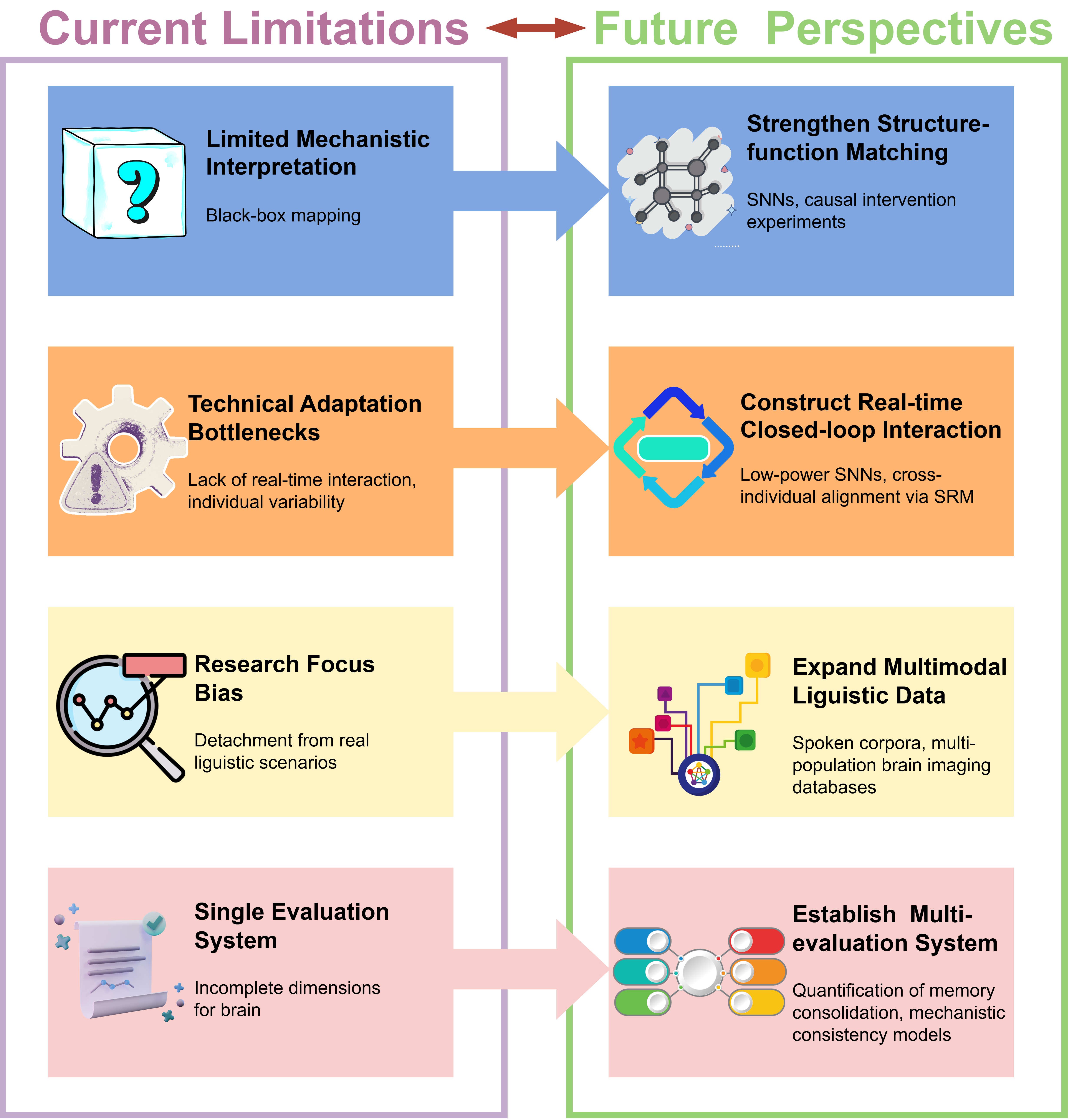}
	\caption{Correspondence between Current Limitations and Future Research Perspectives in the Study of Neural Mechanisms of Language.}
	\label{fig2_3}
\end{figure}
\subsection{Future Research Perspectives}

Despite rapid progress in applying large language models to computational neuroscience, fundamental limitations remain in mechanistic interpretability, ecological validity, and translational applicability. Existing studies are still dominated by correlational analyses and static modeling paradigms, which restrict their ability to capture the dynamic, causal, and cognitively grounded nature of human language processing. Addressing these challenges requires future research to move toward integrated frameworks that jointly consider biological mechanisms, real-time interaction, data diversity across populations and languages, and evaluation standards.

In this context, four complementary research directions are particularly critical. First, strengthening structure--function matching aims to reduce the gap between artificial architectures and biological neural mechanisms through brain-inspired modeling and causal intervention. Second, constructing real-time closed-loop interaction systems emphasizes dynamic, bidirectional coupling between neural activity and language models. Third, expanding multimodal and multi-population linguistic neurodata seeks to improve ecological validity and generalizability. Finally, establishing multi-dimensional evaluation frameworks moves beyond neural predictability toward cognitive and mechanistic validation. Together, these directions define a coherent roadmap for advancing LLM-driven computational neuroscience (Fig.~\ref{fig2_3}).

\subsubsection{Strengthening Structure--Function Matching via Brain-inspired Modeling and Causal Interventions}

Although Transformer-based LLMs can reproduce brain activity patterns at a functional or representational level, substantial differences remain in underlying computational mechanisms and energy efficiency. This structure--function mismatch limits deeper model--brain alignment and weakens claims of mechanistic correspondence. Strengthening structure--function matching therefore constitutes a foundational direction for future research.

SNNs provide a promising pathway toward improved biological plausibility. Inspired by event-driven communication in biological neurons, SNNs encode information through spatiotemporal spike patterns and naturally support temporally precise dynamics \cite{zhang2025neural}. Compared with conventional artificial neural networks, SNNs can achieve orders-of-magnitude improvements in energy efficiency while more faithfully reproducing neural signaling properties \cite{wang2026model}. Recent studies have demonstrated the feasibility of integrating SNNs with large-scale language models through spiking attention mechanisms, hierarchical architectures, and model conversion techniques, enabling improved alignment without prohibitive performance loss \cite{zhou2022spikformer,wang2023spatial,wu2024adaptive}. The recently proposed ``Shunxi 1.0'' spiking large model further illustrates the potential of this approach for ultra-long sequence processing with high efficiency \cite{pan2025spikingbrain}. However, the relevance of SNNs to language--brain alignment should be interpreted with appropriate scope. For many of the studies reviewed in this manuscript, the primary evidence comes from population-level or field-level measures such as fMRI, EEG, and MEG, rather than from spike-resolved recordings. Accordingly, spiking-level detail should not be treated as the universally appropriate level of explanation for all alignment problems discussed here. Instead, the potential advantage of SNNs is likely to be greatest in settings where temporally precise neural dynamics, event-driven computation, online adaptation, or low-power real-time interaction are central concerns, such as speech-related electrophysiology, closed-loop brain--computer interfaces, or causal perturbation paradigms \cite{willett2023high}. In contrast, for many current representational alignment studies based primarily on fMRI or correlation-based correspondence, direct evidence that SNNs provide superior explanatory power over existing LLM-based approaches remains limited. At present, the strongest support for SNNs lies in improved biological plausibility, temporal precision, and computational efficiency, rather than in an already established advantage for explaining language-related neural data. Where spike-based models are specifically concerned, the most direct empirical constraints are likely to come from spike-resolved or microelectrode recordings, while intracranial electrophysiology and high-temporal-resolution field measures can provide intermediate validation at coarser scales.

Beyond architectural similarity, future work should incorporate causal intervention experiments to validate structure--function correspondence. Targeted perturbations of model components, circuit-level ablations, and causal mediation analyses can be combined with neural stimulation or lesion data to test whether homologous structures in models and brains support comparable functional roles. Such causal validation moves beyond correlational alignment and enables stronger inferences about shared computational principles.

\subsubsection{Constructing Real-time Closed-loop Language--Brain Interaction Systems}

Most existing language--brain alignment studies rely on offline analyses of pre-recorded neural data, limiting their relevance to real-time cognition and interaction. Constructing closed-loop systems that dynamically couple neural activity and language models represents a critical future direction.

Low-power, event-driven SNN-based models may be particularly useful for certain real-time applications due to their computational efficiency and temporal precision, especially when online neural recording and adaptive bidirectional interaction are central design requirements. When integrated with online neural recording modalities, such models can support adaptive decoding and encoding of linguistic information, enabling continuous bidirectional interaction between brains and artificial systems. This capability is essential for capturing the dynamic nature of language processing and learning.

A major challenge in closed-loop systems is inter-individual variability in neural representations. Cross-individual alignment methods, such as Shared Response Models (SRM), can be incorporated to establish shared representational spaces across subjects, thereby reducing data requirements for new users. Combined with individual-specific fine-tuning, such hybrid general--specific frameworks may enable scalable and personalized real-time language--brain interaction systems, particularly for brain--computer interface applications.

\subsubsection{Expanding Multimodal and Multi-population Linguistic Neurodata}

Progress in computational language neuroscience is fundamentally constrained by the scope and diversity of available data. Existing datasets predominantly focus on written language and limited participant populations, which restrict ecological validity and generalizability. Expanding multimodal and multi-population linguistic neurodata therefore constitutes a key future priority.

Spoken language corpora aligned with neural recordings are especially important, as speech represents the primary mode of natural language use. Incorporating acoustic, articulatory, and prosodic information alongside neural signals can reveal encoding principles that are not captured by text-based paradigms alone. More broadly, integrating multimodal inputs such as vision and action can better reflect the conditions under which language is acquired and used in natural environments.

In parallel, large-scale brain imaging databases spanning diverse populations, languages, and developmental stages are needed. Such resources would enable systematic investigation of individual differences, cross-linguistic variation, and neurodiversity in language processing, providing a stronger empirical foundation for building and evaluating brain-aligned language models. An important component of this agenda concerns typological diversity. Prior neuroimaging evidence suggests that core properties of the language network are robust across a wide range of languages, but direct evidence from LLM-based alignment studies across typologically diverse languages remains limited \cite{malik2022investigation}. Future work should therefore prioritize datasets that extend beyond a narrow set of high-resource languages to include a broader range of language families, morphosyntactic systems, writing systems, and communicative settings. Such expansion is necessary for determining whether reported correspondences between computational models and neural activity reflect general principles of human language processing or patterns partly shaped by language-specific structural properties and culturally situated language use.

\subsubsection{Establishing Multi-dimensional Evaluation Frameworks for Brain--Language Alignment}

Current evaluation of language--brain alignment relies heavily on neural prediction accuracy, which provides only a partial view of cognitive and mechanistic validity. Establishing multi-dimensional evaluation frameworks is therefore essential for assessing genuine progress.

Future evaluation systems should incorporate metrics related to memory consolidation, including stability--plasticity trade-offs and long-term information retention, to assess whether models capture key properties of human learning. In addition, mechanistic consistency models can be developed to quantify whether internal model dynamics align with known neural principles, such as hierarchical processing, temporal integration, and causal information flow.

By jointly considering behavioral performance, neural predictability, cognitive plausibility, and mechanistic consistency, such multi-evaluation frameworks can provide a more rigorous and interpretable assessment of brain--language alignment, moving the field beyond surface-level correspondence toward deeper explanatory understanding.

\section{Conclusion}

Linguistics and neuroscience have traditionally advanced along largely independent trajectories, with the former emphasizing formal descriptions of language competence and the latter focusing on the biological mechanisms of language processing. Computational neuroscience offers an operational bridge by translating linguistic constructs into computable representations that can be quantitatively compared with neural signals. Historically, interdisciplinary progress has shifted from descriptive fitting toward mechanistic alignment. Early studies relied on simplified models with limited explanatory power, whereas recent advances in deep learning and large language models have enabled large-scale model–brain alignment by capturing both behavioral and neural patterns of language processing. However, such alignment remains largely correlational, as similarities may reflect shared sensitivity to statistical structure rather than shared computational mechanisms. Addressing this ambiguity requires alignment approaches constrained by both functional and structural principles, integrating causal analysis, model dissection, and biologically grounded constraints on neural architecture and dynamics. Looking forward, computational neuroscience may provide a pathway toward a more generative theoretical framework for language, but a substantial gap remains between current alignment-based evidence and fully mechanistic, testable neuro-computational accounts. Progress toward that goal will depend on moving beyond surface-level correspondence toward approaches that more directly address mechanism, causation, and biological implementation.

\section*{Acknowledgments}
This work was supported by The Hong Kong Polytechnic University Start-up Fund (Project ID: P0053210), The Hong Kong Polytechnic University Faculty Reserve Fund (Project ID: P0053738), an internal grant from The Hong Kong Polytechnic University (Project ID: P0048377), The Hong Kong Polytechnic University Departmental Collaborative Research Fund (Project ID: P0056428), The Hong Kong Polytechnic University Collaborative Research with World-leading Research Groups Fund (Project ID: P0058097) and Research Grants Council Collaborative Research Fund (Ref: C5033-24G).
\bibliographystyle{unsrt}  
\bibliography{references}

@article{poeppel2016maps,
	title={The maps problem and the mapping problem: Two challenges for a cognitive neuroscience of speech and language},
	author={Poeppel, David},
	journal={Understanding Cognitive Development},
	pages={34--55},
	year={2016},
	publisher={Psychology Press},
}

@article{sahlgren2008distributional,
	title={The distributional hypothesis},
	author={Sahlgren, Magnus},
	journal={Italian Journal of Linguistics},
	volume={20},
	pages={33--53},
	year={2008}
}

@article{elman1990finding,
	title={Finding structure in time},
	author={Elman, Jeffrey L},
	journal={Cognitive Science},
	volume={14},
	number={2},
	pages={179--211},
	year={1990},
	publisher={Wiley Online Library}
}

@article{hochreiter1997long,
	title={Long short-term memory},
	author={Hochreiter, Sepp and Schmidhuber, J{\"u}rgen},
	journal={Neural Computation},
	volume={9},
	number={8},
	pages={1735--1780},
	year={1997},
	publisher={MIT press}
}

@article{vaswani2017attention,
	title={Attention is all you need},
	author={Vaswani, Ashish and Shazeer, Noam and Parmar, Niki and Uszkoreit, Jakob and Jones, Llion and Gomez, Aidan N and Kaiser, {\L}ukasz and Polosukhin, Illia},
	journal={Advances in Neural Information Processing Systems},
	volume={30},
	year={2017}
}

@article{naselaris2011encoding,
	title={Encoding and decoding in fMRI},
	author={Naselaris, Thomas and Kay, Kendrick N and Nishimoto, Shinji and Gallant, Jack L},
	journal={Neuroimage},
	volume={56},
	number={2},
	pages={400--410},
	year={2011},
	publisher={Elsevier}
}

@book{rumelhart1986parallel,
	title={Parallel distributed processing, volume 1: Explorations in the microstructure of cognition: Foundations},
	author={Rumelhart, David E and McClelland, James L and PDP Research Group and others},
	year={1986},
	publisher={The MIT press}
}

@article{kauf2024lexical,
	title={Lexical-semantic content, not syntactic structure, is the main contributor to ANN-brain similarity of fMRI responses in the language network},
	author={Kauf, Carina and Tuckute, Greta and Levy, Roger and Andreas, Jacob and Fedorenko, Evelina},
	journal={Neurobiology of Language},
	volume={5},
	number={1},
	pages={7--42},
	year={2024},
	publisher={MIT Press One Broadway, 12th Floor, Cambridge, Massachusetts 02142, USA~…}
}

@article{rabovsky2018modelling,
	title={Modelling the N400 brain potential as change in a probabilistic representation of meaning},
	author={Rabovsky, Milena and Hansen, Steven S and McClelland, James L},
	journal={Nature Human Behaviour},
	volume={2},
	number={9},
	pages={693--705},
	year={2018},
	publisher={Nature Publishing Group UK London}
}

@article{michaelov2024strong,
	title={Strong prediction: Language model surprisal explains multiple N400 effects},
	author={Michaelov, James A and Bardolph, Megan D and Van Petten, Cyma K and Bergen, Benjamin K and Coulson, Seana},
	journal={Neurobiology of Language},
	volume={5},
	number={1},
	pages={107--135},
	year={2024},
	publisher={MIT Press One Broadway, 12th Floor, Cambridge, Massachusetts 02142, USA~…}
}

@article{lopopolo2024tracking,
	title={Tracking lexical and semantic prediction error underlying the n400 using artificial neural network models of sentence processing},
	author={Lopopolo, Alessandro and Rabovsky, Milena},
	journal={Neurobiology of Language},
	volume={5},
	number={1},
	pages={136--166},
	year={2024},
	publisher={MIT Press One Broadway, 12th Floor, Cambridge, Massachusetts 02142, USA~…}
}

@article{lenneberg1967biological,
	title={The biological foundations of language},
	author={Lenneberg, Eric H},
	journal={Hospital Practice},
	volume={2},
	number={12},
	pages={59--67},
	year={1967},
	publisher={Taylor \& Francis}
}

@inproceedings{bhattasali2020alice,
	title={The Alice datasets: fMRI \& EEG observations of natural language comprehension},
	author={Bhattasali, Shohini and Brennan, Jonathan and Luh, Wen-Ming and Franzluebbers, Berta and Hale, John},
	booktitle={Proceedings of the Twelfth Language Resources and Evaluation Conference},
	pages={120--125},
	year={2020}
}

@article{swaab2012language,
	title={Language-related ERP components},
	author={Swaab, Tamara Y and Ledoux, Kerry and Camblin, C Christine and Boudewyn, Megan A},
	journal={The Oxford Handbook of Event-Related Potential Components},
	pages={397--439},
	year={2012},
	publisher={Oxford University Press New York, NY}
}

@article{salmelin2009electromagnetic,
	title={Electromagnetic brain imaging},
	author={Salmelin, Riitta and Baillet, Sylvain},
	journal={Human Brain Mapping},
	volume={30},
	number={6},
	pages={1753},
	year={2009}
}

@article{hale2016information,
	title={Information-theoretical complexity metrics},
	author={Hale, John},
	journal={Language and Linguistics Compass},
	volume={10},
	number={9},
	pages={397--412},
	year={2016},
	publisher={Wiley Online Library}
}

@article{willems2016prediction,
	title={Prediction during natural language comprehension},
	author={Willems, Roel M and Frank, Stefan L and Nijhof, Annabel D and Hagoort, Peter and Van den Bosch, Antal},
	journal={Cerebral Cortex},
	volume={26},
	number={6},
	pages={2506--2516},
	year={2016},
	publisher={Oxford Academic}
}

@article{brennan2016naturalistic,
	title={Naturalistic sentence comprehension in the brain},
	author={Brennan, Jonathan},
	journal={Language and Linguistics Compass},
	volume={10},
	number={7},
	pages={299--313},
	year={2016},
	publisher={Wiley Online Library}
}

@article{brennan2019hierarchical,
	title={Hierarchical structure guides rapid linguistic predictions during naturalistic listening},
	author={Brennan, Jonathan R and Hale, John T},
	journal={PloS One},
	volume={14},
	number={1},
	pages={e0207741},
	year={2019},
	publisher={Public Library of Science San Francisco, CA USA}
}

@article{jelinek1991computation,
	title={Computation of the probability of initial substring generation by stochastic context-free grammars},
	author={Jelinek, Frederick and Lafferty, John},
	journal={Computational Linguistics},
	volume={17},
	number={3},
	pages={315--353},
	year={1991}
}

@inproceedings{wehbe2014aligning,
	title={Aligning context-based statistical models of language with brain activity during reading},
	author={Wehbe, Leila and Vaswani, Ashish and Knight, Kevin and Mitchell, Tom},
	booktitle={Proceedings of the 2014 Conference on Empirical Methods in Natural Language Processing},
	pages={233--243},
	year={2014}
}

@article{toneva2019interpreting,
	title={Interpreting and improving natural-language processing (in machines) with natural language-processing (in the brain)},
	author={Toneva, Mariya and Wehbe, Leila},
	journal={Advances in Neural Information Processing Systems},
	volume={32},
	year={2019}
}

@article{smith2005construct,
	title={On construct validity: Issues of method and measurement.},
	author={Smith, Gregory T},
	journal={Psychological Assessment},
	volume={17},
	number={4},
	pages={396},
	year={2005},
	publisher={American Psychological Association}
}

@article{levenstein2023role,
	title={On the role of theory and modeling in neuroscience},
	author={Levenstein, Daniel and Alvarez, Veronica A and Amarasingham, Asohan and Azab, Habiba and Chen, Zhe S and Gerkin, Richard C and Hasenstaub, Andrea and Iyer, Ramakrishnan and Jolivet, Renaud B and Marzen, Sarah and others},
	journal={Journal of Neuroscience},
	volume={43},
	number={7},
	pages={1074--1088},
	year={2023},
	publisher={Society for Neuroscience}
}

@article{chen2025combined,
	title={Combined evidence from artificial neural networks and human brain-lesion models reveals that language modulates vision in human perception},
	author={Chen, Haoyang and Liu, Bo and Wang, Shuyue and Wang, Xiaosha and Han, Wenjuan and Wang, Xiaochun and Zhu, Yixin and Bi, Yanchao},
	journal={Nature Human Behaviour},
	pages={1--17},
	year={2025},
	publisher={Nature Publishing Group UK London}
}

@article{denk2025text,
	title={Text-to-music generation models capture musical semantic representations in the human brain},
	author={Denk, Timo I and Takagi, Yu and Matsuyama, Takuya and Agostinelli, Andrea and Nakai, Tomoya and Frank, Christian and Nishimoto, Shinji},
	journal={Nature Communications},
	year={2025},
	publisher={Nature Publishing Group UK London}
}

@article{pesnot2025shared,
	title={Shared sensitivity to data distribution during learning in humans and transformer networks},
	author={Pesnot Lerousseau, Jacques and Summerfield, Christopher},
	journal={Nature Human Behaviour},
	pages={1--14},
	year={2025},
	publisher={Nature Publishing Group UK London}
}

@article{llanos2025accented,
	title={Accented speech modulates multiple event-related potential components across multiple levels of language processing},
	author={Llanos, Fernando and Wu, Yunan Charles and Abel, Taylor J and Holt, Lori L},
	journal={Communications Psychology},
	volume={3},
	number={1},
	pages={186},
	year={2025},
	publisher={Nature Publishing Group UK London}
}

@article{zhou2025hierarchical,
	title={Hierarchical linguistic predictions and cross-level information updating during narrative comprehension},
	author={Zhou, Faxin and Zhou, Siyuan and Long, Yuhang and Flinker, Adeen and Lu, Chunming},
	journal={Communications Biology},
	year={2025},
	publisher={Nature Publishing Group UK London}
}

@article{boboqulovna2025linguistic,
	title={The Linguistic Levels: Classification, Characteristics, And Interrelationships},
	author={Boboqulovna, Kholova Madina},
	journal={International Journal of Literature and Languages},
	volume={5},
	number={04},
	pages={26--30},
	year={2025}
}

@book{fitch2010evolution,
	title={The evolution of language},
	author={Fitch, W Tecumseh},
	year={2010},
	publisher={Cambridge University Press}
}

@article{christiansen2008language,
	title={Language as shaped by the brain},
	author={Christiansen, Morten H and Chater, Nick},
	journal={Behavioral and Brain Sciences},
	volume={31},
	number={5},
	pages={489--509},
	year={2008},
	publisher={Cambridge University Press}
}

@article{demonet2005renewal,
	title={Renewal of the neurophysiology of language: Functional neuroimaging},
	author={D{\'e}monet, Jean-Fran{\c{c}}ois and Thierry, Guillaume and Cardebat, Dominique},
	journal={Physiological Reviews},
	volume={85},
	number={1},
	pages={49--95},
	year={2005},
	publisher={American Physiological Society}
}

@article{nasios2019broca,
	title={From Broca and Wernicke to the neuromodulation era: insights of brain language networks for neurorehabilitation},
	author={Nasios, Grigorios and Dardiotis, Efthymios and Messinis, Lambros},
	journal={Behavioural Neurology},
	volume={2019},
	number={1},
	pages={9894571},
	year={2019},
	publisher={Wiley Online Library}
}

@article{lansing2003complex,
	title={Complex adaptive systems},
	author={Lansing, J Stephen},
	journal={Annual Review of Anthropology},
	volume={32},
	number={1},
	pages={183--204},
	year={2003},
	publisher={Annual Reviews 4139 El Camino Way, PO Box 10139, Palo Alto, CA 94303-0139, USA}
}

@article{hopfield1982neural,
	title={Neural networks and physical systems with emergent collective computational abilities},
	author={Hopfield, John J},
	journal={Proceedings of the National Academy of Sciences},
	volume={79},
	number={8},
	pages={2554--2558},
	year={1982}
}

@article{hale2022neurocomputational,
	title={Neurocomputational models of language processing},
	author={Hale, John T and Campanelli, Luca and Li, Jixing and Bhattasali, Shohini and Pallier, Christophe and Brennan, Jonathan R},
	journal={Annual Review of Linguistics},
	volume={8},
	number={1},
	pages={427--446},
	year={2022},
	publisher={Annual Reviews}
}

@misc{lopopolo2024cognitive,
	title={Cognitive computational neuroscience of language: Using computational models to investigate language processing in the brain},
	author={Lopopolo, Alessandro and Fedorenko, Evelina and Levy, Roger and Rabovsky, Milena},
	journal={Neurobiology of Language},
	volume={5},
	number={1},
	pages={1--6},
	year={2024},
	publisher={MIT Press One Broadway, 12th Floor, Cambridge, Massachusetts 02142, USA~…}
}

@article{sun2020neural,
	title={Neural encoding and decoding with distributed sentence representations},
	author={Sun, Jingyuan and Wang, Shaonan and Zhang, Jiajun and Zong, Chengqing},
	journal={IEEE Transactions on Neural Networks and Learning Systems},
	volume={32},
	number={2},
	pages={589--603},
	year={2020},
	publisher={IEEE}
}

@article{shain2020fmri,
	title={fMRI reveals language-specific predictive coding during naturalistic sentence comprehension},
	author={Shain, Cory and Blank, Idan Asher and van Schijndel, Marten and Schuler, William and Fedorenko, Evelina},
	journal={Neuropsychologia},
	volume={138},
	pages={107307},
	year={2020},
	publisher={Elsevier}
}

@article{wang2020fine,
	title={Fine-grained neural decoding with distributed word representations},
	author={Wang, Shaonan and Zhang, Jiajun and Wang, Haiyan and Lin, Nan and Zong, Chengqing},
	journal={Information Sciences},
	volume={507},
	pages={256--272},
	year={2020},
	publisher={Elsevier}
}

@article{deniz2019representation,
	title={The representation of semantic information across human cerebral cortex during listening versus reading is invariant to stimulus modality},
	author={Deniz, Fatma and Nunez-Elizalde, Anwar O and Huth, Alexander G and Gallant, Jack L},
	journal={Journal of Neuroscience},
	volume={39},
	number={39},
	pages={7722--7736},
	year={2019},
	publisher={Society for Neuroscience}
}

@article{anderson2017predicting,
	title={Predicting neural activity patterns associated with sentences using a neurobiologically motivated model of semantic representation},
	author={Anderson, Andrew James and Binder, Jeffrey R and Fernandino, Leonardo and Humphries, Colin J and Conant, Lisa L and Aguilar, Mario and Wang, Xixi and Doko, Donias and Raizada, Rajeev DS},
	journal={Cerebral Cortex},
	volume={27},
	number={9},
	pages={4379--4395},
	year={2017},
	publisher={Oxford University Press}
}

@book{koehler2025compositional,
	title={Compositional intelligence: Architectural typology through generative AI},
	author={Koehler, Daniel},
	year={2025},
	publisher={Taylor \& Francis}
}

@book{dayan2005theoretical,
	title={Theoretical neuroscience: Computational and mathematical modeling of neural systems},
	author={Dayan, Peter and Abbott, Laurence F},
	year={2005},
	publisher={MIT press}
}

@incollection{healey2015toward,
	title={Toward a theoretical framework for organizational neuroscience},
	author={Healey, Mark P and Hodgkinson, Gerard P},
	booktitle={Organizational Neuroscience},
	volume={7},
	pages={51--81},
	year={2015},
	publisher={Emerald Group Publishing Limited}
}

@book{rieke1999spikes,
	title={Spikes: Exploring the neural code},
	author={Rieke, Fred and Warland, David and Van Steveninck, Rob de Ruyter and Bialek, William},
	year={1999},
	publisher={MIT press}
}

@article{olshausen2004sparse,
	title={Sparse coding of sensory inputs},
	author={Olshausen, Bruno A and Field, David J},
	journal={Current Opinion in Neurobiology},
	volume={14},
	number={4},
	pages={481--487},
	year={2004},
	publisher={Elsevier}
}

@article{pouget2013probabilistic,
	title={Probabilistic brains: Knowns and unknowns},
	author={Pouget, Alexandre and Beck, Jeffrey M and Ma, Wei Ji and Latham, Peter E},
	journal={Nature Neuroscience},
	volume={16},
	number={9},
	pages={1170--1178},
	year={2013},
	publisher={Nature Publishing Group US New York}
}

@article{frank2015erp,
	title={The ERP response to the amount of information conveyed by words in sentences},
	author={Frank, Stefan L and Otten, Leun J and Galli, Giulia and Vigliocco, Gabriella},
	journal={Brain and Language},
	volume={140},
	pages={1--11},
	year={2015},
	publisher={Elsevier}
}

@article{barlow1961possible,
	title={Possible principles underlying the transformation of sensory messages},
	author={Barlow, Horace B and others},
	journal={Sensory Communication},
	volume={1},
	number={01},
	pages={217--233},
	year={1961},
	publisher={Cambridge, MA: MIT Press}
}

@article{simoncelli2001natural,
	title={Natural image statistics and neural representation},
	author={Simoncelli, Eero P and Olshausen, Bruno A},
	journal={Annual Review of Neuroscience},
	volume={24},
	number={1},
	pages={1193--1216},
	year={2001},
	publisher={Annual Reviews 4139 El Camino Way, PO Box 10139, Palo Alto, CA 94303-0139, USA}
}

@article{knill2004bayesian,
	title={The Bayesian brain: The role of uncertainty in neural coding and computation},
	author={Knill, David C and Pouget, Alexandre},
	journal={Trends in Neurosciences},
	volume={27},
	number={12},
	pages={712--719},
	year={2004},
	publisher={Elsevier}
}

@article{kording2004bayesian,
	title={Bayesian integration in sensorimotor learning},
	author={K{\"o}rding, Konrad P and Wolpert, Daniel M},
	journal={Nature},
	volume={427},
	number={6971},
	pages={244--247},
	year={2004},
	publisher={Nature Publishing Group UK London}
}

@article{heilbron2022hierarchy,
	title={A hierarchy of linguistic predictions during natural language comprehension},
	author={Heilbron, Micha and Armeni, Kristijan and Schoffelen, Jan-Mathijs and Hagoort, Peter and De Lange, Floris P},
	journal={Proceedings of the National Academy of Sciences},
	volume={119},
	number={32},
	pages={e2201968119},
	year={2022},
	publisher={National Academy of Sciences}
}

@article{markram2006blue,
	title={The blue brain project},
	author={Markram, Henry},
	journal={Nature Reviews Neuroscience},
	volume={7},
	number={2},
	pages={153--160},
	year={2006},
	publisher={Nature Publishing Group UK London}
}

@article{eliasmith2012large,
	title={A large-scale model of the functioning brain},
	author={Eliasmith, Chris and Stewart, Terrence C and Choo, Xuan and Bekolay, Trevor and DeWolf, Travis and Tang, Yichuan and Rasmussen, Daniel},
	journal={Science},
	volume={338},
	number={6111},
	pages={1202--1205},
	year={2012},
	publisher={American Association for the Advancement of Science}
}

@article{brennan2016abstract,
	title={Abstract linguistic structure correlates with temporal activity during naturalistic comprehension},
	author={Brennan, Jonathan R and Stabler, Edward P and Van Wagenen, Sarah E and Luh, Wen-Ming and Hale, John T},
	journal={Brain and Language},
	volume={157},
	pages={81--94},
	year={2016},
	publisher={Elsevier}
}

@article{kriegeskorte2018cognitive,
	title={Cognitive computational neuroscience},
	author={Kriegeskorte, Nikolaus and Douglas, Pamela K},
	journal={Nature Neuroscience},
	volume={21},
	number={9},
	pages={1148--1160},
	year={2018},
	publisher={Nature Publishing Group US New York}
}

@article{schrimpf2021neural,
	title={The neural architecture of language: Integrative modeling converges on predictive processing},
	author={Schrimpf, Martin and Blank, Idan Asher and Tuckute, Greta and Kauf, Carina and Hosseini, Eghbal A and Kanwisher, Nancy and Tenenbaum, Joshua B and Fedorenko, Evelina},
	journal={Proceedings of the National Academy of Sciences},
	volume={118},
	number={45},
	pages={e2105646118},
	year={2021},
	publisher={National Academy of Sciences}
}

@article{goldstein2022shared,
	title={Shared computational principles for language processing in humans and deep language models},
	author={Goldstein, Ariel and Zada, Zaid and Buchnik, Eliav and Schain, Mariano and Price, Amy and Aubrey, Bobbi and Nastase, Samuel A and Feder, Amir and Emanuel, Dotan and Cohen, Alon and others},
	journal={Nature Neuroscience},
	volume={25},
	number={3},
	pages={369--380},
	year={2022},
	publisher={Nature Publishing Group US New York}
}

@article{caucheteux2022brains,
	title={Brains and algorithms partially converge in natural language processing},
	author={Caucheteux, Charlotte and King, Jean-R{\'e}mi},
	journal={Communications Biology},
	volume={5},
	number={1},
	pages={134},
	year={2022},
	publisher={Nature Publishing Group UK London}
}

@article{an2025hierarchical,
	title={Hierarchical Frequency Tagging Probe (HFTP): A unified approach to investigate syntactic structure representations in large language models and the human brain},
	author={An, Jingmin and Song, Yilong and Yang, Ruolin and Ding, Nai and Lu, Lingxi and Wang, Yuxuan and Wang, Wei and Zhuang, Chu and Wang, Qian and Fang, Fang},
	journal={ArXiv Preprint ArXiv:2510.13255},
	year={2025}
}

@article{wang2024computational,
	title={Computational models to study language processing in the human brain: A survey},
	author={Wang, Shaonan and Sun, Jingyuan and Zhang, Yunhao and Lin, Nan and Moens, Marie-Francine and Zong, Chengqing},
	journal={ArXiv Preprint ArXiv:2403.13368},
	year={2024}
}

@inproceedings{cai2024large,
	title={Do large language models resemble humans in language use?},
	author={Cai, Zhenguang and Duan, Xufeng and Haslett, David and Wang, Shuqi and Pickering, Martin},
	booktitle={Proceedings of the Workshop on Cognitive Modeling and Computational Linguistics},
	pages={37--56},
	year={2024}
}

@article{caucheteux2023evidence,
	title={Evidence of a predictive coding hierarchy in the human brain listening to speech},
	author={Caucheteux, Charlotte and Gramfort, Alexandre and King, Jean-R{\'e}mi},
	journal={Nature Human Behaviour},
	volume={7},
	number={3},
	pages={430--441},
	year={2023},
	publisher={Nature Publishing Group UK London}
}

@inproceedings{radford2018improving,
	title={Improving language understanding by generative pre-training},
	author={Alec Radford and Karthik Narasimhan},
	year={2018},
}

@article{oniani2024emerging,
	title={Emerging opportunities of using large language models for translation between drug molecules and indications},
	author={Oniani, David and Hilsman, Jordan and Zang, Chengxi and Wang, Junmei and Cai, Lianjin and Zawala, Jan and Wang, Yanshan},
	journal={Scientific Reports},
	volume={14},
	number={1},
	pages={10738},
	year={2024},
	publisher={Nature Publishing Group UK London}
}

@article{doerig2025high,
	title={High-level visual representations in the human brain are aligned with large language models},
	author={Doerig, Adrien and Kietzmann, Tim C and Allen, Emily and Wu, Yihan and Naselaris, Thomas and Kay, Kendrick and Charest, Ian},
	journal={Nature Machine Intelligence},
	volume={7},
	number={8},
	pages={1220--1234},
	year={2025},
	publisher={Nature Publishing Group UK London}
}

@article{hickok2007cortical,
	title={The cortical organization of speech processing},
	author={Hickok, Gregory and Poeppel, David},
	journal={Nature Reviews Neuroscience},
	volume={8},
	number={5},
	pages={393--402},
	year={2007},
	publisher={Nature Publishing Group UK London}
}

@article{friederici2013language,
	title={The language network},
	author={Friederici, Angela D and Gierhan, Sarah ME},
	journal={Current Opinion in Neurobiology},
	volume={23},
	number={2},
	pages={250--254},
	year={2013},
	publisher={Elsevier}
}

@article{turker2023cortical,
	title={Cortical, subcortical, and cerebellar contributions to language processing: A meta-analytic review of 403 neuroimaging experiments},
	author={Turker, Sabrina and Kuhnke, Philipp and Eickhoff, Simon B and Caspers, Svenja and Hartwigsen, Gesa},
	journal={Psychological Bulletin},
	volume={149},
	number={11-12},
	pages={699},
	year={2023},
	publisher={American Psychological Association}
}

@article{karunathilake2025neural,
	title={Neural dynamics of the processing of speech features: Evidence for a progression of features from acoustic to sentential processing},
	author={Karunathilake, IM Dushyanthi and Brodbeck, Christian and Bhattasali, Shohini and Resnik, Philip and Simon, Jonathan Z},
	journal={Journal of Neuroscience},
	volume={45},
	number={11},
	year={2025},
	publisher={Society for Neuroscience}
}

@article{gross2019magnetoencephalography,
	title={Magnetoencephalography in cognitive neuroscience: A primer},
	author={Gross, Joachim},
	journal={Neuron},
	volume={104},
	number={2},
	pages={189--204},
	year={2019},
	publisher={Elsevier}
}

@article{worsley1996unified,
	title={A unified statistical approach for determining significant signals in images of cerebral activation},
	author={Worsley, Keith J and Marrett, Sean and Neelin, Peter and Vandal, Alain C and Friston, Karl J and Evans, Alan C},
	journal={Human Brain Mapping},
	volume={4},
	number={1},
	pages={58--73},
	year={1996},
	publisher={Wiley Online Library}
}

@article{ogawa1990brain,
	title={Brain magnetic resonance imaging with contrast dependent on blood oxygenation},
	author={Ogawa, Seiji and Lee, Tso-Ming and Kay, Alan R and Tank, David W},
	journal={Proceedings of the National Academy of Sciences},
	volume={87},
	number={24},
	pages={9868--9872},
	year={1990}
}

@article{binder1997human,
	title={Human brain language areas identified by functional magnetic resonance imaging},
	author={Binder, Jeffrey R and Frost, Julie A and Hammeke, Thomas A and Cox, Robert W and Rao, Stephen M and Prieto, Thomas},
	journal={Journal of Neuroscience},
	volume={17},
	number={1},
	pages={353--362},
	year={1997},
	publisher={Society for Neuroscience}
}

@article{price2012review,
	title={A review and synthesis of the first 20 years of PET and fMRI studies of heard speech, spoken language and reading},
	author={Price, Cathy J},
	journal={Neuroimage},
	volume={62},
	number={2},
	pages={816--847},
	year={2012},
	publisher={Elsevier}
}

@article{blank2014functional,
	title={A functional dissociation between language and multiple-demand systems revealed in patterns of BOLD signal fluctuations},
	author={Blank, Idan and Kanwisher, Nancy and Fedorenko, Evelina},
	journal={Journal of Neurophysiology},
	volume={112},
	number={5},
	pages={1105--1118},
	year={2014},
	publisher={American Physiological Society Bethesda, MD}
}

@article{vigneau2006meta,
	title={Meta-analyzing left hemisphere language areas: phonology, semantics, and sentence processing},
	author={Vigneau, Mathieu and Beaucousin, Virginie and Herv{\'e}, Pierre-Yves and Duffau, Hugues and Crivello, Fabrice and Houde, Olivier and Mazoyer, Bernard and Tzourio-Mazoyer, Nathalie},
	journal={Neuroimage},
	volume={30},
	number={4},
	pages={1414--1432},
	year={2006},
	publisher={Elsevier}
}

@article{glover1999deconvolution,
	title={Deconvolution of impulse response in event-related BOLD fMRI1},
	author={Glover, Gary H},
	journal={Neuroimage},
	volume={9},
	number={4},
	pages={416--429},
	year={1999},
	publisher={Elsevier}
}

@article{matthews2004functional,
	title={Functional magnetic resonance imaging},
	author={Matthews, Paul M and Jezzard, Peter},
	journal={Journal of Neurology, Neurosurgery \& Psychiatry},
	volume={75},
	number={1},
	pages={6--12},
	year={2004},
	publisher={BMJ Publishing Group Ltd}
}

@book{nunez2006electric,
	title={Electric fields of the brain: The neurophysics of EEG},
	author={Nunez, Paul L and Srinivasan, Ramesh},
	year={2006},
	publisher={Oxford university press}
}

@article{makeig1995independent,
	title={Independent component analysis of electroencephalographic data},
	author={Makeig, Scott and Bell, Anthony and Jung, Tzyy-Ping and Sejnowski, Terrence J},
	journal={Advances in Neural Information Processing Systems},
	volume={8},
	year={1995}
}

@article{hamalainen1993magnetoencephalography,
	title={Magnetoencephalography—theory, instrumentation, and applications to noninvasive studies of the working human brain},
	author={H{\"a}m{\"a}l{\"a}inen, Matti and Hari, Riitta and Ilmoniemi, Risto J and Knuutila, Jukka and Lounasmaa, Olli V},
	journal={Reviews of Modern Physics},
	volume={65},
	number={2},
	pages={413},
	year={1993},
	publisher={APS}
}

@book{hansen2010meg,
	title={MEG: An introduction to methods},
	author={Hansen, Peter and Kringelbach, Morten and Salmelin, Riitta},
	year={2010},
	publisher={Oxford university press}
}

@article{cohen1968magnetoencephalography,
	title={Magnetoencephalography: Evidence of magnetic fields produced by alpha-rhythm currents},
	author={Cohen, David},
	journal={Science},
	volume={161},
	number={3843},
	pages={784--786},
	year={1968},
	publisher={American Association for the Advancement of Science}
}

@article{schmidt2009second,
	title={Second language research using magnetoencephalography: A review},
	author={Schmidt, Gwen L and Roberts, Timothy PL},
	journal={Second Language Research},
	volume={25},
	number={1},
	pages={135--166},
	year={2009},
	publisher={SAGE Publications Sage UK: London, England}
}

@article{kutas1980reading,
	title={Reading senseless sentences: Brain potentials reflect semantic incongruity},
	author={Kutas, Marta and Hillyard, Steven A},
	journal={Science},
	volume={207},
	number={4427},
	pages={203--205},
	year={1980},
	publisher={American Association for the Advancement of Science}
}

@article{osterhout1992event,
	title={Event-related brain potentials elicited by syntactic anomaly},
	author={Osterhout, Lee and Holcomb, Phillip J},
	journal={Journal of Memory and Language},
	volume={31},
	number={6},
	pages={785--806},
	year={1992},
	publisher={Elsevier}
}

@book{luck2014introduction,
	title={An introduction to the event-related potential technique},
	author={Luck, Steven J},
	year={2014},
	publisher={MIT press}
}

@article{kutas2011thirty,
	title={Thirty years and counting: Finding meaning in the N400 component of the event-related brain potential (ERP)},
	author={Kutas, Marta and Federmeier, Kara D},
	journal={Annual Review of Psychology},
	volume={62},
	number={1},
	pages={621--647},
	year={2011},
	publisher={Annual Reviews}
}

@article{kriegeskorte2008representational,
	title={Representational similarity analysis-connecting the branches of systems neuroscience},
	author={Kriegeskorte, Nikolaus and Mur, Marieke and Bandettini, Peter A},
	journal={Frontiers in Systems Neuroscience},
	volume={2},
	pages={249},
	year={2008},
	publisher={Frontiers}
}

@article{harris1954distributional,
	title={Distributional structure},
	author={Harris, Zellig S},
	journal={Word},
	volume={10},
	number={2-3},
	pages={146--162},
	year={1954},
	publisher={Taylor \& Francis}
}

@article{mikolov2013efficient,
	title={Efficient estimation of word representations in vector space},
	author={Mikolov, Tomas and Chen, Kai and Corrado, Greg and Dean, Jeffrey},
	journal={ArXiv Preprint ArXiv:1301.3781},
	year={2013}
}

@inproceedings{pennington2014glove,
	title={Glove: Global vectors for word representation},
	author={Pennington, Jeffrey and Socher, Richard and Manning, Christopher D},
	booktitle={Proceedings of the 2014 Conference on Empirical Methods in Natural Language Processing},
	pages={1532--1543},
	year={2014}
}

@article{church2017word2vec,
	title={Word2Vec},
	author={Church, Kenneth Ward},
	journal={Natural Language Engineering},
	volume={23},
	number={1},
	pages={155--162},
	year={2017},
	publisher={Cambridge University Press}
}

@article{koroteev2021bert,
	title={BERT: A review of applications in natural language processing and understanding},
	author={Koroteev, Mikhail V},
	journal={ArXiv Preprint ArXiv:2103.11943},
	year={2021}
}

@article{mitchell2010composition,
	title={Composition in distributional models of semantics},
	author={Mitchell, Jeff and Lapata, Mirella},
	journal={Cognitive Science},
	volume={34},
	number={8},
	pages={1388--1429},
	year={2010},
	publisher={Wiley Online Library}
}

@inproceedings{pagliardini2018unsupervised,
	title={Unsupervised learning of sentence embeddings using compositional n-gram features},
	author={Pagliardini, Matteo and Gupta, Prakhar and Jaggi, Martin},
	booktitle={Proceedings of the 2018 Conference of the North American Chapter of the Association for Computational Linguistics: Human Language Technologies, Volume 1},
	pages={528--540},
	year={2018}
}

@article{cichy2019deep,
	title={Deep neural networks as scientific models},
	author={Cichy, Radoslaw M and Kaiser, Daniel},
	journal={Trends in Cognitive Sciences},
	volume={23},
	number={4},
	pages={305--317},
	year={2019},
	publisher={Elsevier}
}

@article{smolensky1990tensor,
	title={Tensor product variable binding and the representation of symbolic structures in connectionist systems},
	author={Smolensky, Paul},
	journal={Artificial Intelligence},
	volume={46},
	number={1-2},
	pages={159--216},
	year={1990},
	publisher={Elsevier}
}

@inproceedings{mitchell2008vector,
	title={Vector-based models of semantic composition},
	author={Mitchell, Jeff and Lapata, Mirella},
	booktitle={Proceedings of ACL-08: HLT},
	pages={236--244},
	year={2008}
}

@article{milajevs2014evaluating,
	title={Evaluating neural word representations in tensor-based compositional settings},
	author={Milajevs, Dmitrijs and Kartsaklis, Dimitri and Sadrzadeh, Mehrnoosh and Purver, Matthew},
	journal={ArXiv Preprint ArXiv:1408.6179},
	year={2014}
}

@inproceedings{clark2008compositional,
	title={A compositional distributional model of meaning},
	author={Clark, Stephen and Coecke, Bob and Sadrzadeh, Mehrnoosh},
	booktitle={Proceedings of the Second Quantum Interaction Symposium},
	pages={133--140},
	year={2008},
	organization={Oxford}
}

@article{conneau2017supervised,
	title={Supervised learning of universal sentence representations from natural language inference data},
	author={Conneau, Alexis and Kiela, Douwe and Schwenk, Holger and Barrault, Lo{\"\i}c and Bordes, Antoine},
	journal={ArXiv Preprint ArXiv:1705.02364},
	year={2017}
}

@article{naik2018stress,
	title={Stress test evaluation for natural language inference},
	author={Naik, Aakanksha and Ravichander, Abhilasha and Sadeh, Norman and Rose, Carolyn and Neubig, Graham},
	journal={ArXiv Preprint ArXiv:1806.00692},
	year={2018}
}

@article{belinkov2019analysis,
	title={Analysis methods in neural language processing: A survey},
	author={Belinkov, Yonatan and Glass, James},
	journal={Transactions of the Association for Computational Linguistics},
	volume={7},
	pages={49--72},
	year={2019},
	publisher={MIT Press One Rogers Street, Cambridge, MA 02142-1209, USA journals-info~…}
}

@article{landauer1997solution,
	title={A solution to Plato's problem: The latent semantic analysis theory of acquisition, induction, and representation of knowledge},
	author={Landauer, Thomas K and Dumais, Susan T},
	journal={Psychological Review},
	volume={104},
	number={2},
	pages={211},
	year={1997},
	publisher={American Psychological Association}
}

@inproceedings{devlin2019bert,
	title={Bert: Pre-training of deep bidirectional transformers for language understanding},
	author={Devlin, Jacob and Chang, Ming-Wei and Lee, Kenton and Toutanova, Kristina},
	booktitle={Proceedings of the 2019 Conference of the North American Chapter of the Association for Computational Linguistics: Human Language Technologies, Volume 1 },
	pages={4171--4186},
	year={2019}
}

@inproceedings{baroni2010nouns,
	title={Nouns are vectors, adjectives are matrices: Representing adjective-noun constructions in semantic space},
	author={Baroni, Marco and Zamparelli, Roberto},
	booktitle={Proceedings of the 2010 Conference on Empirical Methods in Natural Language Processing},
	pages={1183--1193},
	year={2010}
}

@article{mitchell2008predicting,
	title={Predicting human brain activity associated with the meanings of nouns},
	author={Mitchell, Tom M and Shinkareva, Svetlana V and Carlson, Andrew and Chang, Kai-Min and Malave, Vicente L and Mason, Robert A and Just, Marcel Adam},
	journal={Science},
	volume={320},
	number={5880},
	pages={1191--1195},
	year={2008},
	publisher={American Association for the Advancement of Science}
}

@article{kietzmann2017deep,
	title={Deep neural networks in computational neuroscience},
	author={Kietzmann, Tim C and McClure, Patrick and Kriegeskorte, Nikolaus},
	journal={BioRxiv},
	pages={133504},
	year={2017},
	publisher={Cold Spring Harbor Laboratory}
}

@article{huth2016natural,
	title={Natural speech reveals the semantic maps that tile human cerebral cortex},
	author={Huth, Alexander G and De Heer, Wendy A and Griffiths, Thomas L and Theunissen, Fr{\'e}d{\'e}ric E and Gallant, Jack L},
	journal={Nature},
	volume={532},
	number={7600},
	pages={453--458},
	year={2016},
	publisher={Nature Publishing Group UK London}
}

@article{naselaris2009bayesian,
	title={Bayesian reconstruction of natural images from human brain activity},
	author={Naselaris, Thomas and Prenger, Ryan J and Kay, Kendrick N and Oliver, Michael and Gallant, Jack L},
	journal={Neuron},
	volume={63},
	number={6},
	pages={902--915},
	year={2009},
	publisher={Elsevier}
}

@article{ettinger2020bert,
	title={What BERT is not: Lessons from a new suite of psycholinguistic diagnostics for language models},
	author={Ettinger, Allyson},
	journal={Transactions of the Association for Computational Linguistics},
	volume={8},
	pages={34--48},
	year={2020},
	publisher={MIT Press One Rogers Street, Cambridge, MA 02142-1209, USA journals-info~…}
}

@article{cuskley2024limitations,
	title={The limitations of large language models for understanding human language and cognition},
	author={Cuskley, Christine and Woods, Rebecca and Flaherty, Molly},
	journal={Open Mind},
	volume={8},
	pages={1058--1083},
	year={2024},
	publisher={MIT Press 255 Main Street, 9th Floor, Cambridge, Massachusetts 02142, USA~…}
}

@article{da2020channel,
	title={Channel capacity in brain--computer interfaces},
	author={da Silva Costa, Thiago Bulhoes and Uribe, Luisa Fernanda Suarez and de Carvalho, Sarah Negreiros and Soriano, Diogo Coutinho and Castellano, Gabriela and Suyama, Ricardo and Attux, Romis and Panazio, Cristiano},
	journal={Journal of Neural Engineering},
	volume={17},
	number={1},
	pages={016060},
	year={2020},
	publisher={IOP Publishing}
}

@article{zada2025linguistic,
	title={Linguistic coupling between neural systems for speech production and comprehension during real-time dyadic conversations},
	author={Zada, Zaid and Nastase, Samuel A and Speer, Sebastian and Mwilambwe-Tshilobo, Laetitia and Tsoi, Lily and Burns, Shannon M and Falk, Emily and Hasson, Uri and Tamir, Diana I},
	journal={Neuron},
	year={2025},
	publisher={Elsevier}
}

@article{brown2020language,
	title={Language models are few-shot learners},
	author={Brown, Tom and Mann, Benjamin and Ryder, Nick and Subbiah, Melanie and Kaplan, Jared D and Dhariwal, Prafulla and Neelakantan, Arvind and Shyam, Pranav and Sastry, Girish and Askell, Amanda and others},
	journal={Advances in Neural Information Processing Systems},
	volume={33},
	pages={1877--1901},
	year={2020}
}

@article{radford2019language,
	title={Language models are unsupervised multitask learners},
	author={Radford, Alec and Wu, Jeffrey and Child, Rewon and Luan, David and Amodei, Dario and Sutskever, Ilya and others},
	journal={OpenAI Blog},
	volume={1},
	number={8},
	pages={9},
	year={2019}
}

@article{ouyang2022training,
	title={Training language models to follow instructions with human feedback},
	author={Ouyang, Long and Wu, Jeffrey and Jiang, Xu and Almeida, Diogo and Wainwright, Carroll and Mishkin, Pamela and Zhang, Chong and Agarwal, Sandhini and Slama, Katarina and Ray, Alex and others},
	journal={Advances in Neural Information Processing Systems},
	volume={35},
	pages={27730--27744},
	year={2022}
}

@article{achiam2023gpt,
	title={Gpt-4 technical report},
	author={Achiam, Josh and Adler, Steven and Agarwal, Sandhini and Ahmad, Lama and Akkaya, Ilge and Aleman, Florencia Leoni and Almeida, Diogo and Altenschmidt, Janko and Altman, Sam and Anadkat, Shyamal and others},
	journal={ArXiv Preprint ArXiv:2303.08774},
	year={2023}
}

@article{sun2019ernie,
	title={Ernie: Enhanced representation through knowledge integration},
	author={Sun, Yu and Wang, Shuohuan and Li, Yukun and Feng, Shikun and Chen, Xuyi and Zhang, Han and Tian, Xin and Zhu, Danxiang and Tian, Hao and Wu, Hua},
	journal={ArXiv Preprint ArXiv:1904.09223},
	year={2019}
}

@inproceedings{du2022glm,
	title={Glm: General language model pretraining with autoregressive blank infilling},
	author={Du, Zhengxiao and Qian, Yujie and Liu, Xiao and Ding, Ming and Qiu, Jiezhong and Yang, Zhilin and Tang, Jie},
	booktitle={Proceedings of the 60th Annual Meeting of the Association for Computational Linguistics, Volume 1},
	pages={320--335},
	year={2022}
}

@article{zeng2025glm,
	title={Glm-4.5: Agentic, reasoning, and coding (arc) foundation models},
	author={Zeng, Aohan and Lv, Xin and Zheng, Qinkai and Hou, Zhenyu and Chen, Bin and Xie, Chengxing and Wang, Cunxiang and Yin, Da and Zeng, Hao and Zhang, Jiajie and others},
	journal={ArXiv Preprint ArXiv:2508.06471},
	year={2025}
}

@article{hong2025glm,
	title={GLM-4.1 V-Thinking: Towards versatile multimodal reasoning with scalable reinforcement learning},
	author={Hong, Wenyi and Yu, Wenmeng and Gu, Xiaotao and Wang, Guo and Gan, Guobing and Tang, Haomiao and Cheng, Jiale and Qi, Ji and Ji, Junhui and Pan, Lihang and others},
	journal={ArXiv Preprint ArXiv:2507.01006},
	year={2025}
}

@article{liu2019roberta,
	title={Roberta: A robustly optimized bert pretraining approach},
	author={Liu, Yinhan and Ott, Myle and Goyal, Naman and Du, Jingfei and Joshi, Mandar and Chen, Danqi and Levy, Omer and Lewis, Mike and Zettlemoyer, Luke and Stoyanov, Veselin},
	journal={ArXiv Preprint ArXiv:1907.11692},
	year={2019}
}

@article{lan2019albert,
	title={Albert: A lite bert for self-supervised learning of language representations},
	author={Lan, Zhenzhong and Chen, Mingda and Goodman, Sebastian and Gimpel, Kevin and Sharma, Piyush and Soricut, Radu},
	journal={ArXiv Preprint ArXiv:1909.11942},
	year={2019}
}

@article{joshi2020spanbert,
	title={Spanbert: Improving pre-training by representing and predicting spans},
	author={Joshi, Mandar and Chen, Danqi and Liu, Yinhan and Weld, Daniel S and Zettlemoyer, Luke and Levy, Omer},
	journal={Transactions of the Association for Computational Linguistics},
	volume={8},
	pages={64--77},
	year={2020},
	publisher={MIT Press One Rogers Street, Cambridge, MA 02142-1209, USA journals-info~…}
}

@article{raffel2020exploring,
	title={Exploring the limits of transfer learning with a unified text-to-text transformer},
	author={Raffel, Colin and Shazeer, Noam and Roberts, Adam and Lee, Katherine and Narang, Sharan and Matena, Michael and Zhou, Yanqi and Li, Wei and Liu, Peter J},
	journal={Journal of Machine Learning Research},
	volume={21},
	number={140},
	pages={1--67},
	year={2020}
}

@inproceedings{xue2021mt5,
	title={mT5: A massively multilingual pre-trained text-to-text transformer},
	author={Xue, Linting and Constant, Noah and Roberts, Adam and Kale, Mihir and Al-Rfou, Rami and Siddhant, Aditya and Barua, Aditya and Raffel, Colin},
	booktitle={Proceedings of the 2021 Conference of the North American Chapter of the Association for Computational Linguistics: Human Language Technologies},
	pages={483--498},
	year={2021}
}

@article{team2023gemini,
	title={Gemini: A family of highly capable multimodal models},
	author={Team, Gemini and Anil, Rohan and Borgeaud, Sebastian and Alayrac, Jean-Baptiste and Yu, Jiahui and Soricut, Radu and Schalkwyk, Johan and Dai, Andrew M and Hauth, Anja and Millican, Katie and others},
	journal={ArXiv Preprint ArXiv:2312.11805},
	year={2023}
}

@article{team2024gemini,
	title={Gemini 1.5: Unlocking multimodal understanding across millions of tokens of context},
	author={Team, Gemini and Georgiev, Petko and Lei, Ving Ian and Burnell, Ryan and Bai, Libin and Gulati, Anmol and Tanzer, Garrett and Vincent, Damien and Pan, Zhufeng and Wang, Shibo and others},
	journal={ArXiv Preprint ArXiv:2403.05530},
	year={2024}
}

@article{comanici2025gemini,
	title={Gemini 2.5: Pushing the frontier with advanced reasoning, multimodality, long context, and next generation agentic capabilities},
	author={Comanici, Gheorghe and Bieber, Eric and Schaekermann, Mike and Pasupat, Ice and Sachdeva, Noveen and Dhillon, Inderjit and Blistein, Marcel and Ram, Ori and Zhang, Dan and Rosen, Evan and others},
	journal={ArXiv Preprint ArXiv:2507.06261},
	year={2025}
}

@inproceedings{alkhamissi2025language,
	title={From language to cognition: How llms outgrow the human language network},
	author={AlKhamissi, Badr and Tuckute, Greta and Tang, Yingtian and Binhuraib, Taha Osama A and Bosselut, Antoine and Schrimpf, Martin},
	booktitle={Proceedings of the 2025 Conference on Empirical Methods in Natural Language Processing},
	pages={24332--24350},
	year={2025}
}

@article{xiao2025exploring,
	title={Exploring similarity between neural and LLM trajectories in language processing},
	author={Xiao, Xin and Wei, Kaiwen and Zhong, Jiang and Yin, Dongshuo and Tian, Yu and Wei, Xuekai and Zhou, Mingliang},
	journal={ArXiv Preprint ArXiv:2509.24307},
	year={2025}
}

@article{proietti2025fine,
	title={Fine-grained analysis of brain-LLM alignment through input attribution},
	author={Proietti, Michela and Capobianco, Roberto and Toneva, Mariya},
	journal={ArXiv Preprint ArXiv:2510.12355},
	year={2025}
}

@article{maini2025beyondweb,
	title={Beyondweb: Lessons from scaling synthetic data for trillion-scale pretraining},
	author={Maini, Pratyush and Dorna, Vineeth and Doshi, Parth and Carranza, Aldo and Pan, Fan and Urbanek, Jack and Burstein, Paul and Fang, Alex and Deng, Alvin and Abbas, Amro and others},
	journal={ArXiv Preprint ArXiv:2508.10975},
	year={2025}
}

@article{xing2025pretrainzero,
	title={PretrainZero: Reinforcement Active Pretraining},
	author={Xing, Xingrun and Fan, Zhiyuan and Lou, Jie and Li, Guoqi and Zhang, Jiajun and Zhang, Debing},
	journal={ArXiv Preprint ArXiv:2512.03442},
	year={2025}
}

@inproceedings{wang2025multilingual,
	title={Multilingual language model pretraining using machine-translated data},
	author={Wang, Jiayi and Lu, Yao and Weber, Maurice and Ryabinin, Max and Adelani, David Ifeoluwa and Chen, Yihong and Tang, Raphael and Stenetorp, Pontus},
	booktitle={Proceedings of the 2025 Conference on Empirical Methods in Natural Language Processing},
	pages={28075--28095},
	year={2025}
}

@article{shahriar2025inceptive,
	title={Inceptive transformers: Enhancing contextual representations through multi-scale feature learning across domains and languages},
	author={Shahriar, Asif and Shahriyar, Rifat and Rahman, M Saifur},
	journal={ArXiv Preprint ArXiv:2505.20496},
	year={2025}
}

@book{beinborn2023cognitive,
	title={Cognitive plausibility in natural language processing},
	author={Beinborn, Lisa and Hollenstein, Nora},
	year={2023},
	publisher={Springer}
}

@article{wang2023aligning,
	title={Aligning large language models with human: A survey},
	author={Wang, Yufei and Zhong, Wanjun and Li, Liangyou and Mi, Fei and Zeng, Xingshan and Huang, Wenyong and Shang, Lifeng and Jiang, Xin and Liu, Qun},
	journal={ArXiv Preprint ArXiv:2307.12966},
	year={2023}
}

@inproceedings{hu2023fine,
	title={A fine-grained comparison of pragmatic language understanding in humans and language models},
	author={Hu, Jennifer and Floyd, Sammy and Jouravlev, Olessia and Fedorenko, Evelina and Gibson, Edward},
	booktitle={Proceedings of the 61st Annual Meeting of the Association for Computational Linguistics, Volume 1},
	pages={4194--4213},
	year={2023}
}

@article{singh2025evaluating,
	title={Evaluating scientific theories as predictive models in language neuroscience},
	author={Singh, Chandan and Antonello, Richard J and Guo, Sihang and Mischler, Gavin and Gao, Jianfeng and Mesgarani, Nima and Huth, Alexander G},
	journal={BioRxiv},
	year={2025}
}

@article{ye2025generative,
	title={Generative language reconstruction from brain recordings},
	author={Ye, Ziyi and Ai, Qingyao and Liu, Yiqun and de Rijke, Maarten and Zhang, Min and Lioma, Christina and Ruotsalo, Tuukka},
	journal={Communications Biology},
	volume={8},
	number={1},
	pages={346},
	year={2025},
	publisher={Nature Publishing Group UK London}
}

@article{lei2025large,
	title={Do large language models think like the brain? Sentence-level evidence from fMRI and hierarchical embeddings},
	author={Lei, Yu and Ge, Xingyang and Zhang, Yi and Yang, Yiming and Ma, Bolei},
	journal={ArXiv Preprint ArXiv:2505.22563},
	year={2025}
}

@article{lopez2025brain,
	title={Brain-language model alignment: Insights into the platonic hypothesis and intermediate-layer advantage},
	author={L{\'o}pez-Cardona, {\'A}ngela and Idesis, Sebasti{\'a}n and Masias-Bruns, Mireia and Abadal, Sergi and Arapakis, Ioannis},
	journal={ArXiv Preprint ArXiv:2510.17833},
	year={2025}
}

@book{whorf2012language,
	title={Language, thought, and reality: Selected writings of Benjamin Lee Whorf},
	author={Whorf, Benjamin Lee},
	year={2012},
	publisher={MIT press}
}

@book{berlin1991basic,
	title={Basic color terms: Their universality and evolution},
	author={Berlin, Brent and Kay, Paul},
	year={1991},
	publisher={Univ of California Press}
}

@article{webster2012color,
	title={Color categories and color appearance},
	author={Webster, Michael A and Kay, Paul},
	journal={Cognition},
	volume={122},
	number={3},
	pages={375--392},
	year={2012},
	publisher={Elsevier}
}

@article{kay2006language,
	title={Language, thought and color: Recent developments},
	author={Kay, Paul and Regier, Terry},
	journal={Trends in Cognitive Sciences},
	volume={10},
	number={2},
	pages={51--54},
	year={2006},
	publisher={Elsevier}
}

@article{gilbert2006whorf,
	title={Whorf hypothesis is supported in the right visual field but not the left},
	author={Gilbert, Aubrey L and Regier, Terry and Kay, Paul and Ivry, Richard B},
	journal={Proceedings of the National Academy of Sciences},
	volume={103},
	number={2},
	pages={489--494},
	year={2006},
	publisher={National Academy of Sciences}
}

@article{button2013power,
	title={Power failure: Why small sample size undermines the reliability of neuroscience},
	author={Button, Katherine S and Ioannidis, John PA and Mokrysz, Claire and Nosek, Brian A and Flint, Jonathan and Robinson, Emma SJ and Munaf{\`o}, Marcus R},
	journal={Nature Reviews Neuroscience},
	volume={14},
	number={5},
	pages={365--376},
	year={2013},
	publisher={Nature Publishing Group UK London}
}

@article{thierry2009unconscious,
	title={Unconscious effects of language-specific terminology on preattentive color perception},
	author={Thierry, Guillaume and Athanasopoulos, Panos and Wiggett, Alison and Dering, Benjamin and Kuipers, Jan-Rouke},
	journal={Proceedings of the National Academy of Sciences},
	volume={106},
	number={11},
	pages={4567--4570},
	year={2009},
	publisher={National Academy of Sciences}
}

@inproceedings{yang2025clip,
	title={CLIP-MSM: A multi-semantic mapping brain representation for human high-level visual cortex},
	author={Yang, Guoyuan and Xue, Mufan and Mao, Ziming and Zheng, Haofang and Xu, Jia and Sheng, Dabin and Sun, Ruotian and Yang, Ruoqi and Li, Xuesong},
	booktitle={Proceedings of the AAAI Conference on Artificial Intelligence},
	pages={9184--9192},
	year={2025}
}

@article{liu2025data,
	title={Data or language supervision: What makes CLIP better than DINO?},
	author={Liu, Yiming and Zhang, Yuhui and Ghosh, Dhruba and Schmidt, Ludwig and Yeung-Levy, Serena},
	journal={ArXiv Preprint ArXiv:2510.11835},
	year={2025}
}

@article{chen2025language,
	title={Language modulates vision: Evidence from neural networks and human brain-lesion models},
	author={Chen, Haoyang and Liu, Bo and Wang, Shuyue and Wang, Xiaosha and Han, Wenjuan and Zhu, Yixin and Wang, Xiaochun and Bi, Yanchao},
	journal={ArXiv Preprint ArXiv:2501.13628},
	year={2025}
}

@incollection{emmorey2014language,
	title={Language in our hands: Neural underpinnings of sign language and co-speech gesture},
	author={Emmorey, Karen and Ozyurek, Asli},
	booktitle={The Cognitive Neurosciences},
	pages={657--666},
	year={2014},
	publisher={MIT Press}
}

@book{patel2010music,
	title={Music, language, and the brain},
	author={Patel, Aniruddh D},
	year={2010},
	publisher={Oxford university press}
}

@article{koelsch2014brain,
	title={Brain correlates of music-evoked emotions},
	author={Koelsch, Stefan},
	journal={Nature Reviews Neuroscience},
	volume={15},
	number={3},
	pages={170--180},
	year={2014},
	publisher={Nature Publishing Group UK London}
}

@article{heard2020shared,
	title={Shared neural resources of rhythm and syntax: An ALE meta-analysis},
	author={Heard, Matthew and Lee, Yune S},
	journal={Neuropsychologia},
	volume={137},
	pages={107284},
	year={2020},
	publisher={Elsevier}
}

@article{hickok2004dorsal,
	title={Dorsal and ventral streams: A framework for understanding aspects of the functional anatomy of language},
	author={Hickok, Gregory and Poeppel, David},
	journal={Cognition},
	volume={92},
	number={1-2},
	pages={67--99},
	year={2004},
	publisher={Elsevier}
}

@article{giglio2022commonalities,
	title={Commonalities and asymmetries in the neurobiological infrastructure for language production and comprehension},
	author={Giglio, Laura and Ostarek, Markus and Weber, Kirsten and Hagoort, Peter},
	journal={Cerebral Cortex},
	volume={32},
	number={7},
	pages={1405--1418},
	year={2022},
	publisher={Oxford Academic}
}

@article{stephens2010speaker,
	title={Speaker--listener neural coupling underlies successful communication},
	author={Stephens, Greg J and Silbert, Lauren J and Hasson, Uri},
	journal={Proceedings of the National Academy of Sciences},
	volume={107},
	number={32},
	pages={14425--14430},
	year={2010},
	publisher={National Academy of Sciences}
}

@article{redcay2019using,
	title={Using second-person neuroscience to elucidate the mechanisms of social interaction},
	author={Redcay, Elizabeth and Schilbach, Leonhard},
	journal={Nature Reviews Neuroscience},
	volume={20},
	number={8},
	pages={495--505},
	year={2019},
	publisher={Nature Publishing Group UK London}
}

@article{aili2025cortical,
	title={The cortical architecture representing the linguistic hierarchy of the conversational speech},
	author={Aili, Ruhuiya and Zhou, Siyuan and Xu, Xinran and He, Xiangyu and Lu, Chunming},
	journal={Neuroimage},
	volume={311},
	pages={121180},
	year={2025}
}

@article{goldstein2025temporal,
	title={Temporal structure of natural language processing in the human brain corresponds to layered hierarchy of large language models},
	author={Goldstein, Ariel and Ham, Eric and Schain, Mariano and Nastase, Samuel A and Aubrey, Bobbi and Zada, Zaid and Grinstein-Dabush, Avigail and Gazula, Harshvardhan and Feder, Amir and Doyle, Werner and others},
	journal={Nature Communications},
	volume={16},
	number={1},
	pages={10529},
	year={2025},
	publisher={Nature Publishing Group UK London}
}

@article{okada2018neural,
	title={Neural evidence for predictive coding in auditory cortex during speech production},
	author={Okada, Kayoko and Matchin, William and Hickok, Gregory},
	journal={Psychonomic Bulletin \& Review},
	volume={25},
	number={1},
	pages={423--430},
	year={2018},
	publisher={Springer}
}

@article{pickering2013integrated,
	title={An integrated theory of language production and comprehension},
	author={Pickering, Martin J and Garrod, Simon},
	journal={Behavioral and Brain Sciences},
	volume={36},
	number={4},
	pages={329--347},
	year={2013},
	publisher={Cambridge University Press}
}

@incollection{fujii2023neural,
	title={Neural basis of language, a comprehensive update for neurosurgeons},
	author={Fujii, Masazumi and Bakhit, Mudathir},
	booktitle={Functional Anatomy of the Brain: A View from the Surgeon's Eye},
	pages={135--173},
	year={2023},
	publisher={Springer}
}

@article{levelt1999theory,
	title={A theory of lexical access in speech production},
	author={Levelt, Willem JM and Roelofs, Ardi and Meyer, Antje S},
	journal={Behavioral and Brain Sciences},
	volume={22},
	number={1},
	pages={1--38},
	year={1999},
	publisher={Cambridge University Press}
}

@article{hirsch2018cross,
	title={A cross-brain neural mechanism for human-to-human verbal communication},
	author={Hirsch, Joy and Adam Noah, J and Zhang, Xian and Dravida, Swethasri and Ono, Yumie},
	journal={Social Cognitive and Affective Neuroscience},
	volume={13},
	number={9},
	pages={907--920},
	year={2018},
	publisher={Oxford University Press}
}

@article{dumas2010inter,
	title={Inter-brain synchronization during social interaction},
	author={Dumas, Guillaume and Nadel, Jacqueline and Soussignan, Robert and Martinerie, Jacques and Garnero, Line},
	journal={PloS One},
	volume={5},
	number={8},
	pages={e12166},
	year={2010},
	publisher={Public Library of Science San Francisco, USA}
}

@book{chomsky2014aspects,
	title={Aspects of the theory of syntax},
	author={Chomsky, Noam},
	number={11},
	year={2014},
	publisher={MIT press}
}

@article{friederici2011brain,
	title={The brain basis of language processing: From structure to function},
	author={Friederici, Angela D},
	journal={Physiological Reviews},
	volume={91},
	number={4},
	pages={1357--1392},
	year={2011},
	publisher={American Physiological Society Bethesda, MD}
}

@article{golestani2014brain,
	title={Brain structural correlates of individual differences at low-to high-levels of the language processing hierarchy: A review of new approaches to imaging research},
	author={Golestani, Narly},
	journal={International Journal of Bilingualism},
	volume={18},
	number={1},
	pages={6--34},
	year={2014},
	publisher={SAGE Publications Sage UK: London, England}
}

@article{linzen2016assessing,
	title={Assessing the ability of LSTMs to learn syntax-sensitive dependencies},
	author={Linzen, Tal and Dupoux, Emmanuel and Goldberg, Yoav},
	journal={Transactions of the Association for Computational Linguistics},
	volume={4},
	pages={521--535},
	year={2016},
	publisher={MIT Press One Rogers Street, Cambridge, MA 02142-1209, USA journals-info~…}
}

@article{ding2016cortical,
	title={Cortical tracking of hierarchical linguistic structures in connected speech},
	author={Ding, Nai and Melloni, Lucia and Zhang, Hang and Tian, Xing and Poeppel, David},
	journal={Nature Neuroscience},
	volume={19},
	number={1},
	pages={158--164},
	year={2016},
	publisher={Nature Publishing Group}
}

@article{pallier2011cortical,
	title={Cortical representation of the constituent structure of sentences},
	author={Pallier, Christophe and Devauchelle, Anne-Dominique and Dehaene, Stanislas},
	journal={Proceedings of the National Academy of Sciences},
	volume={108},
	number={6},
	pages={2522--2527},
	year={2011},
	publisher={National Academy of Sciences}
}

@article{dehaene2011experimental,
	title={Experimental and theoretical approaches to conscious processing},
	author={Dehaene, Stanislas and Changeux, Jean-Pierre},
	journal={Neuron},
	volume={70},
	number={2},
	pages={200--227},
	year={2011},
	publisher={Elsevier}
}

@article{yu2024predicting,
	title={Predicting the next sentence (not word) in large language models: What model-brain alignment tells us about discourse comprehension},
	author={Yu, Shaoyun and Gu, Chanyuan and Huang, Kexin and Li, Ping},
	journal={Science Advances},
	volume={10},
	number={21},
	pages={eadn7744},
	year={2024},
	publisher={American Association for the Advancement of Science}
}

@article{wang2023predictive,
	title={Predictive coding across the left fronto-temporal hierarchy during language comprehension},
	author={Wang, Lin and Schoot, Lotte and Brothers, Trevor and Alexander, Edward and Warnke, Lena and Kim, Minjae and Khan, Sheraz and H{\"a}m{\"a}l{\"a}inen, Matti and Kuperberg, Gina R},
	journal={Cerebral Cortex},
	volume={33},
	number={8},
	pages={4478--4497},
	year={2023},
	publisher={Oxford University Press}
}

@article{lake2017building,
	title={Building machines that learn and think like people},
	author={Lake, Brenden M and Ullman, Tomer D and Tenenbaum, Joshua B and Gershman, Samuel J},
	journal={Behavioral and Brain Sciences},
	volume={40},
	pages={e253},
	year={2017},
	publisher={Cambridge University Press}
}

@article{frankish2010dual,
	title={Dual-process and dual-system theories of reasoning},
	author={Frankish, Keith},
	journal={Philosophy Compass},
	volume={5},
	number={10},
	pages={914--926},
	year={2010},
	publisher={Wiley Online Library}
}

@article{kumar2021semantic,
	title={Semantic memory: A review of methods, models, and current challenges},
	author={Kumar, Abhilasha A},
	journal={Psychonomic Bulletin \& Review},
	volume={28},
	number={1},
	pages={40--80},
	year={2021},
	publisher={Springer}
}

@article{russin2025parallel,
	title={Parallel trade-offs in human cognition and neural networks: The dynamic interplay between in-context and in-weight learning},
	author={Russin, Jacob and Pavlick, Ellie and Frank, Michael J},
	journal={Proceedings of the National Academy of Sciences},
	volume={122},
	number={35},
	pages={e2510270122},
	year={2025},
	publisher={National Academy of Sciences}
}

@article{dherin2025learning,
	title={Learning without training: The implicit dynamics of in-context learning},
	author={Dherin, Benoit and Munn, Michael and Mazzawi, Hanna and Wunder, Michael and Gonzalvo, Javier},
	journal={ArXiv Preprint ArXiv:2507.16003},
	year={2025}
}

@inproceedings{rafiuddin2025learning,
	title={Learning what to remember: Adaptive probabilistic memory retention for memory-efficient language models},
	author={Rafiuddin, SM and Khan, Muntaha Nujat},
	booktitle={Findings of the Association for Computational Linguistics},
	pages={3969--3981},
	year={2025}
}

@inproceedings{muttakhiroh2025tackling,
	title={Tackling distribution shift in LLM via KILO: Knowledge-instructed learning for continual adaptation},
	author={Muttakhiroh, Ling and Fevens, Thomas},
	booktitle={2025 IEEE 35th International Workshop on Machine Learning for Signal Processing},
	pages={1--6},
	year={2025},
	organization={IEEE}
}

@article{kuhl2004early,
	title={Early language acquisition: cracking the speech code},
	author={Kuhl, Patricia K},
	journal={Nature Reviews Neuroscience},
	volume={5},
	number={11},
	pages={831--843},
	year={2004},
	publisher={Nature Publishing Group}
}

@article{grossberg2000complementary,
	title={The complementary brain: Unifying brain dynamics and modularity},
	author={Grossberg, Stephen},
	journal={Trends in Cognitive Sciences},
	volume={4},
	number={6},
	pages={233--246},
	year={2000},
	publisher={Elsevier}
}

@article{kraljic2006generalization,
	title={Generalization in perceptual learning for speech},
	author={Kraljic, Tanya and Samuel, Arthur G},
	journal={Psychonomic Bulletin \& Review},
	volume={13},
	number={2},
	pages={262--268},
	year={2006},
	publisher={Springer}
}

@article{clarke2004rapid,
	title={Rapid adaptation to foreign-accented English},
	author={Clarke, Constance M and Garrett, Merrill F},
	journal={The Journal of the Acoustical Society of America},
	volume={116},
	number={6},
	pages={3647--3658},
	year={2004},
	publisher={Acoustical Society of America}
}

@article{mcclelland1986trace,
	title={The TRACE model of speech perception},
	author={McClelland, James L and Elman, Jeffrey L},
	journal={Cognitive Psychology},
	volume={18},
	number={1},
	pages={1--86},
	year={1986},
	publisher={Elsevier}
}

@article{lipkin2022probabilistic,
	title={Probabilistic atlas for the language network based on precision fMRI data from> 800 individuals},
	author={Lipkin, Benjamin and Tuckute, Greta and Affourtit, Josef and Small, Hannah and Mineroff, Zachary and Kean, Hope and Jouravlev, Olessia and Rakocevic, Lara and Pritchett, Brianna and Siegelman, Matthew and others},
	journal={Scientific Data},
	volume={9},
	number={1},
	pages={529},
	year={2022},
	publisher={Nature Publishing Group UK London}
}

@article{spalding2025shared,
	title={Shared latent representations of speech production for cross-patient speech decoding},
	author={Spalding, Z and Duraivel, S and Rahimpour, S and Wang, C and Barth, K and Schmitz, C and Lad, SP and Friedman, AH and Southwell, DG and Viventi, J and others},
	journal={BioRxiv},
	year={2025}
}

@article{singh2025transfer,
	title={Transfer learning via distributed brain recordings enables reliable speech decoding},
	author={Singh, Aditya and Thomas, Tessy and Li, Jinlong and Hickok, Greg and Pitkow, Xaq and Tandon, Nitin},
	journal={Nature Communications},
	volume={16},
	number={1},
	pages={8749},
	year={2025},
	publisher={Nature Publishing Group UK London}
}

@article{willett2023high,
	title={A high-performance speech neuroprosthesis},
	author={Willett, Francis R and Kunz, Erin M and Fan, Chaofei and Avansino, Donald T and Wilson, Guy H and Choi, Eun Young and Kamdar, Foram and Glasser, Matthew F and Hochberg, Leigh R and Druckmann, Shaul and others},
	journal={Nature},
	volume={620},
	number={7976},
	pages={1031--1036},
	year={2023},
	publisher={Nature Publishing Group UK London}
}

@article{metzger2023high,
	title={A high-performance neuroprosthesis for speech decoding and avatar control},
	author={Metzger, Sean L and Littlejohn, Kaylo T and Silva, Alexander B and Moses, David A and Seaton, Margaret P and Wang, Ran and Dougherty, Maximilian E and Liu, Jessie R and Wu, Peter and Berger, Michael A and others},
	journal={Nature},
	volume={620},
	number={7976},
	pages={1037--1046},
	year={2023},
	publisher={Nature Publishing Group UK London}
}

@article{wang2026model,
	title={Model-agnostic linear-memory online learning in spiking neural networks},
	author={Wang, Chaoming and Dong, Xingsi and Ji, Zilong and Xiao, Mingqing and Jiang, Jiedong and Liu, Xiao and Huan, Yuxiang and Wu, Si},
	journal={Nature Communications},
	year={2026},
	publisher={Nature Publishing Group UK London}
}

@article{pan2025spikingbrain,
	title={Spikingbrain technical report: Spiking brain-inspired large models},
	author={Pan, Yuqi and Feng, Yupeng and Zhuang, Jinghao and Ding, Siyu and Liu, Zehao and Sun, Bohan and Chou, Yuhong and Xu, Han and Qiu, Xuerui and Deng, Anlin and others},
	journal={ArXiv Preprint ArXiv:2509.05276},
	year={2025}
}

@article{bhaya2026shared,
	title={Shared and language-specific phonological processing in the human temporal lobe},
	author={Bhaya-Grossman, Ilina and Leonard, Matthew K and Zhang, Yizhen and Gwilliams, Laura and Johnson, Keith and Lu, Junfeng and Chang, Edward F},
	journal={Nature},
	volume={649},
	number={8095},
	pages={140--151},
	year={2026},
	publisher={Nature Publishing Group}
}

@article{quaresima2012brief,
	title={A brief review on the use of functional near-infrared spectroscopy (fNIRS) for language imaging studies in human newborns and adults},
	author={Quaresima, Valentina and Bisconti, Silvia and Ferrari, Marco},
	journal={Brain and Language},
	volume={121},
	number={2},
	pages={79--89},
	year={2012},
	publisher={Elsevier}
}

@book{tsunashima2012measurement,
	title={Measurement of brain function using near-infrared spectroscopy (NIRS)},
	author={Tsunashima, Hitoshi and Yanagisawa, Kazuki and Iwadate, Masako},
	year={2012},
	publisher={InTech}
}

@article{villringer1997non,
	title={Non-invasive optical spectroscopy and imaging of human brain function},
	author={Villringer, Arno and Chance, Britton},
	journal={Trends in neurosciences},
	volume={20},
	number={10},
	pages={435--442},
	year={1997},
	publisher={Elsevier}
}

@article{scholkmann2014review,
	title={A review on continuous wave functional near-infrared spectroscopy and imaging instrumentation and methodology},
	author={Scholkmann, Felix and Kleiser, Stefan and Metz, Andreas Jaakko and Zimmermann, Raphael and Pavia, Juan Mata and Wolf, Ursula and Wolf, Martin},
	journal={Neuroimage},
	volume={85},
	pages={6--27},
	year={2014},
	publisher={Elsevier}
}

@article{ferrari2012brief,
	title={A brief review on the history of human functional near-infrared spectroscopy (fNIRS) development and fields of application},
	author={Ferrari, Marco and Quaresima, Valentina},
	journal={Neuroimage},
	volume={63},
	number={2},
	pages={921--935},
	year={2012},
	publisher={Elsevier}
}

@article{leon2012functional,
	title={Functional near-infrared spectroscopy (fNIRS): principles and neuroscientific applications},
	author={Le{\'o}n-Carri{\'o}n, Jos{\'e} and Le{\'o}n-Dom{\'\i}nguez, Umberto},
	journal={Neuroimaging Methods},
	volume={97},
	pages={48--74},
	year={2012},
	publisher={IntechOpen}
}

@article{pinti2020present,
	title={The present and future use of functional near-infrared spectroscopy (fNIRS) for cognitive neuroscience},
	author={Pinti, Paola and Tachtsidis, Ilias and Hamilton, Antonia and Hirsch, Joy and Aichelburg, Clarisse and Gilbert, Sam and Burgess, Paul W},
	journal={Annals of the new York Academy of Sciences},
	volume={1464},
	number={1},
	pages={5--29},
	year={2020},
	publisher={Wiley Online Library}
}

@article{li2025tale,
  title={A tale of single-channel electroencephalogram: Devices, datasets, signal processing, applications, and future directions},
  author={Li, Yueyang and Zeng, Weiming and Dong, Wenhao and Han, Di and Chen, Lei and Chen, Hongyu and Kang, Zijian and Gong, Shengyu and Yan, Hongjie and Siok, Wai Ting and others},
  journal={IEEE Transactions on Instrumentation and Measurement},
  year={2025},
  publisher={IEEE}
}

@article{evans2009myth,
  title={The myth of language universals: Language diversity and its importance for cognitive science},
  author={Evans, Nicholas and Levinson, Stephen C},
  journal={Behavioral and Brain Sciences},
  volume={32},
  number={5},
  pages={429--448},
  year={2009},
  publisher={Cambridge University Press}
}

@book{chomsky2002syntactic,
  title={Syntactic structures},
  author={Chomsky, Noam},
  year={2002},
  publisher={Walter de Gruyter}
}

@book{jackendoff1972semantic,
  title={Semantic interpretation in generative grammar.},
  author={Jackendoff, Ray S},
  year={1972},
  publisher={ERIC}
}

@article{lasnik2002minimalist,
  title={The minimalist program in syntax},
  author={Lasnik, Howard},
  journal={Trends in Cognitive Sciences},
  volume={6},
  number={10},
  pages={432--437},
  year={2002},
  publisher={Elsevier}
}

@misc{urban1981language,
  title={Language as social semiotic: The social interpretation of language and meaning},
  author={Urban, Greg},
  year={1981},
  publisher={JSTOR}
}

@book{halliday2013halliday,
  title={Halliday's introduction to functional grammar},
  author={Halliday, Michael Alexander Kirkwood and Matthiessen, Christian MIM},
  year={2013},
  publisher={Routledge}
}

@book{eggins2004introduction,
  title={Introduction to systemic functional linguistics},
  author={Eggins, Suzanne},
  year={2004},
  publisher={A\&c Black}
}

@book{lakoff2008metaphors,
  title={Metaphors we live by},
  author={Lakoff, George and Johnson, Mark},
  year={2008},
  publisher={University of Chicago press}
}

@book{kovecses2010metaphor,
  title={Metaphor: A practical introduction},
  author={Kovecses, Zoltan},
  year={2010},
  publisher={Oxford university press}
}

@article{barsalou2010grounded,
  title={Grounded cognition: Past, present, and future},
  author={Barsalou, Lawrence W},
  journal={Topics in Cognitive Science},
  volume={2},
  number={4},
  pages={716--724},
  year={2010},
  publisher={Wiley Online Library}
}

@book{newmeyer2005possible,
  title={Possible and probable languages: A generative perspective on linguistic typology},
  author={Newmeyer, Frederick J},
  year={2005},
  publisher={OUP Oxford}
}

@book{chomsky1993lectures,
  title={Lectures on government and binding: The Pisa lectures},
  author={Chomsky, Noam},
  number={9},
  year={1993},
  publisher={Walter de Gruyter}
}

@book{croft2002typology,
  title={Typology and universals},
  author={Croft, William},
  year={2002},
  publisher={Cambridge university press}
}

@book{bartlett2017routledge,
  title={The Routledge handbook of systemic functional linguistics},
  author={Bartlett, Tom and O'Grady, Gerard},
  year={2017},
  publisher={Routledge London}
}

@book{kovecses2005metaphor,
  title={Metaphor in culture: Universality and variation},
  author={K{\"o}vecses, Zolt{\'a}n},
  year={2005},
  publisher={Cambridge university press}
}

@book{christiansen2018creating,
  title={Creating language: Integrating evolution, acquisition, and processing},
  author={Christiansen, Morten H and Chater, Nick},
  year={2018},
  publisher={Mit Press}
}

@article{van2023cleaning,
  title={Cleaning up the brickyard: How theory and methodology shape experiments in cognitive neuroscience of language},
  author={Van Der Burght, Constantijn L and Friederici, Angela D and Maran, Matteo and Papitto, Giorgio and Pyatigorskaya, Elena and Schro{\"e}n, Jo{\"e}lle AM and Trettenbrein, Patrick C and Zaccarella, Emiliano},
  journal={Journal of Cognitive Neuroscience},
  volume={35},
  number={12},
  pages={2067--2088},
  year={2023},
  publisher={MIT Press One Broadway, 12th Floor, Cambridge, Massachusetts 02142, USA~…}
}

@article{dambacher2006frequency,
  title={Frequency and predictability effects on event-related potentials during reading},
  author={Dambacher, Michael and Kliegl, Reinhold and Hofmann, Markus and Jacobs, Arthur M},
  journal={Brain Research},
  volume={1084},
  number={1},
  pages={89--103},
  year={2006},
  publisher={Elsevier}
}

@article{brothers2015effects,
  title={Effects of prediction and contextual support on lexical processing: Prediction takes precedence},
  author={Brothers, Trevor and Swaab, Tamara Y and Traxler, Matthew J},
  journal={Cognition},
  volume={136},
  pages={135--149},
  year={2015},
  publisher={Elsevier}
}

@article{verbeke2024role,
  title={On the role of ecological validity in language and speech research.},
  author={Verbeke, Gil and Buysschaert, J and Lef{\`e}vre, A},
  journal={Taalkunde nu. Gent: Skribis. Series Studia Germanica Gandensia (Libri) \& Spieghel Historiael},
  pages={69--95},
  year={2024}
}

@book{cutler2005twenty,
  title={Twenty-first century psycholinguistics: Four cornerstones},
  author={Cutler, Anne},
  year={2005},
  publisher={Psychology Press}
}

@article{drijvers2025language,
  title={Language is widely distributed throughout the brain},
  author={Drijvers, Linda and Small, Steven L and Skipper, Jeremy I},
  journal={Nature Reviews Neuroscience},
  volume={26},
  number={3},
  pages={189--189},
  year={2025},
  publisher={Nature Publishing Group UK London}
}

@article{pickering2004toward,
  title={Toward a mechanistic psychology of dialogue},
  author={Pickering, Martin J and Garrod, Simon},
  journal={Behavioral and Brain Sciences},
  volume={27},
  number={2},
  pages={169--190},
  year={2004},
  publisher={Cambridge University Press}
}

@article{pickering2018predicting,
  title={Predicting while comprehending language: A theory and review.},
  author={Pickering, Martin J and Gambi, Chiara},
  journal={Psychological Bulletin},
  volume={144},
  number={10},
  pages={1002},
  year={2018},
  publisher={American Psychological Association}
}

@article{siok2004biological,
  title={Biological abnormality of impaired reading is constrained by culture},
  author={Siok, Wai Ting and Perfetti, Charles A and Jin, Zhen and Tan, Li Hai},
  journal={Nature},
  volume={431},
  number={7004},
  pages={71--76},
  year={2004},
  publisher={Nature Publishing Group UK London}
}

@article{zhou2022spikformer,
  title={Spikformer: When spiking neural network meets transformer},
  author={Zhou, Zhaokun and Zhu, Yuesheng and He, Chao and Wang, Yaowei and Yan, Shuicheng and Tian, Yonghong and Yuan, Li},
  journal={ArXiv Preprint ArXiv:2209.15425},
  year={2022}
}

@inproceedings{wang2023spatial,
  title={Spatial-temporal Self-attention for asynchronous spiking neural networks.},
  author={Wang, Yuchen and Shi, Kexin and Lu, Chengzhuo and Liu, Yuguo and Zhang, Malu and Qu, Hong},
  booktitle={IJCAI},
  pages={3085--3093},
  year={2023}
}

@article{wu2024adaptive,
  title={Adaptive spatiotemporal neural networks through complementary hybridization},
  author={Wu, Yujie and Shi, Bizhao and Zheng, Zhong and Zheng, Hanle and Yu, Fangwen and Liu, Xue and Luo, Guojie and Deng, Lei},
  journal={Nature Communications},
  volume={15},
  number={1},
  pages={7355},
  year={2024},
  publisher={Nature Publishing Group UK London}
}

@article{hauser2002faculty,
  title={The faculty of language: what is it, who has it, and how did it evolve?},
  author={Hauser, Marc D and Chomsky, Noam and Fitch, W Tecumseh},
  journal={Science},
  volume={298},
  number={5598},
  pages={1569--1579},
  year={2002},
  publisher={American Association for the Advancement of Science}
}

@incollection{chomsky2014minimal,
  title={Minimal recursion: Exploring the prospects},
  author={Chomsky, Noam},
  booktitle={Recursion: Complexity in cognition},
  pages={1--15},
  year={2014},
  publisher={Springer}
}

@article{wu2023occupation,
  title={Occupation-modulated language networks and its lateralization: A resting-state fMRI study of seafarers},
  author={Wu, Huijun and Peng, Deyuan and Yan, Hongjie and Yang, Yang and Xu, Min and Zeng, Weiming and Chang, Chunqi and Wang, Nizhuan},
  journal={Frontiers in Human Neuroscience},
  volume={17},
  pages={1095413},
  year={2023},
  publisher={Frontiers Media SA}
}

@article{indefrey2004spatial,
  title={The spatial and temporal signatures of word production components},
  author={Indefrey, Peter and Levelt, Willem JM},
  journal={Cognition},
  volume={92},
  number={1-2},
  pages={101--144},
  year={2004},
  publisher={Elsevier}
}

@article{silbert2014coupled,
  title={Coupled neural systems underlie the production and comprehension of naturalistic narrative speech},
  author={Silbert, Lauren J and Honey, Christopher J and Simony, Erez and Poeppel, David and Hasson, Uri},
  journal={Proceedings of the National Academy of Sciences},
  volume={111},
  number={43},
  pages={E4687--E4696},
  year={2014},
  publisher={National Academy of Sciences}
}

@article{hill2012recording,
  title={Recording human electrocorticographic (ECoG) signals for neuroscientific research and real-time functional cortical mapping},
  author={Hill, N Jeremy and Gupta, Disha and Brunner, Peter and Gunduz, Aysegul and Adamo, Matthew A and Ritaccio, Anthony and Schalk, Gerwin},
  journal={Journal of Visualized Experiments},
  number={64},
  pages={e3993},
  year={2012}
}

@article{kanth2020electrocorticogram,
  title={Electrocorticogram (ECoG) is highly informative in primate visual cortex},
  author={Kanth, Sidrat Tasawoor and Ray, Supratim},
  journal={Journal of Neuroscience},
  volume={40},
  number={12},
  pages={2430--2444},
  year={2020},
  publisher={Society for Neuroscience}
}

@article{de2025intraoperative,
  title={Intraoperative techniques for language mapping in brain surgery: A comparison between direct electrical stimulation (DES) and electrocorticography (ECoG)},
  author={de Camargo, Patricia Silva and Santos e Souza, Giovanna de Oliveira and Ar{\'e}valo, Anal{\'\i}a and Lepski, Guilherme},
  journal={Brain and Behavior},
  volume={15},
  number={10},
  pages={e70900},
  year={2025},
  publisher={Wiley Online Library}
}

@article{youngerman2019stereoelectroencephalography,
  title={Stereoelectroencephalography in epilepsy, cognitive neurophysiology, and psychiatric disease: Safety, efficacy, and place in therapy},
  author={Youngerman, Brett E and Khan, Farhan A and McKhann, Guy M},
  journal={Neuropsychiatric Disease and Treatment},
  pages={1701--1716},
  year={2019},
  publisher={Taylor \& Francis}
}

@article{aron2021language,
  title={Language mapping using stereo electroencephalography: a review and expert opinion},
  author={Aron, Olivier and Jonas, Jacques and Colnat-Coulbois, Sophie and Maillard, Louis},
  journal={Frontiers in Human Neuroscience},
  volume={15},
  pages={619521},
  year={2021},
  publisher={Frontiers Media SA}
}

@article{zhao2024spatiotemporal,
  title={Spatiotemporal neural network for sublexical information processing: An intracranial SEEG study},
  author={Zhao, Chunyu and Liu, Yi and Zeng, Jiahong and Luo, Xiangqi and Sun, Weijin and Luan, Guoming and Liu, Yuxin and Zhang, Yumei and Shi, Gaofeng and Guan, Yuguang and others},
  journal={Journal of Neuroscience},
  volume={44},
  number={45},
  year={2024},
  publisher={Society for Neuroscience}
}

@article{parvizi2018promises,
  title={Promises and limitations of human intracranial electroencephalography},
  author={Parvizi, Josef and Kastner, Sabine},
  journal={Nature Neuroscience},
  volume={21},
  number={4},
  pages={474--483},
  year={2018},
  publisher={Nature Publishing Group US New York}
}

@article{oganian2019speech,
  title={A speech envelope landmark for syllable encoding in human superior temporal gyrus},
  author={Oganian, Yulia and Chang, Edward F},
  journal={Science Advances},
  volume={5},
  number={11},
  pages={eaay6279},
  year={2019},
  publisher={American Association for the Advancement of Science}
}

@article{moses2016neural,
  title={Neural speech recognition: continuous phoneme decoding using spatiotemporal representations of human cortical activity},
  author={Moses, David A and Mesgarani, Nima and Leonard, Matthew K and Chang, Edward F},
  journal={Journal of Neural Engineering},
  volume={13},
  number={5},
  pages={056004},
  year={2016},
  publisher={IOP Publishing}
}

@article{pasley2012reconstructing,
  title={Reconstructing speech from human auditory cortex},
  author={Pasley, Brian N and David, Stephen V and Mesgarani, Nima and Flinker, Adeen and Shamma, Shihab A and Crone, Nathan E and Knight, Robert T and Chang, Edward F},
  journal={PLoS Biology},
  volume={10},
  number={1},
  pages={e1001251},
  year={2012},
  publisher={Public Library of Science San Francisco, USA}
}

@article{broderick2018electrophysiological,
  title={Electrophysiological correlates of semantic dissimilarity reflect the comprehension of natural, narrative speech},
  author={Broderick, Michael P and Anderson, Andrew J and Di Liberto, Giovanni M and Crosse, Michael J and Lalor, Edmund C},
  journal={Current Biology},
  volume={28},
  number={5},
  pages={803--809},
  year={2018},
  publisher={Elsevier}
}

@InProceedings{pmlr-v139-radford21a,
  title = 	 {Learning Transferable Visual Models From Natural Language Supervision},
  author =       {Radford, Alec and Kim, Jong Wook and Hallacy, Chris and Ramesh, Aditya and Goh, Gabriel and Agarwal, Sandhini and Sastry, Girish and Askell, Amanda and Mishkin, Pamela and Clark, Jack and Krueger, Gretchen and Sutskever, Ilya},
  booktitle = 	 {Proceedings of the 38th International Conference on Machine Learning},
  pages = 	 {8748--8763},
  year = 	 {2021},
  editor = 	 {Meila, Marina and Zhang, Tong},
  volume = 	 {139},
  series = 	 {Proceedings of Machine Learning Research},
  month = 	 {18--24 Jul},
  publisher =    {PMLR}
}

@article{bhattacharjee2026aligning,
  title={Aligning brains into a shared space improves their alignment with large language models},
  author={Bhattacharjee, Arnab and Zada, Zaid and Wang, Haocheng and Aubrey, Bobbi and Doyle, Werner and Dugan, Patricia and Friedman, Daniel and Devinsky, Orrin and Flinker, Adeen and Ramadge, Peter J and others},
  journal={Nature Computational Science},
  volume={6},
  number={2},
  pages={169--178},
  year={2026},
  publisher={Nature Publishing Group US New York}
}

@article{zhang2025neural,
  title={Neural heterogeneity as a unifying mechanism for efficient learning in spiking neural networks},
  author={Zhang, Fudong and Cui, Jingjing},
  journal={Frontiers in Computational Neuroscience},
  volume={19},
  pages={1661070},
  year={2025},
  publisher={Frontiers Media SA}
}

@article{nili2014toolbox,
  title={A toolbox for representational similarity analysis},
  author={Nili, Hamed and Wingfield, Cai and Walther, Alexander and Su, Li and Marslen-Wilson, William and Kriegeskorte, Nikolaus},
  journal={PLoS computational biology},
  volume={10},
  number={4},
  pages={e1003553},
  year={2014},
  publisher={Public Library of Science San Francisco, USA}
}

@Book{martin2009speech,
  author = {Daniel Jurafsky and James H. Martin},
  title ={Speech and language processing: An introduction to natural language processing, 
  		  computational linguistics, and speech recognition,
		   with language models},
  year =         {2026},

  }

@article{bellec2021fitting,
  title={Fitting summary statistics of neural data with a differentiable spiking network simulator},
  author={Bellec, Guillaume and Wang, Shuqi and Modirshanechi, Alireza and Brea, Johanni and Gerstner, Wulfram},
  journal={Advances in Neural Information Processing Systems},
  volume={34},
  pages={18552--18563},
  year={2021}
}

@article{mischler2024contextual,
  title={Contextual feature extraction hierarchies converge in large language models and the brain},
  author={Mischler, Gavin and Li, Yinghao Aaron and Bickel, Stephan and Mehta, Ashesh D and Mesgarani, Nima},
  journal={Nature Machine Intelligence},
  volume={6},
  number={12},
  pages={1467--1477},
  year={2024},
  publisher={Nature Publishing Group UK London}
}

@article{goldstein2025unified,
  title={A unified acoustic-to-speech-to-language embedding space captures the neural basis of natural language processing in everyday conversations},
  author={Goldstein, Ariel and Wang, Haocheng and Niekerken, Leonard and Schain, Mariano and Zada, Zaid and Aubrey, Bobbi and Sheffer, Tom and Nastase, Samuel A and Gazula, Harshvardhan and Singh, Aditi and others},
  journal={Nature Human Behaviour},
  volume={9},
  number={5},
  pages={1041--1055},
  year={2025},
  publisher={Nature Publishing Group UK London}
}

@article{gnanateja2025cortical,
  title={Cortical processing of discrete prosodic patterns in continuous speech},
  author={Gnanateja, G Nike and Rupp, Kyle and Llanos, Fernando and Hect, Jasmine and German, James S and Teichert, Tobias and Abel, Taylor J and Chandrasekaran, Bharath},
  journal={Nature Communications},
  volume={16},
  number={1},
  pages={1947},
  year={2025},
  publisher={Nature Publishing Group UK London}
}

@article{khanna2024single,
  title={Single-neuronal elements of speech production in humans},
  author={Khanna, Arjun R and Mu{\~n}oz, William and Kim, Young Joon and Kfir, Yoav and Paulk, Angelique C and Jamali, Mohsen and Cai, Jing and Mustroph, Martina L and Caprara, Irene and Hardstone, Richard and others},
  journal={Nature},
  volume={626},
  number={7999},
  pages={603--610},
  year={2024},
  publisher={Nature Publishing Group UK London}
}

@article{malik2022investigation,
  title={An investigation across 45 languages and 12 language families reveals a universal language network},
  author={Malik-Moraleda, Saima and Ayyash, Dima and Gall{\'e}e, Jeanne and Affourtit, Josef and Hoffmann, Malte and Mineroff, Zachary and Jouravlev, Olessia and Fedorenko, Evelina},
  journal={Nature neuroscience},
  volume={25},
  number={8},
  pages={1014--1019},
  year={2022},
  publisher={Nature Publishing Group US New York}
}

@inproceedings{joshi2020state,
  title={The state and fate of linguistic diversity and inclusion in the NLP world},
  author={Joshi, Pratik and Santy, Sebastin and Budhiraja, Amar and Bali, Kalika and Choudhury, Monojit},
  booktitle={Proceedings of the 58th annual meeting of the association for computational linguistics},
  pages={6282--6293},
  year={2020}
}

@article{pylkkanen2019neural,
  title={The neural basis of combinatory syntax and semantics},
  author={Pylkk{\"a}nen, Liina},
  journal={Science},
  volume={366},
  number={6461},
  pages={62--66},
  year={2019},
  publisher={American Association for the Advancement of Science}
}

@inproceedings{gao2024measuring,
  title={Measuring meaning composition in the human brain with composition scores from large language models},
  author={Gao, Changjiang and Li, Jixing and Chen, Jiajun and Huang, Shujian},
  booktitle={Proceedings of the 62nd Annual Meeting of the Association for Computational Linguistics (Volume 1: Long Papers)},
  pages={11295--11308},
  year={2024}
}

@article{gao2025increasing,
  title={Increasing alignment of large language models with language processing in the human brain},
  author={Gao, Changjiang and Ma, Zhengwu and Chen, Jiajun and Li, Ping and Huang, Shujian and Li, Jixing},
  journal={Nature computational science},
  volume={5},
  number={11},
  pages={1080--1090},
  year={2025},
  publisher={Nature Publishing Group US New York}
}

@article{li2022petit,
  title={Le Petit Prince multilingual naturalistic fMRI corpus},
  author={Li, Jixing and Bhattasali, Shohini and Zhang, Shulin and Franzluebbers, Berta and Luh, Wen-Ming and Spreng, R Nathan and Brennan, Jonathan R and Yang, Yiming and Pallier, Christophe and Hale, John},
  journal={Scientific data},
  volume={9},
  number={1},
  pages={530},
  year={2022},
  publisher={Nature Publishing Group UK London}
}

@article{chen2026bilingual,
  title={Bilingual language processing relies on shared semantic representations that are modulated by each language},
  author={Chen, Catherine and Gong, Xue L and Tseng, Christine and Klein, Daniel L and Gallant, Jack L and Deniz, Fatma},
  journal={Proceedings of the National Academy of Sciences},
  volume={123},
  number={9},
  pages={e2503721123},
  year={2026},
  publisher={National Academy of Sciences}
}

@article{brennan2012syntactic,
  title={Syntactic structure building in the anterior temporal lobe during natural story listening},
  author={Brennan, Jonathan and Nir, Yuval and Hasson, Uri and Malach, Rafael and Heeger, David J and Pylkk{\"a}nen, Liina},
  journal={Brain and language},
  volume={120},
  number={2},
  pages={163--173},
  year={2012},
  publisher={Elsevier}
}

@article{hasson2012brain,
  title={Brain-to-brain coupling: a mechanism for creating and sharing a social world},
  author={Hasson, Uri and Ghazanfar, Asif A and Galantucci, Bruno and Garrod, Simon and Keysers, Christian},
  journal={Trends in cognitive sciences},
  volume={16},
  number={2},
  pages={114--121},
  year={2012},
  publisher={Elsevier}
}

\end{document}